\documentclass[10pt]{iopart}
\usepackage{graphicx}
\usepackage{cite}

\def\be{\begin{equation}}
\def\ee{\end{equation}}
\def\bea{\begin{eqnarray}}
\def\eea{\end{eqnarray}}

\begin{document}

\topical{Rare region effects at classical, quantum, and nonequilibrium phase transitions}

\author{Thomas Vojta}

\address{Department of Physics, University of Missouri - Rolla, Rolla, MO 65409, USA}

\begin{abstract}
Rare regions, i.e., rare large spatial disorder fluctuations, can dramatically change the
properties of a phase transition in a quenched disordered system. In generic classical
equilibrium systems, they lead to an essential singularity, the so-called Griffiths
singularity, of the free energy in the vicinity of the phase transition. Stronger effects
can be observed at zero-temperature quantum phase transitions, at nonequilibrium phase
transitions, and in systems with correlated disorder. In some cases, rare regions can
actually completely destroy the sharp phase transition by smearing.

This topical review presents a unifying framework for rare region effects at weakly
disordered classical, quantum, and nonequilibrium phase transitions based on the
effective dimensionality of the rare regions. Explicit examples include disordered
classical Ising and Heisenberg models, insulating and metallic random quantum magnets,
and the disordered contact process.

\end{abstract}


\submitto{\JPA}
%\maketitle

\tableofcontents \markboth{Rare region effects at classical, quantum, and nonequilibrium
phase transitions}{}

%%%%%%%%%%%%%%%%%%%%%%%%%%%%%%%%%%%%%%%%%%%%%%%%%%%%%%%%%%%%%%%%%%%%%%%%%%%%%%%%%%%%%%%%%%%%
\section{Preamble}
\label{sec:preamble}
%%%%%%%%%%%%%%%%%%%%%%%%%%%%%%%%%%%%%%%%%%%%%%%%%%%%%%%%%%%%%%%%%%%%%%%%%%%%%%%%%%%%%%%%%%%%

Phase transitions belong to the most fascinating phenomena in nature. The large scale
structure of the universe is the result of a series of phase transitions during the early
stages of its development. At low temperatures, quantum phase transitions between
different ground states lead to unconventional behavior and exotic new phases of matter.
Phase transitions also occur between nonequilibrium states, e.g., during growth processes
or chemical reactions. Even our everyday life is unimaginable without the never ending
transformations of water between ice, liquid and vapor.

Under normal conditions, the phase transitions of water are so-called first-order
transitions. At these transitions, the two phases coexist at the transition point, and a
finite amount of heat (the latent heat) is released when going from one phase to the
other. Transitions that do not involve phase coexistence and latent heat are called
continuous transitions or critical points. They are particularly interesting because the
typical length and time scales of fluctuations diverge when approaching the transition
point. Understanding the resulting singularities of physical observables, the so-called
\emph{critical behavior} has been a prime endeavor in physics, and the concepts
established during this process, viz., scaling and the renormalization group, now belong
to the central paradigms of modern physics.

Many realistic systems contain a certain amount of quenched (i.e., time-independent or
frozen in) disorder. This disorder can take the form of vacancies or impurity atoms in a
crystal lattice, or it can consist of extended defects such as dislocations or grain
boundaries. The question of how quenched disorder influences phase transitions and
critical points is both conceptually very interesting and of fundamental importance for
practical applications. It is therefore not surprising that this question has attracted
considerable attention over the course of the last thirty years or so. Early work often
focused on the average behavior of the disorder on larger and larger length scales.

However, it has become increasingly clear that an important role is played by \emph{rare
but large} disorder fluctuations and the \emph{rare spatial regions} where they occur.
These regions can lead to a singularity (the so-called Griffiths singularity) in the free
energy in an entire parameter region close to the phase transition that is now known as
the Griffiths region or Griffiths phase. In generic classical systems, the thermodynamic
Griffiths singularity is very weak and probably unobservable in experiment. In recent
years, rare region effects have reattracted considerable attention, in particular in
connection with nonequilibrium phase transitions and with zero-temperature quantum phase
transitions. At these transitions, rare regions can have much stronger effects, ranging
from from strong power-law singularities in the free energy to a complete destruction of
the phase transition.

The goal of this review is to provide a unifying framework for rare region effects on
classical, quantum, and nonequilibrium phase transitions with quenched disorder. The core
of the discussion will be limited to (order-disorder) phase transitions between
conventional phases, i.e., we assume that the bulk phases are not changed qualitatively
by the presence of the disorder. The microscopic effect of the impurities and defects
then consists in locally favoring one of the two phases over the other. This type of
quenched disorder is sometimes referred to as \emph{weak} disorder, or random-$T_c$
disorder (because it changes the local critical temperature at a thermal transition), or,
from analogy with quantum field theory, random-mass disorder.

The structure of this review is as follows. In section \ref{sec:pt}, we collect the basic
concepts of phase transitions and critical phenomena necessary for the purpose of this
review. In section \ref{sec:disorder} we give an introduction to the influence of
quenched disorder on phase transitions with particular emphasis on rare region effects.
We then put forward a general classification of weakly disordered phase transitions
according to the effective dimensionality of the rare regions. In the bulk of the review,
consisting of sections \ref{sec:classical}, \ref{sec:quantum}, and \ref{sec:noneq}, we
illustrate these general ideas by discussing in detail rare regions effects at several
classical, quantum, and nonequilibrium phase transitions, respectively. We conclude in
section \ref{sec:conclusions} with a summary and a discussion of experiments and open
questions.

%%%%%%%%%%%%%%%%%%%%%%%%%%%%%%%%%%%%%%%%%%%%%%%%%%%%%%%%%%%%%%%%%%%%%%%%%%%%%%%%%%%%%%%%%%%%
\section{Phase transitions and critical behavior}
\label{sec:pt}
%%%%%%%%%%%%%%%%%%%%%%%%%%%%%%%%%%%%%%%%%%%%%%%%%%%%%%%%%%%%%%%%%%%%%%%%%%%%%%%%%%%%%%%%%%%%

In this section, we briefly collect the basic theoretical concepts of phase transitions
and critical phenomena to the extent necessary for the purpose of this review. More
details can be found in textbooks on this topic such as those by Ma \cite{Ma_book76} or
Goldenfeld \cite{Goldenfeld_book92}.

\subsection{Landau theory}
\label{subsec:Landau}

Most modern theories of phase transitions are built on the foundation of Landau theory
\cite{Landau37a,Landau37b,Landau37c,Landau37d}. Landau introduced the general concept of
an \emph{order parameter}, a thermodynamic quantity that vanishes in one phase (the
disordered phase\footnote{Here "disordered phase" refers to a phase without long-range
order. This is not to be confused with quenched disorder in the form of impurities and
defects.}) and is non-zero in the other phase (the ordered phase). Very often the choice
of an order parameter for a particular transition is obvious, e.g., for the ferromagnetic
transition where the total magnetization is an order parameter. Sometimes, however,
finding an appropriate order parameter is a complicated problem in itself, e.g., for the
disorder-driven localization-delocalization transition of non-interacting electrons.

Landau theory can be understood as the unification of earlier mean-field theories such as
the van-der-Waals theory of the liquid gas transition \cite{Waals73} or Weiss' molecular
field theory of ferromagnetism \cite{Weiss07}. It is based on the crucial assumption that
the free energy is an analytic function of the order parameter $m$ and can thus be
expanded in a power series,
\be
F = F_L(m) = F_0 + r \, m^2 +v\, m^3 + u\, m^4 +O(m^5)~.
\label{eq:Landau}
\ee
Here $r,v,u$ are parameters that depend on all degrees of freedom other then the order
parameter $m$. The physical value of $m$ is the one that minimizes $F_L$.
For sufficiently large $r$, the minimum free energy is always located at $m=0$ (the system is
in the disordered phase), while for sufficiently small $r$, the minimum as at some $m \ne
0$ (ordered phase). If $v\ne 0$, the transition from $m=0$ to $m\ne 0$ occurs
discontinuously, i.e., it is of first order. If $v=0$ (as is often the case by symmetry),
the theory describes a continuous transition or critical point at $r=0$. Thus, $r$ is
measuring the distance from the critical point, $r\propto (T-T_c)$ for a thermal
transition.

Within Landau theory, the qualitative behavior of all critical points is identical.
Specifically, the order parameter vanishes as $m = (-r/2u)^{1/2}$ when the critical
point is approached from the ordered phase. This is an example of the so-called
super-universality of Landau theory: The critical exponent $\beta$ which describes
the singularity of the order parameter at the critical point via $m \propto |r|^\beta$,
is predicted to have the mean-field value $1/2$ for all critical points.
Universality of the critical exponents is actually observed in experiments but it is
weaker than the super-universality predicted by Landau theory, and the exponent values
are in general different from what Landau theory predicts. Moreover, the exponent values
turn out to be dimensionality dependent. For instance, all three-dimensional Ising
ferromagnets have a common $\beta\approx 0.32$ while two-dimensional Ising magnets have
$\beta=1/8$. Three-dimensional ferromagnets with $O(3)$ Heisenberg symmetry also have a
common value of $\beta\approx 0.35$ but it is different from the one in Ising magnets.

The reason for the failure of Landau theory to correctly describe the critical behavior
is that it does not adequately include the fluctuations of the order parameter about its
average value. The effects of these fluctuations in general decrease with increasing
dimensionality and with increasing number of order parameter components. This explains
why the critical behavior of Ising magnets (one order parameter component) deviates more
strongly from Landau theory than that of Heisenberg magnets (three components), and why
the three-dimensional exponents are closer to the mean-field values than those of
two-dimensional systems.

This suggests that Landau theory might actually be correct for systems with sufficiently
high dimensionality $d$. In fact, the fluctuations lead to two different critical
dimensionalities, $d_c^+$ and $d_c^-$ for a given phase transition. If the dimensionality
$d$ is larger than the upper critical dimension $d_c^+$, fluctuations are unimportant for
the leading critical behavior, and Landau theory gives the correct answers. If $d$ is
between the upper and the lower critical dimensions, $d_c^+>d>d_c^-$, a phase transition
still exists but the critical behavior is different from mean-field theory. For
dimensionalities below the lower critical dimension, fluctuations become so strong that
they completely destroy the ordered phase. For the ferromagnetic transition at nonzero
temperature, $d_c^+=4$, and $d_c^-=2$ or 1 for Heisenberg and Ising symmetries,
respectively.

%%%%%%%%%%%%%%%%%%%%%%%%%%%%%%%%%%%%%%%%%%%%%%%%%%%%%%%%%%%%%%%%%%%%%%%%%%%%%%%%%%%%%%%%%%%%
\subsection{Scaling and the renormalization group}
\label{subsec:scaling}

As indicated in the last subsection, below the upper critical dimension $d_c^+$,
order parameter fluctuations play a crucial role in determining the critical behavior.
To include them, one has to generalize the Landau free energy (\ref{eq:Landau}) by writing
the partition function $Z$ as a functional integral
\be
Z=e^{-F/T} = \int D[\phi] e^{-S[\phi]}~,
\ee
with the action or Landau-Ginzburg-Wilson (LGW) functional $S$ given by
\be
S[\phi] = \frac 1 T \int d^d x ~ \left[ c(\nabla\phi(\mathbf{x}))^2 +
F_L(\phi(\mathbf{x})) -h \phi(\mathbf{x})\right]~. \label{eq:LGW}
\ee
$\phi$ is a
fluctuating field whose average value is equal to the order parameter, $m=\langle \phi
\rangle$. (Here, $\langle \ldots \rangle$ denotes the average with respect to the
statistical weight $e^{-S[\phi]}$). We have also included an external field $h$ conjugate
to the order parameter.

In the disordered phase, away from the critical point, the correlation function of the
order parameter fluctuations, $G(\mathbf{x}-\mathbf{y}) =\langle
\phi(\mathbf{x})\phi(\mathbf{y})\rangle$ is generically short-ranged.  When the critical
point is approached, the correlations become long-ranged.  Their typical length scale,
the correlation length $\xi$, diverges when the distance $r$ from the critical point
vanishes\footnote{One needs to distinguish between the bare value of $r$ which appears in
the LGW functional (\ref{eq:LGW}) and the physical or renormalized value which measures
the distance from the true critical point. We will explicitly make this distinction when
necessary but otherwise suppress it.},
\be
  \xi \propto |r|^{-\nu}~.
  \label{eq:xi}
\ee
Here, $\nu$ is the correlation length critical exponent. Close to the critical point, the
correlation length $\xi$ is the only relevant length scale in the system. Therefore,
the physical properties must be unchanged, if
we rescale all lengths in the system by a common factor $b$,
and at the same time adjust the external parameters in such a way
that the correlation length retains its old value. This gives rise to
the homogeneity relation for the free energy density $f=-(T/V)\ln Z$,
\begin{equation}
 f(r,h) = b^{-d} f(r\, b^{1/\nu}, h\, b^{y_h}).
\label{eq:widom}
\end{equation}
The scale factor $b$ is an arbitrary positive number, and $y_h$ is another critical
exponent. Analogous homogeneity relations for other thermodynamic quantities can be
obtained by differentiating $f$. These homogeneity laws were first obtained
phenomenologically by Widom \cite{Widom65} and are sometimes summarily called the scaling
hypothesis. Within the framework of the modern renormalization group theory of phase
transitions \cite{WilsonKogut74} the scaling laws can be derived from first principles.

In addition to the critical exponents $\nu$ and $y_h$ defined above,
a number of other exponents is in common use. They describe the
dependence of the order parameter and its correlations on the
distance from the
critical point and on the field conjugate to the order parameter.
The definitions of the most commonly used
critical exponents are summarized in Table \ref{table:exponents}.
\begin{table}
\renewcommand{\arraystretch}{1.2}
\begin{tabular*}{\textwidth}{c@{\extracolsep\fill}ccc}
\hline
\hline
&exponent& definition & conditions \\
\hline
specific heat &$\alpha$& $c \propto |r|^{-\alpha}$ & $r \to 0, h=0$\\
order parameter& $\beta$ & $m \propto (-r)^\beta$ & $r \to 0-$, $h=0$\\
susceptibility& $\gamma$ & $\chi \propto |r|^{-\gamma}$ & $r \to 0, h=0$\\
critical isotherm & $\delta$ & $h \propto |m|^\delta {\rm sign}(m)$ & $h \to 0, r=0$\\
\hline
correlation length& $\nu$ & $\xi \propto |r|^{-\nu}$ & $r \to 0, h=0$\\
correlation function& $\eta$ & $G(\mathbf{x}) \propto |\mathbf{x}|^{-d+2-\eta}$ & $r=0, h=0$\\
\hline
dynamical& $z$ & $\xi_t \propto \xi^{z}$ & $r \to 0, h=0$\\
activated dynamical  & $\psi$ & $\ln \xi_t \propto \xi^{\psi}$ & $r \to 0, h=0$\\
\hline
\hline
\end{tabular*}
\caption{Definitions of the commonly used critical exponents. $m=\langle \phi \rangle$ is the
  order parameter, and $h$ is the conjugate field. $r$ denotes the
  distance from the critical point and $d$ is the space dimensionality.
  (The exponent $y_h$ defined in (\ref{eq:widom}) is related to $\delta$
  by $y_h=d \, \delta /(1+\delta)$.)}
\label{table:exponents}
\end{table}
Note that not all the exponents defined in Table \ref{table:exponents}
are independent from each other.
The four thermodynamic exponents $\alpha, \beta,\gamma,\delta$ can
all be obtained from the free energy (\ref{eq:widom}) which contains
only two independent exponents.
They are therefore connected by the so-called scaling relations
\begin{eqnarray}
2- \alpha &=&  2 \beta +\gamma~,  \\
2 - \alpha &=& \beta ( \delta + 1)~.
\end{eqnarray}
Analogously, the exponents of the correlation length and correlation
function are connected by the relations
\begin{eqnarray}
2- \alpha &=&  d\,\nu~,  \\
\gamma &=& (2-\eta) \nu~.
\end{eqnarray}
Note that scaling relations explicitly involving the dimensionality, the so called
hyperscaling relations, only hold below the upper critical dimension $d_c^+$.
Their breakdown above $d_c^+$ is caused by dangerously irrelevant variables,
see, e.g., Ref.\ \cite{Goldenfeld_book92}.

In addition to the diverging length scale $\xi$, a critical point is characterized by a
diverging time scale, the correlation time $\xi_t$. It leads to the phenomenon of
\emph{critical slowing down}, i.e., very slow relaxation towards equilibrium near a
critical point. At generic critical points, the divergence of the correlation time
follows a power law $\xi_t \propto \xi^z$ where $z$ is the dynamical critical exponent.
At some transitions, in particular in the presence of quenched disorder, the divergence
can be exponential, $\ln \xi_t \propto \xi^\psi$. The latter case is referred to as
activated dynamical scaling in contrast to the generic power-law dynamical scaling.

As mentioned above, the critical exponents display the remarkable phenomenon of
universality, i.e., they are the same for entire classes of phase transitions which
 may occur in very different physical systems.
These classes, the so-called universality classes, are determined only
by the symmetries of the Hamiltonian and the spatial dimensionality
of the system. The physical mechanism behind universality is the divergence of the correlation
length. Close to the critical point the system effectively averages over
large volumes rendering the microscopic details of the Hamiltonian
unimportant.

%%%%%%%%%%%%%%%%%%%%%%%%%%%%%%%%%%%%%%%%%%%%%%%%%%%%%%%%%%%%%%%%%%%%%%%%%%%%%%%%%%%%%%%%%%%%
\subsection{Classical versus quantum phase transitions}
\label{subsec:CPTQPT}

In classical statistical mechanics, statics and dynamics decouple. This can be seen by
considering a classical Hamiltonian $H(p_i,q_i)=H_{\rm kin}(p_i) +H_{\rm pot}(q_i)$
consisting of a kinetic part $H_{\rm kin}$ that only depends on the momenta $p_i$ and a
potential part $H_{\rm pot}$ that only depends on the coordinates $q_i$. The classical
partition function of such a system,
\be
 Z= \int \prod dp_i e^{-H_{\rm kin}/ T} ~\int \prod dq_i e^{-H_{\rm pot}/ T}
   = Z_{\rm kin}  Z_{\rm pot}~,
\label{eq:classical Z}
\ee
factorizes in kinetic and potential parts which are completely independent from each
other. The kinetic contribution to the free energy
density  will usually not display any singularities,
since it derives from a product of simple Gaussian integrals. Therefore
one can study the \emph{thermodynamic} critical behavior in classical systems using time-independent
theories like the Landau-Ginzburg-Wilson theory discussed in the last subsection.
As a result of this decoupling, the dynamical critical exponent is completely independent from
the thermodynamic ones.

In quantum statistical mechanics, the situation is different. The Hamiltonian $H$ and its
parts $H_{\rm kin}$ and $H_{\rm pot}$ are now operators, and $H_{\rm kin}$ and $H_{\rm pot}$
in general do not commute. Consequently, the partition function $Z= \Tr e^{-H/T}$ does
not factorize, and one must solve for the dynamics together with the thermodynamics.
Quantum mechanical analogs of the Landau-Ginzburg-Wilson theory (\ref{eq:LGW}) therefore need
to be formulated in terms of space and time dependent fields; they can be derived, e.g., via
a functional integral representation of the partition function. The simplest example of such
a quantum Landau-Ginzburg-Wilson functional takes the form
\begin{equation}
\fl S[\phi] = \int_0^{1/T} d\tau \int d^d x ~ \left[
a(\partial_\tau\phi(\mathbf{x,\tau}))^2 + c(\nabla\phi(\mathbf{x,\tau}))^2 +
F_L(\phi(\mathbf{x,\tau})) -h \phi(\mathbf{x,\tau})\right]
\label{eq:QLGW}
\end{equation}
with $\tau$ being the imaginary time. This functional also illustrates another remarkable
feature of quantum statistical mechanics. The imaginary time variable $\tau$ effectively
acts as an additional coordinate. For nonzero temperatures, $1/T<\infty$, the extra
dimension extends only over a finite interval. If one is sufficiently close to the
critical point so that the condition $\xi_t >1/T$ is fulfilled, the extra dimension will
not affect the leading critical behavior. We thus conclude that the asymptotic critical
behavior at any transition with a nonzero critical temperature is purely classical.
However, a quantum phase transition at $T=0$ is described by a theory that effectively is
in a different (higher) dimension. Therefore, it will be in a different universality
class. At finite temperatures and in the vicinity of a quantum phase transition, the
interplay between classical and quantum fluctuations leads to nontrivial crossover
phenomena \cite{SGCS97,Sachdev_book99,Vojta_review00,BelitzKirkpatrickVojta05}.

Using the equivalence between imaginary time and an additional dimension,
the homogeneity law for the free energy can be easily generalized to
the quantum case (see, e.g., \cite{Sachdev_book99}). For conventional power law dynamical
scaling it takes the form
\begin{equation}
 f(r,h,T) = b^{-(d+z)} f(r\, b^{1/\nu}, h\, b^{y_h},T\, b^z)~.
\label{eq:qptscaling}
\end{equation}
This relation implies that the scaling behavior of a zero-temperature quantum phase
transition in $d$ dimensions can be mapped onto that of some classical transition in
$d+z$ spatial dimensions. If space and time enter the theory symmetrically the dynamical
exponent is $z=1$, but in general, it can take any positive value. Note that this quantum
to classical mapping is valid for the thermodynamics only. Other properties like the real
time dynamics at finite temperatures require more careful considerations
\cite{Sachdev_book99}. Moreover, some quantum phase transitions, such as the
metal-insulator transition, do not have a classical counterpart and thus the quantum to
classical mapping does not directly apply.

Let us also emphasize that the entire concept of a Landau-Ginzburg-Wilson theory in terms
of the order parameter field relies on the order parameter fluctuations being the only
soft or gapless mode in the system. If there are other soft modes, e.g., due to
conservation laws or broken continuous symmetries, they lead to long-range power-law
correlations of various quantities even \emph{away from the critical point}. This
phenomenon is called generic scale invariance
\cite{DorfmanKirkpatrickSengers94,LawNieuwoudt89,Nagel92}. If one insists on deriving a
Landau-Ginzburg-Wilson theory for a phase transition in the presence of generic scale
invariance, the additional soft modes lead to coefficients that are singular functions of
space and time, severely limiting the usefulness of the theory
\cite{VBNK96,VBNK97,BelitzKirkpatrickVojta97}. In such a situation one should instead
work with a coupled theory that keeps all soft modes at the same footing. This question
has been explored in detail in Ref.\ \cite{BelitzKirkpatrickVojta05}. In practise,
complications due to generic scale invariance are more common for quantum phase
transitions than for classical ones because (i) there are more soft modes at $T=0$ than
at $T>0$, and (ii) phenomena that classically only affect the dynamics influence the
static critical behavior of a quantum phase transition as well. In addition to this
mechanism, there are other reasons for a quantum phase transition being more complicated
than one might naively expect, including topological effects. In many of these cases, a
description of the transition in terms of a Landau-Ginzburg-Wilson theory is not
possible.

%%%%%%%%%%%%%%%%%%%%%%%%%%%%%%%%%%%%%%%%%%%%%%%%%%%%%%%%%%%%%%%%%%%%%%%%%%%%%%%%%%%%%%%%%%%%
\subsection{Finite-size scaling}
\label{subsec:FSS}

We now turn to the question of how a finite system size influences the behavior at a
critical point, a question that will be very important later on for our discussion of rare region
effects. In general, a sharp phase transition can only exist in the thermodynamic limit, i.e., in
an infinite system. A finite system size results in a rounding and shifting of the
critical singularities. If the extension of the system is finite only in some directions
and infinite in others, a sharp phase transition can still exist, but the critical point
will be shifted compared to the bulk value. Moreover, the universality class will change.

The behavior of finite size systems in the vicinity of a critical point is described by
finite-size scaling theory \cite{FisherBarber72,Barber_review83,Cardy_book88}.
Finite-size scaling starts from the observation that the inverse system size acts as an
additional parameter that takes the system away from the critical point. Because the
correlation length of the infinite system $\xi_\infty$ is the only relevant length scale close
to the critical point, finite-size effects in a system of linear size $L$ must be controlled
by the ratio $L/\xi_\infty$ only. We can therefore generalize the classical homogeneity
relation (\ref{eq:widom}) for the free energy density by including the system size
\be
 f(r,h,L) = b^{-d} f(r\, b^{1/\nu}, h\, b^{y_h}, L\, b^{-1}).
\label{eq:widom_FSS}
\ee
The quantum version (\ref{eq:qptscaling}) of the homogeneity law can be generalized in the same way:
\be
 f(r,h,T,L) = b^{-(d+z)} f(r\, b^{1/\nu}, h\, b^{y_h}, T\, b^z, L\, b^{-1})~.
\label{eq:qptscaling_FSS} \ee We note in passing that the temperature scaling in the
quantum homogeneity law can be viewed as finite-size scaling in imaginary time direction
with the inverse temperature playing the role of the system size.

The shift of the critical temperature or critical coupling as a function of $L$ for
geometries that still allow a sharp phase transition at finite $L$ follows directly from
(\ref{eq:widom_FSS}). Setting $b=L$, $h=0$, we obtain
\be
f(r,L) = L^d F(r\,L^{1/\nu})
\label{eq:rL_FSS}
\ee
where $F(x)$ is a dimensionless scaling function. The quantum case can be treated
analogously. The finite-$L$ phase transition corresponds to a singularity in the scaling
function at some nonzero argument $x_c$. The transition thus occurs at $r_c\,L^{1/\nu} =
x_c$, and the transition temperature $T_c(L)$ of the finite-size system is shifted from the
bulk value $T_c^0$ by
\be
T_c(L) -T_c^0 \sim r_c = x_c \,L^{-1/\nu}~.
\label{eq:shift_FSS}
\ee

Note that the simple form of finite-size scaling summarized above is only valid below the
upper critical dimension $d_c^+$ of the phase transition. Finite-size scaling can be generalized
to dimensions above $d_c^+$, but this requires taking dangerously irrelevant variables
into account. One important consequence is that the shift of the critical temperature,
$T_c(L) -T_c^0 \sim L^{-\phi}$,
is controlled by an exponent $\phi$ which in general is different from $1/\nu$.

%%%%%%%%%%%%%%%%%%%%%%%%%%%%%%%%%%%%%%%%%%%%%%%%%%%%%%%%%%%%%%%%%%%%%%%%%%%%%%%%%%%%%%%%%%%%
\section{Quenched disorder effects}
\label{sec:disorder}
%%%%%%%%%%%%%%%%%%%%%%%%%%%%%%%%%%%%%%%%%%%%%%%%%%%%%%%%%%%%%%%%%%%%%%%%%%%%%%%%%%%%%%%%%%%%

In this section, we give a brief introduction into the effects of quenched disorder
on phase transitions and critical points. We focus on what is arguably the simplest and
most benign type of disorder, viz., impurities and defects that lead to spatial variations of
the coupling strength (i.e., spatial variations in the tendency towards the ordered phase)
but not to frustration or random external fields. This also implies that the two bulk phases
(that are separated by the transition) are not changed qualitatively by the presence of the
quenched disorder. This type of quenched disorder is sometimes referred to as weak disorder,
random-$T_c$ disorder, or, in analogy to quantum field theory,  random-mass
disorder.

In ferromagnetic materials, random-$T_c$ disorder can be achieved, e.g., by diluting the
lattice, i.e., by replacing magnetic atoms with nonmagnetic ones. Within a
Landau-Ginzburg-Wilson theory such as (\ref{eq:LGW}) or (\ref{eq:QLGW}), random-$T_c$
disorder can be modelled by making the bare distance from the critical point a random
function of spatial position, $r \to r +\delta r (\mathbf{x})$. In this representation,
the character of the impurities and defects is encoded in the statistical properties of
the random-$T_c$ term $\delta r (\mathbf{x})$. Point-like impurities correspond to
short-range correlations of $\delta r (\mathbf{x})$ in all directions while linear and
planar defects lead to perfect correlations of $\delta r (\mathbf{x})$ in one and two
dimensions, respectively. Let us also briefly comment on the disorder probability
distribution $P(\delta r)$. As long as the physics is dominated by long-wavelength
properties and the average behavior of the disorder, the details of the probability
distribution should not play an important role. In theoretical investigations a realistic
distribution is often replaced by a Gaussian because the latter can be easily handled
mathematically. We emphasize, however, that in some cases the dominating physics occurs
in the tail of the probability distribution. In these cases, the form of the tail is
important, and  more careful considerations are required.

Adding weak, random-$T_c$, quenched disorder to a system undergoing a continuous phase transition
naturally leads to the following questions:
\begin{itemize}
\item Will the phase transition remain sharp in the presence of quenched disorder?
\item If so, will the critical behavior change quantitatively (different universality class with
     new exponents) or even qualitatively (exotic non-power-law scaling)?
\item Will only the transition itself be influenced or also the behavior in its vicinity?
\end{itemize}

%%%%%%%%%%%%%%%%%%%%%%%%%%%%%%%%%%%%%%%%%%%%%%%%%%%%%%%%%%%%%%%%%%%%%%%%%%%%%%%%%%%%%%%%%%%%
\subsection{Average disorder and Harris criterion}
\label{subsec:avdis}

The question of how quenched disorder influences phase transitions has a long history.
Initially it was suspected that disorder destroys any critical point because in the
presence of defects, the system divides itself up into spatial regions which
independently undergo the phase transition at different temperatures (see Ref.\
\cite{Grinstein85} and references therein). However, subsequently it became clear that
generically a phase transition remains sharp in the presence of defects, at least for
classical systems with short-range disorder correlations.

Harris \cite{Harris74} derived a criterion for the perturbative stability of a clean
critical point against weak disorder. Harris considered the effective local critical
temperature in blocks of linear size $\xi$ which is given by an average of $r+\delta
r(\mathbf{x})$ over the volume $V=\xi^d$. He observed that a sharp phase transition can
only occur if the variation $\Delta r$ of these local critical temperatures from block to
block is smaller than the global distance from the critical point $r$. For short-range
correlated disorder, the central limit theorem yields $\Delta r \propto \xi^{-d/2}
\propto r^{d\nu/2}$. Thus, a clean critical point is perturbatively stable, if the clean
critical exponents fulfill the inequality $r^{d\nu/2}<r$ for $r\to 0$. This implies the
exponent inequality
\be
d\, \nu > 2 \label{eq:Harris} \ee which is called the Harris criterion.
 Note that the Harris criterion is a necessary condition
for the stability of a clean fixed point, not a sufficient one. It only deals with the
average behavior of the disorder at large length scales. However, effects due to
qualitatively new physics at finite length scales (and finite disorder strength) are not
covered by the Harris criterion.

The Harris criterion can be used as the basis for a classification of critical points
with quenched disorder, based on the behavior of the average disorder strength with
increasing length scale, i.e., under coarse graining. Three classes of critical points
can be distinguished \cite{MMHF00}.

(i) The first class contains systems whose clean critical points fulfill the Harris
criterion $d\, \nu
> 2$. At these phase transitions, the disorder strength {\em decreases} under coarse
graining, and the system becomes asymptotically homogeneous at large length scales.
Consequently, the quenched disorder is asymptotically unimportant, and the critical
behavior of the dirty system is identical to that of the clean system. Technically, this
means the disorder is renormalization group irrelevant at the clean critical fixed point.
In this class of systems, the macroscopic observables are self-averaging at the critical
point, i.e., the relative width of their probability distributions vanishes in the
thermodynamic limit \cite{AharonyHarris96,WisemanDomany98}. A prototypical example in
this class is the three-dimensional classical Heisenberg model whose clean correlation
length exponent is $\nu\approx 0.698$ (see, e.g., \cite{HolmJanke93}), fulfilling the
Harris criterion.

If a clean critical point violates the Harris criterion, it is destabilized by weak
quenched disorder, and the behavior must change. Nonetheless, a sharp critical point can
still exist in the presence of the disorder, but it must fall into one of the two
following classes.

(ii) In the second class, the system remains inhomogeneous at all length scales with the
relative strength of the inhomogeneities approaching a finite value for large length
scales. The resulting critical point still displays conventional power-law scaling but
with new critical exponents which differ from those of the clean system (and fulfill the
inequality $d\nu > 2$). These transitions are controlled by renormalization group fixed
points with a nonzero value of the disorder strength. Macroscopic observables are not
self-averaging, but in the thermodynamic limit, the relative width of their probability
distributions approaches a size-independent constant
\cite{AharonyHarris96,WisemanDomany98}. An example in this class is the classical
three-dimensional Ising model. Its clean correlation length exponent, $\nu\approx 0.627$
(see, e.g. \cite{FerrenbergLandau91}) does not fulfill the Harris criterion. Introduction
of quenched disorder, e.g., via dilution, thus leads to a new critical point with an
exponent of $\nu\approx 0.684$ \cite{BFMM98}.

(iii) At critical points in the third class, the relative magnitude of the
inhomogeneities {\em increases} without limit under coarse graining. The corresponding
renormalization group fixed points are characterized by infinite disorder strength. At
these infinite-randomness critical points, the power-law scaling is replaced by activated
(exponential) scaling. The probability distributions of macroscopic variables become very
broad (even on a logarithmic scale) with the width diverging with system size.
Consequently, their averages are often dominated by rare events, e.g., spatial regions
with atypical disorder configurations. This type of behavior was first found in the
McCoy-Wu model, a two-dimensional Ising model with bond disorder perfectly correlated in
one dimension \cite{McCoyWu68,McCoyWu68a}. However, it was fully understood only when
Fisher \cite{Fisher92,Fisher95} solved the one-dimensional random transverse field Ising
model by a version of the Ma-Dasgupta-Hu real space renormalization group
\cite{MaDasguptaHu79}. Since then, several infinite-randomness critical points have been
identified, mainly at quantum phase transitions since the disorder, being perfectly
correlated in (imaginary) time, has a stronger effect on quantum phase transitions than
on thermal ones. Examples include one-dimensional random quantum spin chains as well as
one-dimensional and two-dimensional random quantum Ising models
\cite{BhattLee82,Fisher94,YoungRieger96,PYRK98,MMHF 00}. Very recently, a similar
infinite-randomness fixed point has been found in the disordered one-dimensional contact
process \cite{HooyberghsIgloiVanderzande03,HooyberghsIgloiVanderzande04,VojtaDickison05}.

%%%%%%%%%%%%%%%%%%%%%%%%%%%%%%%%%%%%%%%%%%%%%%%%%%%%%%%%%%%%%%%%%%%%%%%%%%%%%%%%%%%%%%%%%%%%%%%%%%%
\subsection{Rare regions and Griffiths effects}
\label{sec:rr}

In the last subsection we have discussed scaling scenarios for phase transitions with
quenched disorder based on the \emph{global, i.e., average,} behavior of the disorder
strength under coarse graining. In this subsection, we turn to the main topic of this review,
viz., the effects of \emph{rare} strong spatial disorder fluctuations.

\begin{figure}
\centerline{\includegraphics[width=7cm]{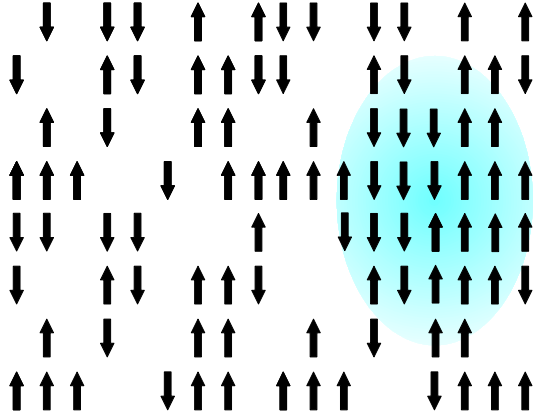}} \caption{Sketch of a diluted
magnet. The shaded region is devoid of impurities and
         therefore acts as a piece of the clean bulk system}
\label{fig:dilutedmagnet}
\end{figure}

Let us start by considering the example of a diluted classical ferromagnet as sketched in
figure \ref{fig:dilutedmagnet}.
The dilution reduces the tendency towards magnetic long-range order and thus reduces
$T_c$ from its clean bulk value $T_c^0$. However, in an infinite system one can find
arbitrarily large spatial regions that are devoid of impurities. For temperatures between
$T_c^0$ and $T_c$, these regions will show local magnetic order even though the bulk
system is globally in the paramagnetic phase. These spatial regions are known as
\emph{rare regions} and the order parameter fluctuations induced by them belong to a
class of excitations known as \emph{local moments}; sometimes they are also referred to
as \emph{instantons}. The dynamics of the rare regions is very slow because flipping them
requires a coherent change of the order parameter over a large volume. Griffiths
\cite{Griffiths69} was the first to show that these rare regions can lead to a
singularity in the free energy, the Griffiths singularity, in the entire temperature
region $T_c<T<T_c^0$ which is now known as the Griffiths region or the Griffiths phase
\cite{RanderiaSethnaPalmer85}. Analogous singularities also exist on the ordered side of
the phase transition.

The probability $w$ for finding a rare region is exponentially small in its volume
$V_{\rm RR}$ and in the impurity concentration $p$; up to pre-exponential factors it is
given by $w \sim \exp(-pV_{\rm RR})$. Rare regions are thus nonperturbative degrees of
freedom that are not accounted for in conventional approaches to phase transitions and
critical points that are based on perturbation theory and the perturbative
renormalization group. They can be viewed as examples of a broader class of rare event
phenomena that range from the well known Lifshitz tails in the density of states of
disordered semiconductors \cite{Lifshitz64,Lifshitz64b} to the slowing down of computer
algorithms in combinatorial optimization \cite{SchwarzMiddleton04}.

In classical systems with uncorrelated or short-range correlated disorder, thermodynamic
Griffiths effects are very weak because the singularity in the free energy is only an
essential one \cite{Wortis74,Harris75,BrayHuifang89}. To the best of our knowledge,
classical thermodynamic Griffiths singularities have therefore not been verified in
experiment (see also \cite{Imry77}). In contrast, the long-time \emph{dynamics} inside
the Griffiths phase is dominated by the rare regions, as is signified by a
non-exponential time-dependence of the spin autocorrelation function
\cite{Dhar83,RanderiaSethnaPalmer85,DharRanderiaSethna88,Bray87,Bray88}.

Long-range disorder correlations can increase the rare region effects
qualitatively. In particular, if the disorder is perfectly correlated in some spatial directions, or --
in the case of a zero-temperature quantum phase transition -- in imaginary time, the rare
regions are extended objects which are infinite in the
correlated space or imaginary time directions. This makes their dynamics even slower and so
increases their effects. This enhancement of rare region effects was first
observed in the abovementioned McCoy-Wu model \cite{McCoyWu68,McCoyWu68a}, a two-dimensional
classical Ising model with linear defects.  More recently, it has been studied in great detail
in various disordered quantum  systems: In one-dimensional and two-dimensional random
transverse field Ising models, the singularity of the free energy in the Griffiths phase takes
a power-law form with nonuniversal continuously varying exponents. Several thermodynamic
observables including the average susceptibility actually diverge in a finite region of the
disordered phase \cite{Fisher92,Fisher95,YoungRieger96,PYRK98,MMHF00} rather than only at the critical
point. Similar phenomena have also been found in quantum Ising spin glasses
\cite{ThillHuse95,GuoBhattHuse96,RiegerYoung96}. In addition, power-law Griffiths
effects, sometimes also called Griffiths-McCoy effects, were also found at the nonequilibrium phase transition in a disordered
one-dimensional contact process
\cite{HooyberghsIgloiVanderzande03,HooyberghsIgloiVanderzande04,VojtaDickison05}.

Recently, it has been shown that rare region effects can become strong enough to
completely destroy the sharp phase transition. This happens, if the dynamics of the rare
regions stops completely, i.e., if a static order parameter develops on an isolated rare
region independent of the bulk system. The global phase transition is then smeared or
rounded, as was found in itinerant quantum magnets \cite{Vojta03a} where the overdamping
of the magnetic excitations by the electronic degrees of freedom suppresses the dynamics
of sufficiently large rare regions \cite{MillisMorrSchmalian01,MillisMorrSchmalian02}.
Later, analogous smeared transitions been found in a variety of systems, ranging from
classical Ising magnets with planar defects \cite{Vojta03b,SknepnekVojta04} to
nonequilibrium spreading transitions in the contact process
\cite{Vojta04,DickisonVojta05}.

%%%%%%%%%%%%%%%%%%%%%%%%%%%%%%%%%%%%%%%%%%%%%%%%%%%%%%%%%%%%%%%%%%%%%%%%%%%%%%%%%%%%%%%%%%%%

\subsection{A classification of rare region effects}
\label{subsec:rrclass}

In the last subsection, we have seen that the influence of rare regions on continuous
phase transitions can range from exponentially weak free energy contributions to a
complete destruction of the sharp phase transition by smearing. It would clearly be very
desirable to determine what conditions lead to which type of rare region effects.
Recently, a general classification of rare region effects at order-disorder phase
transitions in systems with weak, random-mass disorder has been suggested
\cite{VojtaSchmalian05}. This classification is based on the following idea. The
probability $w$ for finding a large spatial region of volume $V_{\rm RR} \sim L_{\rm
RR}^d$ devoid of impurities decreases exponentially with its volume, $w(L_{\rm RR}) \sim
\exp(-pV_{\rm RR})$. The importance of large rare regions thus depends on how rapidly the
contribution of a \emph{single} region to observable quantities \emph{increases} with its
size. It turns out that this question is controlled by the effective dimensionality
$d_{\rm RR}$ of the rare regions, or more precisely, by the relation between $d_{\rm RR}$
and the lower critical dimension $d_c^-$ of the phase transition under
consideration.\footnote{When counting the \emph{effective} dimensionalities for quantum
phase transitions we include the imaginary time direction as one of the dimensions.}

To understand this, let us consider a single isolated region of linear size $L_{\rm RR}$
that is locally in the ordered phase, i.e., the bulk distance from the critical point is
negative, $r=r(\infty)<0$. We now ask: How does the renormalized distance from the
critical point, $r(L_{\rm RR})$, (or equivalently, the energy gap) of the finite size
rare region depend on its linear size $L_{\rm RR}$? Three cases can be distinguished (a
summary is given in table \ref{tab:rrclass}).
\begin{table}
\begin{center}
\begin{tabular}[c]{ccccc}
\hline
 Class &{ RR dimension}  &  ~{Griffiths effects}~   & ~{Dirty critical point}~ & Critical scaling \\
\hline
 A& $d_{\rm RR} < d_c^-$  & weak exponential    & conventional &  power law \\
 B& $d_{\rm RR} = d_c^-$  & strong power-law    &   infinite randomness &  activated \\
 C& $d_{\rm RR} > d_c^-$  & RR become static    &   smeared transition  & no scaling \\
\hline
\end{tabular}
\end{center}
\caption{Classification of rare region (RR) effects at critical points in the presence of
weak quenched disorder according to the effective dimensionality $d_{\rm RR}$ of the rare
regions. For details see text.} \label{tab:rrclass}
\end{table}

\subsubsection{Class A:}
If the effective dimensionality $d_{\rm RR}$ of the rare regions is \emph{below} the
lower critical dimensionality $d_{c}^{-}$ of the problem, an isolated rare region cannot
undergo the phase transition by itself. The renormalized $r(L_{\rm RR})$ is therefore
positive, but it decreases with $L_{\rm RR}$ following a power law.  The leading
contributions of a rare region to thermodynamic quantities are controlled by $r(L_{\rm
RR})$, e.g., the order parameter susceptibility $\chi$ is proportional to $r^{-1}(L_{\rm
RR})$. The contributions of a rare region to observables can therefore at most grow as a
power of its linear size $L_{\rm RR}$. (In the case of a classical magnet, $\chi \sim
S_{\rm eff}^2 \sim L_{\rm RR}^{2d}$). This power-law increase cannot overcome the
exponential drop in the rare region density $w(L_{\rm RR})$.

As a result, for transitions in this class, thermodynamic rare region effects are
exponentially weak and characterized by an essential singularity in the free energy
\cite{Griffiths69,Wortis74,Harris75,BrayHuifang89}. Right at the dirty critical point,
the rare region effects are subleading, and the critical behavior is of conventional
power-law type.  Examples for critical points in class A can be found in generic
classical equilibrium systems with point defects (where the rare regions are finite in
all directions and thus $d_{\rm RR}=0$) but also at some quantum phase transitions such
as the transition in the diluted bilayer Heisenberg quantum antiferromagnet
\cite{Sandvik02,VajkGreven02,SknepnekVojtaVojta04} (here $d_{\rm RR}=1$ but $d_c^-=2$ so
that the condition for class A is fulfilled).

\subsubsection{Class B:}
In this class, the rare regions are exactly \emph{at} the lower critical dimension,
$d_{\rm RR}=d_c^-$. They still cannot undergo the phase transition independently, but the
renormalized distance from the critical point (or equivalently, the energy gap) decreases
\emph{exponentially} with the rare region volume $V_{\rm RR}$. Therefore, the rare region
contribution to observables such as the susceptibility increases exponentially with size
and can potentially overcome the exponential decrease of the rare region density.

Generically, this results in a power-law low-energy density of states leading to strong
power-law Griffiths singularities with a nonuniversal continuously varying exponents. The
scaling behavior at the dirty critical point itself is dominated by the rare regions
resulting in exotic activated (exponential) scaling instead of conventional power-law
scaling. Class B is realized, e.g., in a classical Ising model with linear defects, i.e.,
disorder perfectly correlated in one direction (McCoy-Wu model
\cite{McCoyWu68,McCoyWu68a}), in percolation with linear defects \cite{JuhaszIgloi02},
and in random quantum Ising models (where the disorder correlations are in imaginary time
direction) \cite{Fisher92,Fisher95,YoungRieger96,PYRK98,MMHF00}. In all these systems,
$d_{\rm RR}=d_c^-=1$. Several thermodynamic observables including the average
susceptibility actually diverge in a finite region of the disordered phase rather than
just at the critical point. Class B rare region effects also occur at the nonequilibrium
phase transition of the contact process with point defects. Including the time dimension,
this transition also has $d_{\rm RR}=d_c^-$=1.

\subsubsection{Class C:}
Finally, in class C, the rare regions can undergo the phase transition independently from
the bulk system, i.e., they are effectively \emph{above} the lower critical dimension of
the problem, $d_{\rm RR}>d_c^{-}$. In this case, the dynamics of the locally ordered rare
regions completely freezes, and they develop a truly static order parameter. As a result,
the global phase transition is destroyed by smearing \cite{Vojta03a} because different
spatial parts of the system order at different values of the control parameter. In the
tail of the smeared transition, the order parameter is extremely inhomogeneous, with
statically ordered islands or droplets coexisting with the disordered bulk of the system.
This behavior was identified in itinerant quantum Ising magnets
\cite{MillisMorrSchmalian01,MillisMorrSchmalian02,Vojta03a} where the damping due to the
electronic degrees of freedom leads to an effective long-range ($1/\tau^2$) interaction
in time direction. At this transition $d_{\rm RR}=1$ and $d_c^-=1$ but it nonetheless
belongs to class C because the rare region can undergo a Kosterlitz-Thouless type phase
transition to the ordered phase. Smeared transitions can also be found in classical
magnets with planar defects \cite{Vojta03b,SknepnekVojta04} ($d_{\rm RR}=2>d_c^-=1$), in
nonequilibrium spreading transitions with extended (linear or planar defects)
\cite{Vojta04,DickisonVojta05} where $d_{\rm RR}>1$ but $d_c^-=1$ (including the time
dimension) or in stratified percolation \cite{Obukhov86} ($d_{\rm RR}=2, d_c^-=1$).

\vspace*{3mm}

This classification is expected to apply to all continuous order-disorder transitions
between conventional phases that can be described by Landau-Ginzburg-Wilson type theories
with random-mass disorder and short-range interactions. It also relies on the assumption
that the interaction between the rare regions can be neglected if their density is
sufficiently low (i.e., sufficiently far away from the true disordered critical point, if
any). While this seems a reasonable assumption in the case of short-range spatial
interactions, long-range interactions will likely modify the rare region effects. In the
case of itinerant quantum Heisenberg magnets, it has been argued that the interaction
between the rare regions due to the long-range RKKY interaction strongly enhances the
rare region effects, leading to a smeared transition even though a single rare region
would not independently undergo the phase transition \cite{DobrosavljevicMiranda05}.

The classification of rare region effects provides a general framework for a variety of
existing results, and it helps organizing the search for new phenomena. In the following
sections of this review we will discuss in detail several examples of rare regions
effects at classical, quantum, and nonequilibrium  phase transitions with quenched
disorder and relate them to the above classification.

%%%%%%%%%%%%%%%%%%%%%%%%%%%%%%%%%%%%%%%%%%%%%%%%%%%%%%%%%%%%%%%%%%%%%%%%%%%%%%%%%%%%%%%%%%%%
\section{Classical phase transitions}
\label{sec:classical}
%%%%%%%%%%%%%%%%%%%%%%%%%%%%%%%%%%%%%%%%%%%%%%%%%%%%%%%%%%%%%%%%%%%%%%%%%%%%%%%%%%%%%%%%%%%%

This section is devoted to rare region effects at classical equilibrium phase transitions
in systems with random-$T_c$ type disorder.  We first consider the case of uncorrelated
disorder (point-like defects) which leads to exponentially weak ``classical'' Griffiths
effects. We then show that disorder correlation enhance the rare region effects by
discussing systems with linear and planar defects.

\subsection{Randomly diluted Ising model}
\label{subsec:dil_ising}

\begin{figure}
\centerline{\includegraphics[width=8cm]{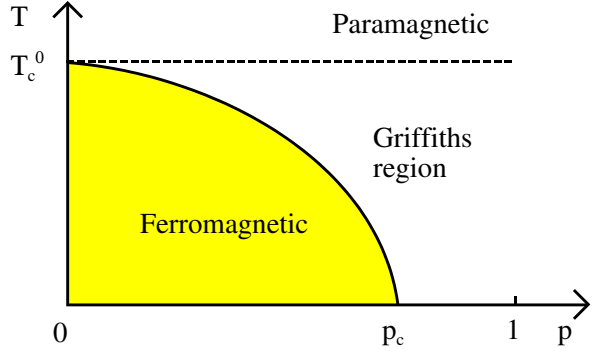}}
\caption{Schematic phase diagram of a randomly diluted Ising model as function of
         temperature $T$ and impurity concentration $p$. $p_c$ is the geometric
         percolation threshold of the lattice, and $T_c^0$ is the clean critical temperature}
\label{fig:pd_dil_ising}
\end{figure}

For definiteness, let us consider a randomly site-diluted classical Ising magnet on a
$d$-dimensional hypercubic lattice of $N=L^d$ sites. The Hamiltonian is given by
\be
 H= -J \sum_{\langle i,j\rangle} \epsilon_i \epsilon_j S_i S_j - h \sum_{i} \epsilon_i S_i
 \label{eq:H_dil_ising}
\ee
Here, $S_i=\pm 1$ is the Ising spin at lattice site $i$, $h$ is the external magnetic field,
$J>0$ is the exchange interaction, and $\langle i,j\rangle$ indicates that the sum is
over pairs of nearest neighbor sites only. The site dilution is introduced by the
quenched random variables $\epsilon_i$ with $\epsilon_i=1$ representing an occupied site
and $\epsilon_i=0$ corresponding to a vacancy. In our case of point defects, the
$\epsilon_i$ are statistically independent with a probability density
\be
 P(\epsilon) = (1-p) \delta(\epsilon-1) + p \delta(\epsilon)~,
\ee i.e., $p$ is the probability for finding a vacancy at a particular site. A schematic
phase diagram of this system as function of $p$ and temperature $T$ is shown in figure
\ref{fig:pd_dil_ising}.
At zero temperature, magnetic long-range order survives for all concentrations $p\leq p_{c}$,
where $p_{c}$ is the geometric percolation threshold of the lattice. If $p>p_{c}$, no long-range order is possible as the
system is decomposed into independent finite-size clusters.  For $p<p_{c}$, magnetic order survives
up to a nonzero temperature $T_{c}(p)$; but at  $p_{c}$, it is destroyed
immediately by thermal fluctuations\footnote{For Ising symmetry, this appears counterintuitive as
the fractal dimension of the critical percolation cluster is larger than the lower critical dimension
$d_c^-=1$. However, for
fractals, topological factors like ramification are important for magnetic ordering
\cite{GefenMandelbrotAharony80}.}, $T_c(p_c)=0$ \cite{Bergstresser77,StephenGrest77,GefenMandelbrotAharony80}.

In a seminal paper, Griffiths \cite{Griffiths69} showed that in the thermodynamic limit,
the magnetization (per site), $m=\frac 1 N M = \frac 1 N \sum_i \epsilon_i S_i$, is a
singular function of the magnetic field $h$ at $h=0$ for any temperature $T$ below the
transition temperature $T_c^0$ of the clean system (below the dashed line in figure
\ref{fig:pd_dil_ising}) independent of $p$. Conventional paramagnetic behavior only
appears above $T_c^0$. This singularity and the corresponding singularities in other
thermodynamic quantities are now known as the Griffiths singularities and the temperature
region over which they occur is called the Griffiths region or the Griffiths phase
\cite{RanderiaSethnaPalmer85}. Griffiths' proof of the singular dependence of the
magnetization on the field relied on an application of the theorem of Yang and Lee on the
complex zeros of the partition function \cite{YangLee52,LeeYang52}, and it did not
immediately give the functional form of the predicted singularity. However, after some
controversy it was soon established that the Griffiths singularity is a weak essential
singularity with the magnetization remaining infinitely differentiable at $h=0$
\cite{Wortis74,Harris75,Imry77}.

\subsection{Optimal fluctuation theory}
\label{subsec:OFT}

The leading singularities in the thermodynamic quantities in the Griffiths
region can be easily estimated using optimal fluctuation theory
 arguments similar to those employed in the study of band
tails of doped semiconductors \cite{HalperinLax66,Lifshitz67}.

In the site diluted Ising model (\ref{eq:H_dil_ising}), the probability $w(L_{\rm RR})$
for finding a region of volume $L_{\rm RR}^d$ devoid of vacancies is (to exponential
accuracy)
\be
 w(L_{\rm RR}) \sim (1-p)^{L_{\rm RR}^d} =\exp (-\tilde{p} L_{\rm RR}^d)
\label{eq:w(L)} \ee with $\tilde{p} =-\ln(1-p)$. Let us now consider a temperature $T$ in
the paramagnetic Griffiths region, $T_c(p) < T < T_c^0$, or equivalently $r_0 =
(T-T_c^0)/T_c^0 <0$. The bare distance $r(L_{\rm RR})$ from the local critical point for
a rare region of size $L_{\rm RR}$ is shifted by finite size effects because the rare
region acts as a finite-size piece of the clean system. If $r_0$ is not too close to zero
(in the Gaussian fluctuation region), the shift can be determined by finite size scaling
(\ref{eq:shift_FSS}) with the mean-field value $\nu=1/2$, leading to
\be
r(L_{\rm RR}) = r_0 + A/L_{\rm RR}^2 \label{eq:r_(L)_mf}
\ee where A is a constant.

If $r(L_{\rm RR})<0$, the rare region is locally in the ordered phase. Thus, at a given
$r_0$, only rare regions larger than  $L_{\rm min}(r_0)=|r_0/A|^{-1/2}$ will show local
order. Since they are of finite size, the locally ordered rare regions do not undergo a
true transition but coherently fluctuate due to thermal fluctuations. To estimate their
contribution $F_{\rm RR}$ to the free energy, we note that the contribution of a single
region varies at most like a power of its size. To leading exponential accuracy, the free
energy of the rare regions can therefore be estimated by summing their probabilities,
\be
 F_{\rm RR} \sim \int_{L_{\rm min}}^\infty dL_{\rm RR}~ w(L_{\rm RR}) \sim  \exp(-\tilde{p} \, |r_0/A|^{-d/2}) ~.
\label{eq:F_RR_ising} \ee Close to the bulk critical point $r_0=0$, i.e., in the clean
bulk critical region, the mean field relation for the finite-size shift of $r$ is not
valid anymore and has to be replaced by finite size scaling with the shift exponent
$\phi=1/\nu$ where $\nu$ is the clean bulk correlation length exponent. The behavior of
the free energy then crosses over to $F_{\rm RR} \sim \exp(-\tilde{p} \,
|r_0/A|^{-d\nu})$.

The response to a small external magnetic field can be derived in a similar way: In the
Griffiths region, an isolated, locally ordered rare region essentially acts as a single
big Ising spin of magnitude $S_{\rm eff} \sim L_{\rm RR}^d$. Thus its magnetization in an
external field $h$ is given by
\be
\fl M(L_{\rm RR}) = S_{\rm eff} \tanh(hS_{\rm eff}/T) \sim \left \{
\begin{array}{cc}
h L_{\rm RR}^{2d}  \quad & (L_{\rm RR} < b\,h^{-1/d})\\ L_{\rm RR}^{d}     \quad &
(L_{\rm RR} > b\,h^{-1/d})
\end{array} \right.
\label{eq:m(L)}
\ee
where $b$ is a constant that depends on $r_0$.
The singularity in the magnetization-field relation arises from the large rare regions
whose magnetization is saturated by the field $h$. To exponential accuracy, the
singular rare region contribution to the total magnetization is therefore given by
\be
M_{\rm RR} = \int_{b\,h^{-1/d}}^\infty dL_{\rm RR} ~w(L_{\rm RR}) \sim \exp(-\tilde{p}
b^d/h) \label{eq:M_RR_ising} \ee in agreement with references \cite{Harris75,Imry77}.
More recently, many of the properties of Griffiths phases have been proven rigorously in
the mathematical physics literature, see e.g.,
\cite{BassalygoDobrushin86,DreyfusKleinPerez95,GielisMaes95} and references therein.

At this point, it is important to emphasize that the Griffiths singularities do \emph{not}
display the same degree of universality as the critical singularities right at a continuous
phase transition. This is caused by the fact the rare regions are phenomena at \emph{finite}
length scales in the \emph{tail} of the disorder probability distribution. Therefore, the
functional form of the Griffiths singularities can depend on the specifics of the disorder
distribution. We will now illustrate this non-universality by comparing the  results
(\ref{eq:F_RR_ising}) and (\ref{eq:M_RR_ising}) for the randomly diluted Ising model with
those of a Landau-Ginzburg-Wilson theory with Gaussian disorder.

\subsection{Random-$T_c$ Landau-Ginzburg-Wilson theory}
\label{subsec:randomTc_LGW}

In this subsection, we consider rare region effects in a classical Landau-Ginzburg-Wilson
(LGW) theory with random-$T_c$ disorder. From a critical phenomena point of view, the
phase transition in this model belongs to the same universality class as the transition
in the randomly site diluted Ising model considered above. Nonetheless, the functional
form of the Griffiths singularities will turn out to be different from
(\ref{eq:F_RR_ising}) and (\ref{eq:M_RR_ising}).

Let us start from a $d$-dimensional LGW theory of the form [see equations
(\ref{eq:Landau}) and (\ref{eq:LGW})]
\be
S[\phi] = \int d^dx ~\left[ (\nabla \phi(\mathbf{x}))^2 + (r_0+\delta r (\mathbf{x})) \phi^2(\mathbf{x})
        + u \phi^4(\mathbf{x}) -h \phi(\mathbf{x}) \right].
\label{eq:LGW_random}
\ee
The disorder is modelled by the random contribution $\delta r(\mathbf{x})$ to the
distance to the critical point. The distribution of $\delta r(\mathbf{x})$ is usually
chosen to be a Gaussian
\be
P[\delta r(\mathbf{x})] \propto \exp \left[- \frac 1 {2\Delta} \int d^dx ~\delta^2
r(\mathbf{x}) \right]~. \label{eq:P_delta_r} \ee In this model, a rare region is a large
spatial region in which the typical value of $r_0+\delta r(\mathbf{x})$ is negative while
the bulk distance $r_0$ from the critical point is positive. One qualitative difference
to the randomly diluted Ising model case is immediately obvious: Because the Gaussian
disorder is unbounded, locally ordered rare regions can exist for any value of the bulk
distance $r_0$ from the critical point. This means, there is no conventional paramagnetic
phase, and the Griffiths region covers the entire high-temperature part of the phase
diagram.

The contribution of these rare regions to the free energy can be estimated using optimal
fluctuation arguments \cite{Dotsenko06} similar to those in the subsection
\ref{subsec:OFT}. The probability for finding a large region of linear size $L_{\rm RR}$
and a typical value $\delta r_{typ}= -r_0-s$ of the disorder variable is exponentially
small,
\be
w(L_{\rm RR},s) \sim \exp \left[- \frac 1 {2\Delta} L_{\rm RR}^d (s+r_0)^2 \right]~.
\label{eq:w(L,s)} \ee Treating the finite-size corrections in mean-field theory (valid
outside the asymptotic bulk critical region) as before, the bare distance from the
critical point for such a rare region is given by
\be
r(L_{\rm RR},s) = r_0 + \delta r_{typ}+ A/L_{\rm RR}^2 = -s + A/L_{\rm RR}^2~.
\label{eq:r_(L)_LGW} \ee If $r(L_{\rm RR},s)<0$, the rare region is locally in the
ordered phase. Thus, the minimum size permitting local order is $L_{\rm min} =
(A/s)^{1/2}$. To leading exponential accuracy, the contribution of the locally ordered
rare regions to the free energy can again be estimated by summing their probabilities,
\be
\fl
 F_{\rm RR} \sim \int_0^\infty ds \int_{L_{\rm min}}^\infty dL_{\rm RR}~ w(L_{\rm RR},s) \sim \int_0^\infty ds \exp [-C s^{-d/2} (s+r_0)^2]~.
\label{eq:F_RR_LGW} \ee where $C$ is a constant. The second equality arises because the
integral over $L$ is dominated by its lower bound. The integral over $s$ can be estimated
using the saddle point method. The leading contribution comes from $s_{\rm SP} = r_0
d/(4-d)$. This implies that with increasing $r_0$, i.e., with increasing distance from
the true transition, the dominating rare region contribution to the free energy comes
from \emph{smaller} and \emph{deeper} disorder fluctuations. This is qualitatively
different from the case of the randomly diluted Ising model where with increasing $r_0$
larger and larger islands dominate. Inserting the saddle point value into the equation
for $F_{\rm RR}$ gives
\be
 F_{\rm RR}  \sim  \exp [-{C^\prime} r_0^{(4-d)/2}]~.
\label{eq:F_RR_LGW2} \ee where $C^\prime$ is another constant. This result has also been
derived more rigorously by considering replica instanton solutions of the replicated and
disorder averaged Landau-Ginzburg-Wilson theory \cite{Dotsenko99,Dotsenko06}.

Comparing the Griffiths singularities of the randomly diluted Ising model
(\ref{eq:F_RR_ising}) and the LGW theory with Gaussian disorder (\ref{eq:F_RR_LGW2}), we
find them both exponentially weak. However, the functional form of the singularity is
\emph{not} universal, it does depend on the specifics of the microscopic disorder
distribution. Dotsenko \cite{Dotsenko06} also calculated the rare region contribution to
the free energy in a nonzero magnetic field for the LGW theory with Gaussian disorder. He
found $\delta F_{\rm RR}(h) \sim \exp[-E h^{-d/3}]$ with $E$ a constant. To exponential
accuracy, the same same singularity should also govern the magnetization. As in the
zero-field case, the functional form of the Griffiths singularity is different from that
of  the randomly site diluted Ising model, equation (\ref{eq:M_RR_ising}).

% However, arguments \cite{Vojta_unpublished} similar to those in subsection \ref{subsec:OFT}
% lead to $M_{\rm RR}(h) \sim \exp[- E h^{-d/(d-1)}]$. Clearly, more work is necessary to clarify
% the nature of this Griffiths singularity.

%%%%%%%%%%%%%%%%%%%%%%%%%%%%%%%%%%%%%%%%%%%%%%%%%%%%%%%%%%%%%%%%%%%%%%%%%%%%%%%%%%%%%%%%%%%%%%
\subsection{Dynamic rare region effects}
\label{subsec:dynamics}

So far, we have focussed on the effects of the rare regions on \emph{thermodynamic}
properties, and we have found them to be exponentially weak in generic classical systems
with short-range correlated disorder. In contrast, rare region effects on the dynamics
are much more dramatic, as will be discussed in this subsection.

For definiteness, let us again consider the dilute ferromagnet (\ref{eq:H_dil_ising}).
The dynamics is assumed to be purely relaxational, corresponding to model A in the
classification of Hohenberg and Halperin \cite{HohenbergHalperin77}. Microscopically,
this type of dynamics can be realized, e.g., via the Glauber \cite{Glauber63} or
Metropolis \cite{MRRT53} algorithms. Because the rare region effects are local in space,
their dynamic properties can be characterized in terms of the average spin
\emph{auto}correlation function $C(t)$ defined by
\be
C(t) = \frac 1 N \sum_i \langle S_i(t) \, S_i(0) \rangle
\label{eq:C(t)_def}
\ee
where $S_i(t)$ is the value of the spin at time $t$ and $\langle \cdot \rangle$ denotes the
thermodynamic average. In the conventional paramagnetic phase $T>T_c^0$ (see figure
\ref{fig:pd_dil_ising}), the average autocorrelation function decays exponentially fast
because none of the rare regions is locally ordered. In contrast, in the Griffiths
region $T_c(p) < T <T_c^0$, we expect the rare regions to display slow dynamics.
Following arguments arguments first made by Dhar \cite{Dhar83} and later refined by
Randeria et al. \cite{RanderiaSethnaPalmer85} and Bray \cite{Bray88,Bray89}, the rare region
part of the autocorrelation function  can be estimated as the sum of independent contributions
from each of the rare regions
\be
C_{\rm RR}(t) \sim \int_{0}^\infty dL_{\rm RR}~ w(L_{\rm RR}) ~ \exp[-t/\xi_t(L_{\rm
RR},r_0)]~. \label{eq:C(t)_RR} \ee Here $w(L_{\rm RR})$ is the probability for finding a
rare region of size $L_{\rm RR}$ as given in (\ref{eq:w(L)}) and $\xi_t(L_{\rm RR},r_0)$
is the relaxation time of such a rare region at reduced temperature $r_0$. Depending on
$r_0$, equation (\ref{eq:C(t)_RR}) predicts different long time behavior of the
autocorrelation function.

(i) In the conventional paramagnetic phase, $T>T_c^0$ ($r_0>0$), the relaxation time
remains finite even for the largest islands.  Close to $T_c^0$ it behaves as
$\xi_t(L_{\rm RR},r_0) \sim \xi^z \sim r_0^{-z\nu}$ for sufficiently large islands
($L_{\rm RR}>\xi$). Here $\xi$ is the clean bulk correlation length and $z$ and $\nu$ are
the clean bulk critical exponents. The asymptotic time dependence of $C(t)$ in this
regime is a simple exponential.
\be
\ln C(t) \sim -t/\xi^z~.
\label{eq:C(t)_pm}
\ee

(ii) At the clean critical temperature $T=T_c^0$ ($r_0=0$), i.e., at the boundary of the
Griffiths region, the relaxation time of the large rare regions is given by finite-size
scaling \cite{Barber_review83}. It diverges as $\xi_t(L_{\rm RR}) \sim L_{\rm RR}^z$.
After inserting this into (\ref{eq:C(t)_RR}), the integral over $L_{\rm RR}$ can be
estimated using the saddle-point method. The asymptotic time dependence of $C(t)$ is a
stretched exponential
\be
\ln C(t) \sim - \tilde{p}^{z/(d+z)} \, t^{d/(d+z)}~.
\label{eq:C(t)_tc0}
\ee

(iii) Inside the Griffiths region, $T_c(p) < T <T_c^0$, the relaxation time is limited by
the time taken to coherently reverse the entire rare region. Such a reversal requires
creating a domain wall whose free energy is $\sigma L_{\rm RR}^{d-1}$. Therefore, the
relaxation time of an island of linear size $L_{\rm RR}$ is given by $\xi_t(L_{\rm RR})
\sim \tau_0 \exp[\sigma L_{\rm RR}^{d-1}]$. The factor $1/T$ in the Arrhenius behavior
has been absorbed into the surface tension $\sigma$, and $\tau_0$ is a microscopic time
scale. For large $t$, the integral (\ref{eq:C(t)_RR}) can be calculated using the saddle
point method leading to an even slower time dependence of the form
\be
\ln C(t) \sim -(\ln t)^{d/(d-1)}~.
\label{eq:C(t)_Griffiths}
\ee
More recently, this asymptotic decay of the autocorrelation function was established
rigorously for disordered discrete lattice spin systems with \emph{bounded} disorder
distributions \cite{CesiMaesMartinelli97a,CesiMaesMartinelli97b}. It should be noted
however, that in Monte-Carlo simulations \cite{Jain88,ColborneBray89,AndreichenkoSelkeTalapov92,Jain95}
the time dependence of $C(t)$ seems to follow a stretched exponential form $\ln C(t) \sim -t^{x}$
rather than the asymptotic form (\ref{eq:C(t)_Griffiths}). This suggests that the
asymptotic regime is only reached after a very long transient time interval.
We also point out that (\ref{eq:C(t)_Griffiths}) is universally valid only for bounded disorder
distributions; unbounded distributions can lead to asymptotic forms different
from (\ref{eq:C(t)_Griffiths}) as was shown in Ref.\ \cite{BrayRodgers88}.

So far our discussion of dynamic rare region effects has focused on Ising systems. For
continuous spin symmetry, the behavior of the autocorrelation function at and above the
clean critical temperature $t_c^0$ remains to be given by relations (\ref{eq:C(t)_pm})
and (\ref{eq:C(t)_tc0}). However, in the Griffiths region, $T_c(p) < T <T_c^0$, the
behavior changes. The relaxation time of a rare region is much shorter than for Ising
spins since there is no free energy barrier hindering relaxation. Rather, the relaxation
occurs by diffusion of the order parameter, driven by thermal noise. Following Bray
\cite{Bray88}, the relaxation time can be estimated as follows: The change of cluster
magnetization during some small time interval $\delta t$ is $\delta \mathbf M \sim L_{\rm
RR}^{d/2}$ since the noise at different sites adds incoherently. Since several of such
steps also act incoherently, the change of magnetization after time $t$ is $\delta
\mathbf M (t) \sim t^{1/2}L_{\rm RR}^{d/2}$. Complete relaxation has occurred for $\delta
\mathbf M (t) \sim \mathbf M \sim L_{\rm RR}^d$ giving $\xi_t \sim L_{\rm RR}^d$. Putting
this into (\ref{eq:C(t)_RR}) and evaluating the integral by the saddle point method gives
\be
\ln C(t) \sim -t^{1/2}~,
\label{eq:C(t)_Griffiths_Heisenberg}
\ee
a much faster decay than the corresponding equation (\ref{eq:C(t)_Griffiths}) for the
Ising case. The same result has also been derived in the large-$N$ limit (where $N$
refers to the number of spin components) \cite{Bray87}. In contrast to the Ising case,
Monte-Carlo simulations of a diluted Heisenberg magnet \cite{ColborneBray89} are in
good agreement with (\ref{eq:C(t)_Griffiths_Heisenberg}) suggesting that the asymptotic regime
is reached much faster in the Heisenberg case than in the Ising case.

\subsection{Disorder correlations and enhanced rare region effects}
\label{subsec:McCoyWu}

In addition to localized defects such as impurity atoms or vacancies, many realistic systems
also contain extended defects such as dislocations or grain boundaries. They can be modelled by quenched disorder
that is perfectly correlated in one (linear defects) or two (planar defects) dimensions.

It is interesting to ask whether such disorder correlations increase or decrease the
influence of the defects. Naively, one might expect that disorder which is perfectly
correlated in $d_{\rm cor}$ dimensions but uncorrelated in the remaining $d_{\rm
ran}=d-d_{\rm cor}$ dimensions is weaker than uncorrelated disorder because the system is
``random only in some of the directions''. However, this naive assumption turns out to be
incorrect. In fact, the opposite is true: Long-range disorder correlations enhance the
influence of the defects because ``it is harder to average out an extended fluctuation
than a local one''.

The first systematic study of a phase transition in the presence of extended defects was
carried out more than 30 years ago by McCoy and Wu. In a series of papers
\cite{McCoyWu68, McCoyWu68a,McCoyWu69,McCoy69,McCoy69a,McCoy70}, these authors studied a
disordered two-dimensional Ising model, now known as McCoy-Wu model, in which all
horizontal bonds are identical while all vertical bonds in each row are the same.
However, the vertical bonds vary from row to row as independent random variables. Thus,
the disorder is perfectly correlated in one dimension (the horizontal direction) but
uncorrelated in the other (the vertical direction). This implies $d_{\rm cor}=d_{\rm
ran}=1$.

The McCoy-Wu model in zero magnetic field is partially exactly solvable with
transfer-matrix type methods (see also \cite{ShankarMurthy87}). McCoy and Wu
\cite{McCoyWu68a} analyzed the free energy and found that there is only an essential
singularity at the critical temperature. No exact results could be obtained for the
spontaneous magnetization, but McCoy \cite{McCoy69,McCoy69a} calculated the properties of the
spins on the surface of a half-plane cut perpendicular to the random direction.
Because of Griffiths' theorem \cite{Griffiths67}, magnetization and susceptibility of
these boundary spins provide lower bounds for the corresponding bulk values. McCoy found
the surprising result that the susceptibility is infinite for an entire range of
temperatures above the critical temperature $T_c$. He also found that there is an even
larger range of temperatures in which the susceptibility exists but the magnetization
is still a nonanalytic function of the field. These unusual results initially gave rise
to some confusion as to whether the phase transition in the McCoy-Wu model is sharp or
smeared.

By now, it has become clear that the transition is actually sharp, i.e., the spontaneous magnetization
develops as a collective effect of the whole system at the critical temperature $T_c$
which is also the (only) temperature at which the spatial correlation length becomes infinite.
The unusual behavior of the susceptibility in the vicinity of the critical temperature
is a rare region effect that can be understood as follows: For simplicity, let us assume
that the random vertical bonds in the McCoy-Wu model have a binary probability distribution
\be
P(J) = (1-p)\, \delta(J-J_0) + p\, \delta(J-cJ_0) \label{eq:binary} \ee with $c<1$. Thus,
$p$ is the probability of a vertical bond being weak. In an infinite sample, one can find
arbitrarily large ``stacks'' of rows with only strong vertical bonds. The probability for
finding such a rare region of width $L_{\rm RR}$ is given by
\be
w(L_{\rm RR}) \sim (1-p)^{L_{\rm RR}} \sim \exp(-\tilde{p} L_{\rm RR}) \label{eq:w(L)_MW}
\ee with $\tilde{p} = -\ln(1-p)$. In the Griffiths region below the transition
temperature $T_c^0$ of the clean system with only strong vertical bonds, these rare
regions are locally in the ordered phase. Since each rare region is infinite in the
horizontal direction and of finite width $L_{\rm RR}$ in the vertical direction, it is
equivalent to a one-dimensional Ising model with an effective interaction proportional to
$L_{\rm RR}$. As such, it cannot undergo a phase transition by itself but its
susceptibility increases exponentially with its width,
\be
\chi(L_{\rm RR}) \sim \frac 1 T \exp(a L_{\rm RR}) \label{eq:chi_l_MW} \ee where the
constant $a$ vanishes at the clean critical temperature $T_C^0$ and increases with
decreasing temperature (see, e.g., \cite{Goldenfeld_book92}).

The rare region contribution to the total susceptibility can be obtained by summing over
all rare regions,
\be
 \chi_{\rm RR} \sim \int dL_{\rm RR}~ w(L_{\rm RR})~ \chi(L_{\rm RR})~.
\label{eq:chi_RR_MW}
\ee
This integral diverges when $a > \tilde{p}$ \emph{before} the true critical point of the dirty system is
reached (the divergence does \emph{not} require coherence between the rare regions).
Thus, in the Griffiths region of the McCoy-Wu model, the exponential rarity of large rare
regions is overcome by an exponential increase of a rare region's susceptibility with
its size. The probability distribution $w_\chi$ of local susceptibilities $\chi_l$ can be obtained by
combining (\ref{eq:w(L)_MW}) and (\ref{eq:chi_l_MW}), leading to
\be
w_\chi(\chi_l) \sim \chi_l^{-1/z'-1} \label{eq:w(chi)_MW} \ee where $z'=a/\tilde{p}$ is
an exponent that varies continuously throughout the Griffiths region. In contrast, the
corresponding probability distribution for the the case of uncorrelated disorder
decreases exponentially for large $\chi_l$ because the susceptibility of a rare region
only increases like a power of $L_{\rm RR}$, see equation (\ref{eq:m(L)}). The perfect
disorder correlations in the McCoy-Wu model therefore enhance the rare region effects
from exponentially weak classical Griffiths singularities to much stronger power-law
singularities that are sometimes called Griffiths-McCoy singularities.

Let us emphasize that none of the peculiar properties of McCoy and Wu's exact solution
are reflected in the standard perturbative renormalization group approach to the problem.
Early renormalization group work \cite{Lubensky75} based on a single expansion in
$\epsilon=4-d$ did not produce a critical fixed point. This was interpreted as a smeared
transition or a first order one \cite{Rudnick78,AndelmanAharony85}. An alternative
approach, the double-$\epsilon$ expansion in powers of $\epsilon=4-d$ and the number
$d_{\rm cor}$ of correlated dimensions \cite{Dorogovtsev80,BoyanovskyCardy82,DeCesare94}
gave rise to a critical fixed point with a conventional power-law singularity in the free
energy rather than the essential singularity found in McCoy and Wu's solution.  However,
the general arguments on the importance of rare regions made in section
\ref{subsec:rrclass} point to a qualitative difference between $d_{\rm cor}=1$ in the
McCoy-Wu model and $d_{\rm cor} \ll 1$ assumed in the double-$\epsilon$ expansion. For
$d_{\rm cor}=1$, the rare regions (whose effective dimension $d_{\rm RR}=d_{\rm cor}$)
are at the lower critical dimension $d_c^-$, leading to the exponential size dependence
(\ref{eq:chi_l_MW}) of the susceptibility. In contrast, for $d_{\rm cor}<1$, the rare
regions are below $d_c^-$. Thus, the susceptibility depends on the rare region size via a
power law, and the Griffiths effects in the double-$\epsilon$ expansion are exponentially
small. Therefore it is not surprising that the double-$\epsilon$ expansion approach
cannot capture the physics of the McCoy-Wu model.

The partition function of the McCoy-Wu model written in terms of the transfer matrix in
horizontal direction is essentially equivalent to the partition function of a quantum
mechanical random transverse-field Ising model \cite{ShankarMurthy87,Sachdev_book99}.
Much insight can therefore be obtained by studying this quantum Ising model. In fact, the
behavior of the McCoy-Wu model remained incompletely understood until Fisher
\cite{Fisher92,Fisher95} applied a version of the Ma-Dasgupta-Hu \cite{MaDasguptaHu79}
real-space renormalization group to the random transverse-field Ising model. We will
discuss these developments in more detail in sections \ref{subsec:rtim} to
\ref{subsec:sdrg}.

\subsection{Smeared phase transition in an Ising model with planar defects}
\label{subsec:planar}

In the last subsections we have seen that perfect disorder correlations in one direction
greatly enhance the rare region effects. In this subsection, we consider planar defects,
i.e., disorder perfectly correlated in two dimensions. Recently, it has been shown
\cite{Vojta03b,SknepnekVojta04} that they have even more dramatic consequences: They completely
destroy the sharp Ising phase transition by smearing.

Most of our arguments below will be rather general. If necessary for definiteness, we
focus on a three-dimensional Ising model with nearest-neighbor interactions on a cubic
lattice. In the clean system, all interactions are identical and have the value $J_0$.
The defects are modelled via 'weak' bonds randomly distributed in one dimension
(uncorrelated direction). The bonds in the remaining two dimensions (correlated
directions) remain equal to $J_0$. The system effectively consists of blocks separated by
parallel planes of weak bonds. Thus, $d_{\rm ran}=1$ and $d_{\rm cor}=2$. The Hamiltonian
of the system is given by:
\begin{eqnarray}
\fl H=&&-\sum_{{i=1,\dots,L_{\rm ran}} \atop {j,k=1,\dots,L_{\rm
cor}}}J_iS_{i,j,k}S_{i+1,j,k} - \sum_{{i=1,\dots,L_{\rm ran}} \atop {j,k=1,\dots,L_{\rm
cor }}}J_0(S_{i,j,k}S_{i,j+1,k}+S_{i,j,k}S_{i,j,k+1}), \label{eq:H_planar}
\end{eqnarray}
where $L_{\rm ran}$($L_{\rm cor}$) is the length in the uncorrelated (correlated)
direction, $i$, $j$ and $k$ are integers counting the sites of the cubic lattice, and
$J_i$ is the random coupling constant in the uncorrelated direction. The $J_i$ are drawn
from a binary distribution:
\be
P(J) = (1-p)\, \delta(J-J_0) + p\, \delta(J-cJ_0)
\label{eq:binary_sm}
\ee
characterized by the concentration $p$ and the relative strength $c$ of the weak bonds ($0<c\le 1$).

Let us now consider rare regions in this system which consist of ``thick slabs''
containing only strong bonds in the uncorrelated direction. As in the case of point or
linear defects, for temperatures below the clean critical temperature $T_c^0$, these rare
regions are locally in the ordered phase even if bulk system is still in the disordered
(paramagnetic) phase. However, the behavior of locally ordered planar rare regions
differs \emph{qualitatively} of that of localized or linear rare regions. Each planar
rare region is infinite in the two correlated dimensions but finite in the uncorrelated
direction. Thus, it is equivalent to a two dimensional Ising model that can undergo a
real phase transition independently of the rest of the system. Thus, each rare region can
independently develop true static order with a non-zero static value of the local
magnetization. (Griffiths theorem \cite{Griffiths67} ensures that the interaction of the
rare region with the paramagnetic bulk system can only \emph{increase} its magnetization
over that of a completely isolated ``slab''.) Once static order has developed, the
magnetizations of different rare regions can be aligned by an infinitesimally small
interaction or external field. The resulting phase transition will thus be markedly
different from a conventional continuous phase transition. At a conventional transition,
a non-zero order parameter develops as a collective effect of the entire system which is
signified by a diverging correlation length of the order parameter fluctuations at the
critical point. In contrast, in a system with planar defects, different parts of the
system (in the uncorrelated direction) will order independently, at different
temperatures. Therefore the global order will develop inhomogeneously and the correlation
length in the uncorrelated direction will remain finite at all temperatures. This defines
a \emph{smeared} or \emph{rounded} phase transition.

\begin{figure}
\centerline{\includegraphics[width=10cm]{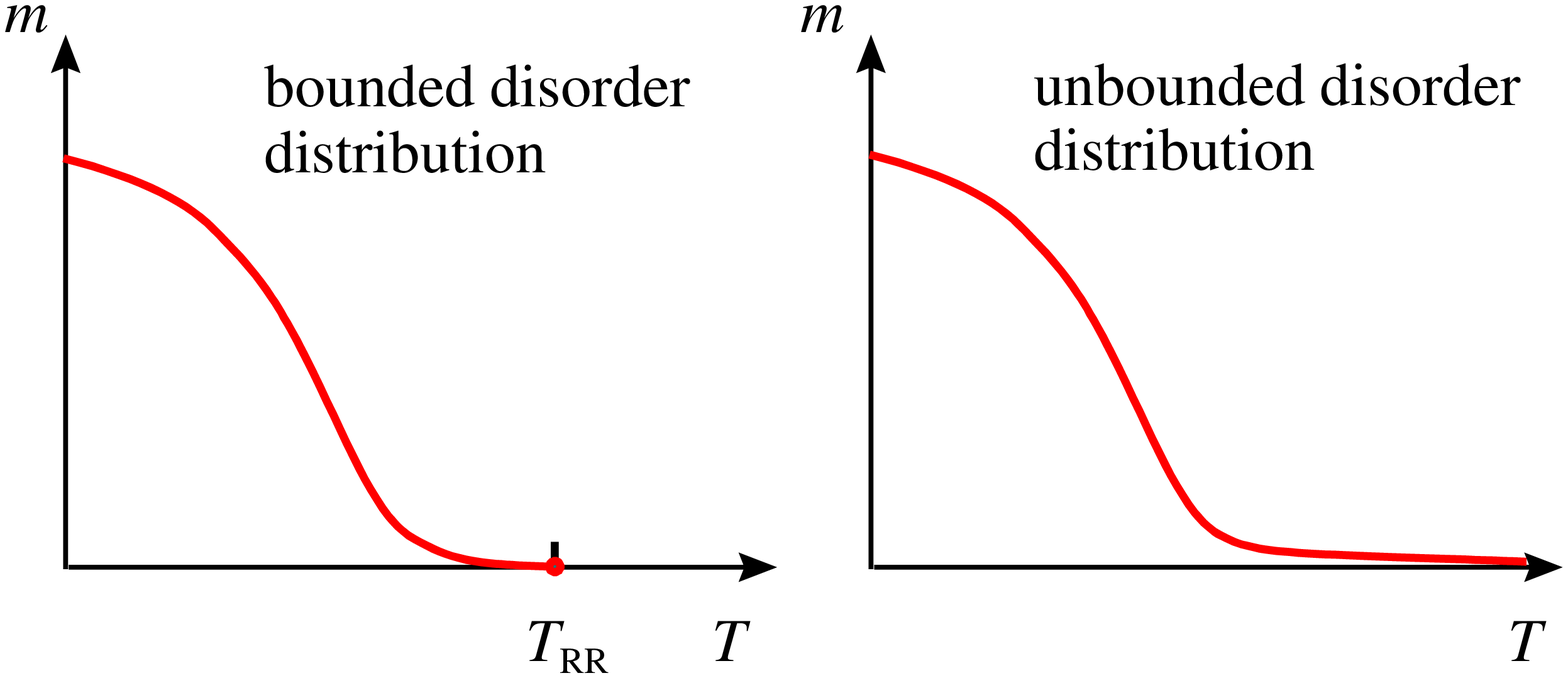}}
\caption{Schematic behavior of the magnetization in the tail of a smeared phase transition
for bounded and unbounded disorder distributions. For details see text.}
\label{fig:mag_smeared}
\end{figure}

Note that similar to the Griffiths effects discussed in the proceeding sections, the properties of
the smeared transition depend on whether or not the disorder distribution is bounded (see figure
\ref{fig:mag_smeared}).
If the disorder distribution is bounded like the binary distribution
(\ref{eq:binary_sm}), there is a true paramagnetic phase with zero total order parameter
at high temperatures. At some temperature $T_{\rm RR}$ (identical to the clean critical
temperature $T_c^0$ for the distribution (\ref{eq:binary_sm})), a non-zero order
parameter starts to develop on the rare regions, accompanied by an essential singularity
in the free energy density (which stems from the probability for finding a rare region).
In contrast, for distributions that are unbounded in the sense that they permit rare
regions with a local $T_c=\infty$ (e.g., a Gaussian distribution), the total order
parameter is non-zero for \emph{all} temperatures, and the free energy density is
analytic.

%%%%%%%%%%%%%%%%%%%%%%%%%%%%%%%%%%%%%%%%%%%%%%%%%%%%%%%%%%%%%%%%%%%%%%%%%%%%%%%%%%%%%%%%%%%
\subsection{Optimal fluctuation theory for smeared phase transition}
\label{subsec:optimalfluc}

The leading thermodynamic behavior in the tail of the smeared transition can be
determined using optimal fluctuation theory similar to section \ref{subsec:OFT}.
Following \cite{Vojta03b}, we develop this theory for the general case of $d_{\rm cor}\ge
2$ correlated dimensions and $d_{\rm ran}=d-d_{\rm cor}$ uncorrelated dimensions. We
first consider bounded disorder distributions like the binary distribution
(\ref{eq:binary_sm}) and later briefly describe the differences for Gaussian disorder.

The probability $w$ of finding a thick slab of width $L_{\rm RR}$ devoid of any weak
bonds is, up to pre-exponential factors, given by
\begin{equation}
w(L_{\rm RR}) \sim \exp( -\tilde{p} L_{\rm RR}^{d_{\rm ran}})~. \label{eq:w(LRR)_planar}
\end{equation}
with $\tilde{p}=-\ln(1-p)$. Such a rare region develops static long-range order at some
temperature $T_c(L_{\rm RR})$ below the clean critical point $T_c^0$. The value of
$T_c(L_{\rm RR})$ varies with the size of the region: The largest islands develop
long-range order closest to the clean critical point. As discussed in section
\ref{subsec:FSS}, finite size scaling for the clean system (because the rare regions are
free of defects) yields
\begin{equation}
T_c^0 - T_c(L_{\rm RR}) =|r_c(L_{\rm RR})| = A \,L_{\rm RR}^{-\phi} \label{eq:FSS}
\end{equation}
where $\phi$ is the finite-size scaling shift exponent and $A$ is the amplitude for the
crossover from $d$ dimensions to a slab geometry infinite in $d_{\rm cor}$ dimensions but
finite in $d_{\rm ran}=d-d_{\rm cor}$ dimensions. If the total dimensionality $d$ is
below the upper critical dimension $d_c^+=4$, hyperscaling is valid and the shift
exponent $\phi=1/\nu$. The reduced temperature $r=T-T_c^0$ measures the distance from the
\emph{clean} critical point. Combining (\ref{eq:w(LRR)_planar}) and (\ref{eq:FSS}) we
obtain the probability for finding a rare region which becomes critical at $r_c$ as
\begin{equation}
w(r_c) \sim \exp  (-B ~|r_c|^{-d_{\rm ran}/\phi}) \qquad (\textrm{for } r\to 0-)
 \label{eq:w-dilute}
\end{equation}
where the constant $B$ is given by $B=\tilde{p}\,A^{d_{\rm ran}/\phi}$.  The total (or
average) order parameter $m$ is obtained by integrating over all rare regions having
$r_c(L_{\rm RR})>r$. Since the functional dependence on $r$ of the order parameter on a
given island is of power-law type it only enters the pre-exponential factors. Therefore
we obtain to exponential accuracy
\begin{eqnarray}
m(r) \sim \exp  (-B ~|r|^{-d_{\rm ran}/\phi}) \qquad (\textrm{for } r\to 0-)~.
\label{eq:m-dilute}
\end{eqnarray}
Once a nonzero density of ordered and aligned rare regions exist, the Ising symmetry is
spontaneously globally broken. The rare regions produce an effective background magnetic field
everywhere in space which cuts off any possible further singularities.

The spatial magnetization distribution in the tail of the smeared transition is very
inhomogeneous. On the already ordered islands, the local order parameter $m(\mathbf{x})$
is of the same order of magnitude as in the clean system. Away from these islands it
decays exponentially with the distance from the nearest island. The {\em typical} local
order parameter $m_{\rm typ}$ can be estimated via the typical distance of any point from
the nearest ordered island. From (\ref{eq:w-dilute}) we obtain
\begin{equation}
x_{\rm typ} \sim \exp  (B ~|r|^{-d_{\rm ran}/\phi}/d_{\rm ran}) ~. \label{eq:xtyp}
\end{equation}
At this distance from an ordered island, the local order parameter has decayed to
\begin{equation}
m_{\rm typ} \sim e^{-x_{\rm typ}/\xi_0} \sim \exp \left[ -C \exp(B ~|r|^{-d_{\rm ran}
/\phi}/d_{\rm ran})\right]~ \label{eq:mtyp}
\end{equation}
where $\xi_0$ is the bulk correlation length (which is finite and changes slowly
throughout the tail region of the smeared transition)  and $C$ is constant. A comparison
with (\ref{eq:m-dilute}) gives the relation between $m_{\rm typ}$ and the thermodynamic
order parameter $m$,
\begin{equation}
|\log m_{\rm typ}| \sim m^{-1/d_{\rm ran}} \label{eq:mtypav}~.
\end{equation}
Thus, $m_{\rm typ}$ decays exponentially with $m$ indicating an extremely broad
distribution of the local order parameter $m_\mathbf{x}=\langle S_\mathbf{x} \rangle$.
Its functional form has been determined in reference \cite{Vojta03b}, it reads
\begin{equation}
 P[\log(m_\mathbf{x})] \sim |\log(m_\mathbf{x})|^{d_{\rm ran}-1}   \qquad (\textrm{for } m_\mathbf{x} \ll 1)~.
 \label{eq:plogm}
\end{equation}

These predictions of the optimal fluctuation theory have been tested by large scale
computer simulations. A  model with infinite-range interactions in the correlated
directions was studied in reference \cite{Vojta03b}. This mean-field type calculation
permitted large system sizes of up to $L_{\rm ran}=10^6$ making it very suitable for
identifying effects of rare events. The results were in excellent agreement with the
theoretical predictions. Note that a similar mean-field model had been investigated
earlier \cite{BBIP98} but for much smaller sizes of up to $L_{\rm ran}=1000$. The raw
data in this paper are very similar to reference \cite{Vojta03b}, but they were
interpreted in terms of power-law critical behavior with an unusually large order
parameter exponent $\beta\approx 3.6$. This may be partially due to an unfortunate choice
of parameters (in particular, a higher impurity concentration of $p=0.5$) which makes it
hard to extract the functional form of the magnetization tail.

\begin{figure}[t]
\centerline{\includegraphics[width=7cm]{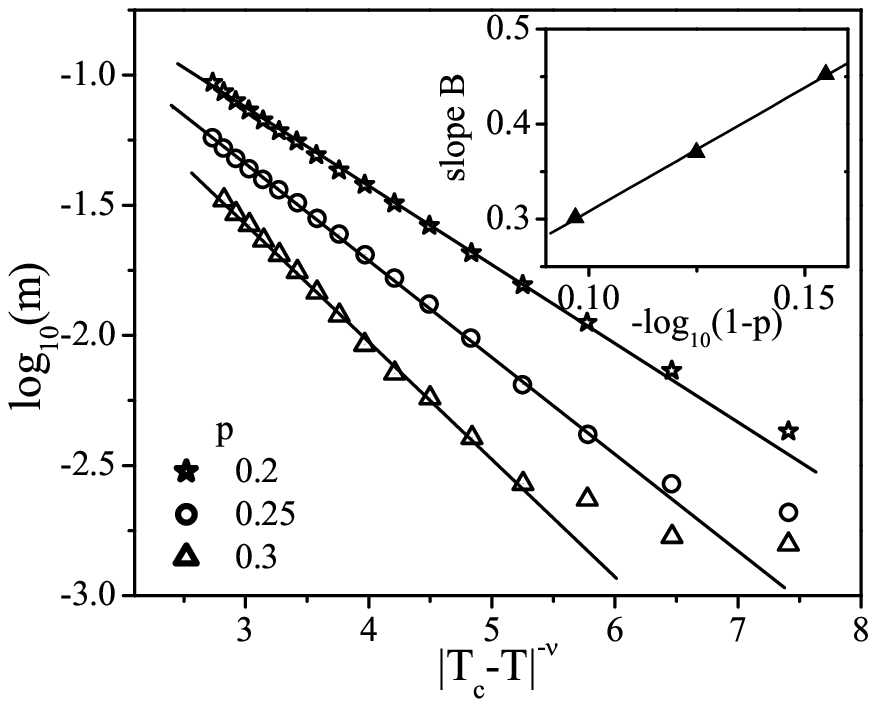}\includegraphics[width=7cm]{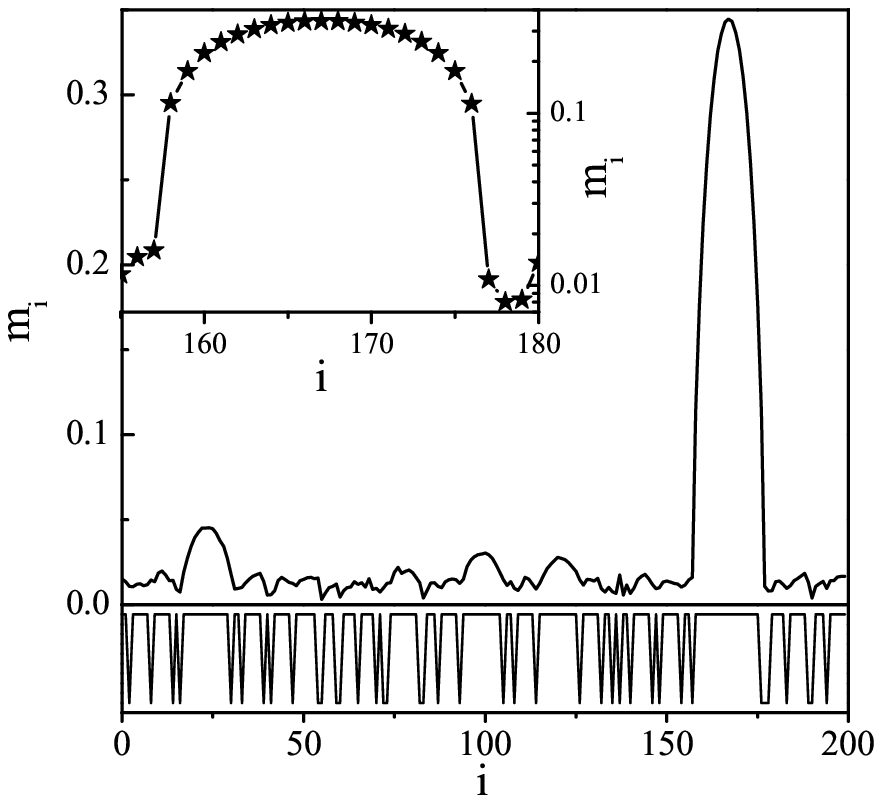}}
\caption{Left: Logarithm of the magnetization $m$ as a function of $|T_c^0-T|^{-\nu}$
($\nu=0.627$) for impurity concentrations $p=0.2, 0.25, 0.3$ and strength $c=0.1$. Inset:
Decay slope $B$ as a function of $-\log(1-p)$. Right: Local magnetization $m_i$ as
function of position $i$ in the uncorrelated direction at temperature $T=4.425$ (one
particular disorder realization). Lower panel: Coupling constant $J_i$ in the
uncorrelated direction as a function of position $i$. Inset: Log-linear plot of the
region in the vicinity of the largest ordered island (from \cite{SknepnekVojta04}).}
\label{fig:logmag1}
\end{figure}

Recently, the short-range Ising model with planar defects (\ref{eq:H_planar}) has been
studied by large scale Monte-Carlo simulations \cite{SknepnekVojta04} using the Wolff
cluster algorithm  \cite{Wolff89}. The results confirm the smeared transition scenario
and the predictions of optimal fluctuation theory for a realistic model. A typical result
for the magnetization in the tail of the smeared transition below $T_c^0\approx 4.511$ is
presented in the left panel of figure \ref{fig:logmag1}. It shows the logarithm of the
total magnetization vs. $|T_c^0-T|^{-\nu}$ averaged over 240 samples for system size
$L_{\rm ran}=200$, $L_{\rm cor}=280$ and three disorder concentrations
$p=\{0.2,0.25,0.3\}$.
Here, $\nu=0.627$ is the three-dimensional clean critical Ising exponent. For all three
concentrations the data follow (\ref{eq:m-dilute})
over more than an order of magnitude in $m$. The deviation from the straight line for
small $m$ is due to the conventional finite size effects. In the inset we show that
the decay constant $B$ depends linearly on $\tilde{p}=-\log(1-p)$.

The right panel of figure\ \ref{fig:logmag1} illustrates the very inhomogeneous character
of the local magnetization in the tail of the smeared transition. It displays the local
magnetization $m_i$ of a particular disorder realization as a function of the position
$i$ in the uncorrelated direction for the size $L_{\rm ran}=200$, $L_{\rm cor}=200$ at a
temperature $T=4.425$ in the tail of the smeared transition. The lower panel shows the
local coupling constant $J_i$ as a function of $i$. The figure shows that a sizable
magnetization has developed on the longest island only (around position $i=160$). One can
also observe that order starts to emerge on the next longest island located close to
$i=25$. Far form these islands the system is still in its disordered phase.

As discussed in the last subsection, the functional form of the tail of the smeared
transition is nonuniversal. In particular, for unbounded disorder distributions, it
differs from (\ref{eq:m-dilute}). In the case of Gaussian disorder, the magnetization for
large $r$ behaves as $m \sim \exp(-B r^{2-d_{\rm ran}/\phi})$, i.e., he magnetization
tail reaches all the way to $T=\infty$ \cite{Vojta03b}.

\subsection{Dynamics at a smeared phase transition}
\label{subsec:dynsmeared}

The static magnetic order on the rare regions in an Ising model with planar defects also
modifies the dynamics. The leading long-time behavior can be determined by generalizing
the approach of section \ref{subsec:dynamics} to the case of extended defects, as was
done in reference \cite{FendlerSknepnekVojta05}. The behavior above and at the clean
critical temperature $T_c^0$ turns out to be identical to the conventional dynamic Griffiths
effects discussed in section \ref{subsec:dynamics}.  This is not surprising because above
$T_c^0$ there is no difference between the Griffiths and the smearing scenarios: All rare
regions are locally still in the disordered phase. Specifically, the asymptotic time
dependence of the spin autocorrelation function $C(t)$ right at the clean critical temperature
is given by
\be
\ln C(t) \sim -\tilde{p}^{z/(d_{\rm ran}+z)}\, t^{d_{\rm ran}/(d_{\rm ran}+z)}
\label{eq:C(t)_planar} \ee where $z$ is the clean critical exponent of the
$d$-dimensional system. For temperatures below $T_c^0$, the behavior at a smeared
transition differs from the dynamic Griffiths effects of section \ref{subsec:dynamics}.
Here, at short times before a crossover time $t_x$, the decay of $C(t)$ is still given by
by the stretched exponential (\ref{eq:C(t)_planar}). For $t>t_x$, the system realizes
that some of the rare regions have developed static order and contribute to a non-zero
equilibrium value $C(\infty)$ of the autocorrelation function $C(t)$. The approach of
$C(t)$ to this equilibrium value follows a nonuniversal power law
\begin{equation}
C(t) - C(\infty) \sim t^{-\psi}~. \label{eq:power}
\end{equation}
The value of $\psi$ could not be found since it depends on the pre-exponential factors
neglected in the optimal fluctuation theory.

\begin{figure}[t]
\centerline{\includegraphics[width=6.6cm]{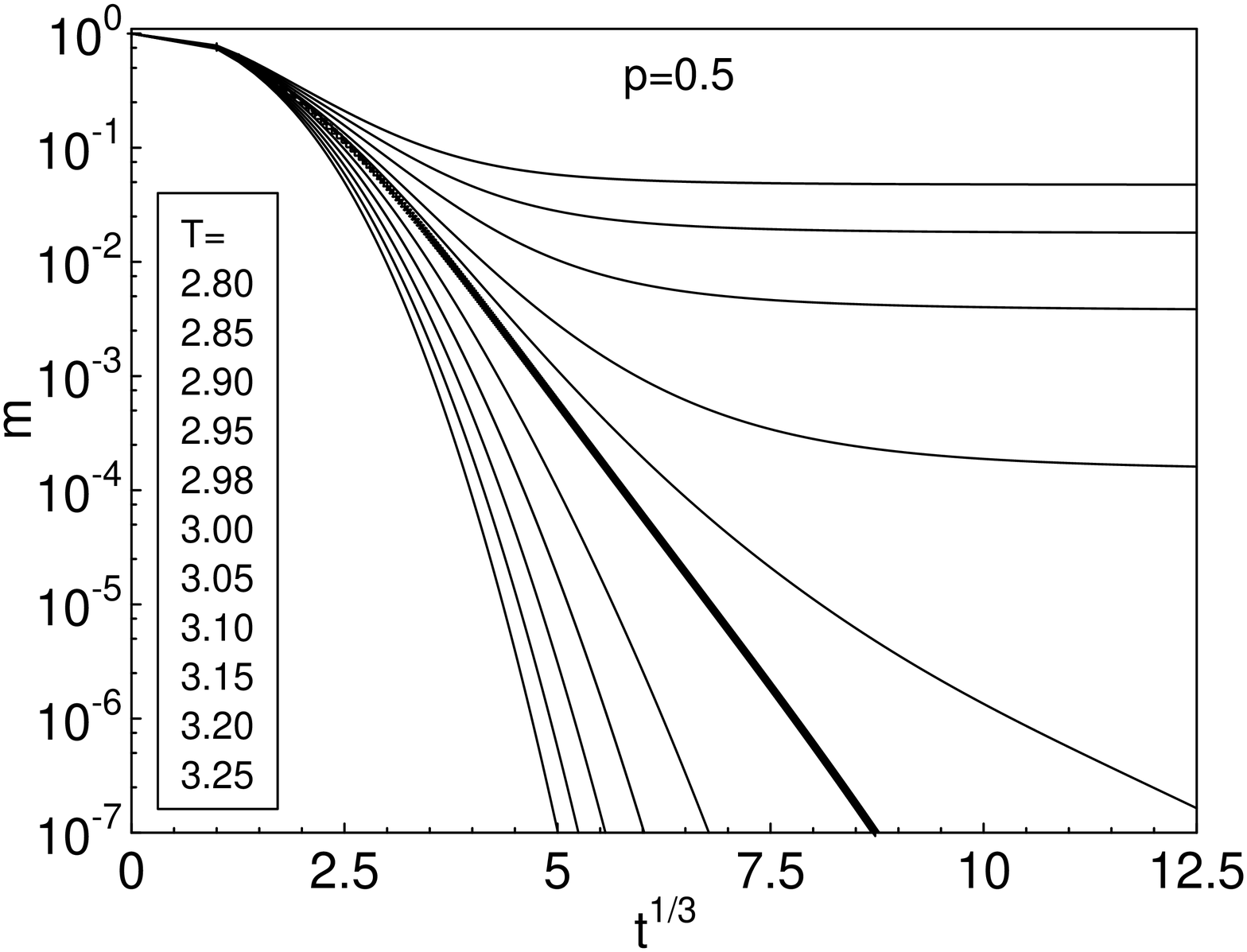}\includegraphics[width=6.6cm]{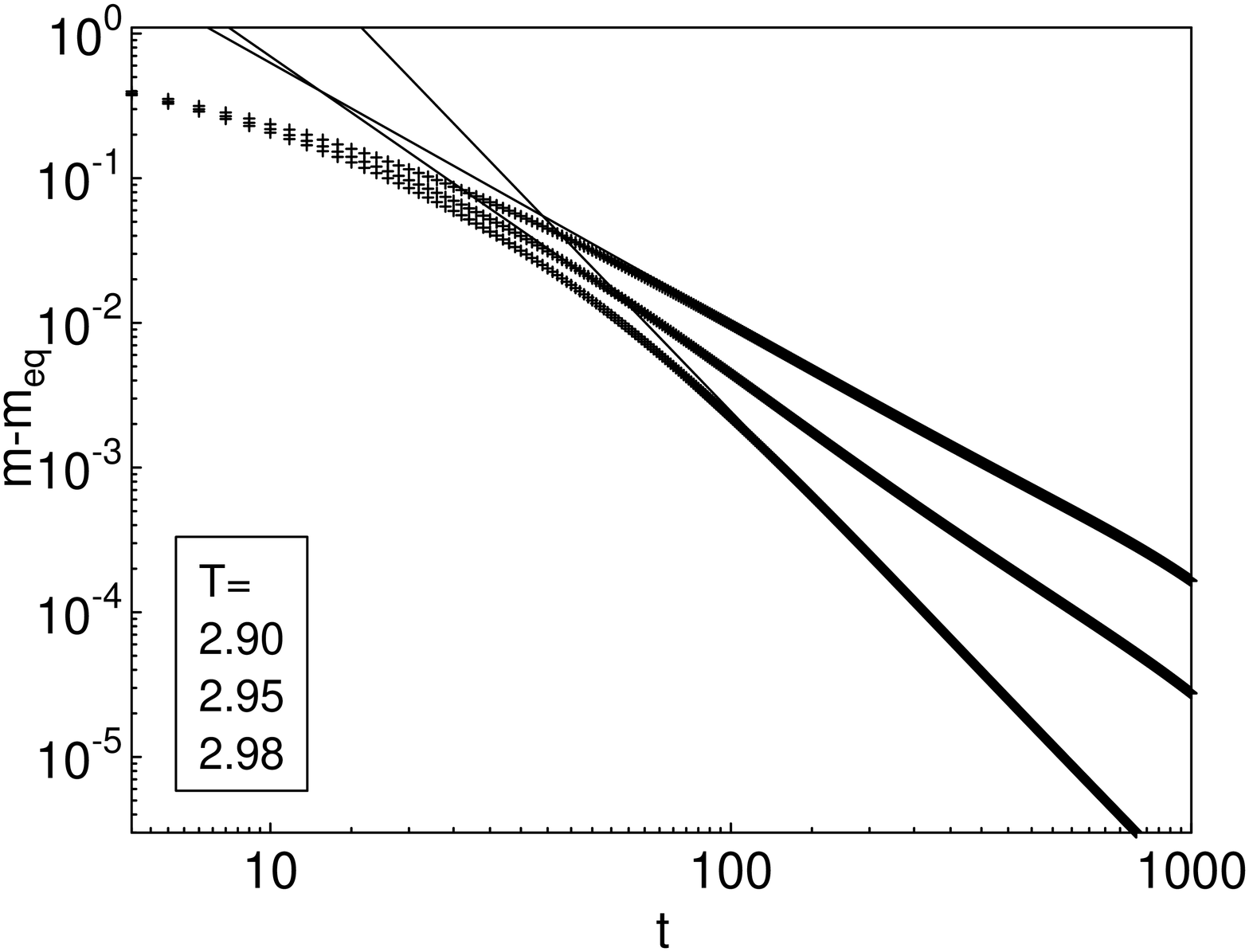}}
\caption{Left: Time evolution of the magnetization for several temperatures from $T=2.8$ to 3.25
(top to bottom) in the vicinity of the clean critical point $T_c^0=3$. The impurity concentration
is $p=0.5$ their strength is $c=0.2$. The runs start from a
fully polarized state at $t_0=0$ and go to $t_{max}=2000$. Right: Double-logarithmic plot of the
approach of the magnetization to the equilibrium value for $T=2.90$, 2.95 and 2.98 (top to bottom).
The straight lines are fits to power laws yielding exponents of 1.78,
2.20, and 3.31, respectively
(from \cite{FendlerSknepnekVojta05}).} \label{fig:dyn_planar}
\end{figure}

Reference \cite{FendlerSknepnekVojta05} also discussed the rare region contribution to the
nonequilibrium relaxation of the magnetization in the tail of the smeared transition.
It turned out to be controlled by decay laws with the same functional form as those of
the autocorrelation function. These results were confirmed by computer simulations of a
infinite-range model similar to that in reference \cite{Vojta03b}. Figure
\ref{fig:dyn_planar} shows a typical result of these simulations.
The left panel shows the time evolution of the magnetization of a large system of $L_{\rm
ran} =10^5$ sites, averaged over 25 disorder realizations, for several temperatures from
$T=2.8$ to 3.25 in the vicinity of the clean critical point which is $T_c^0=3$ for the
infinite-range model. The behavior at $T_c^0$ follows the stretched exponential $\ln m(t)
\sim -t^{1/3}$ as expected from the analog of (\ref{eq:C(t)_planar}) because the
mean-field value of the clean dynamical exponent is $z=2$,  and $d_{\rm ran}=1$. For
$T>T_c^0$, the magnetization decay is faster, and a detailed analysis
\cite{FendlerSknepnekVojta05} shows it to be simple exponential. For $T<T_c^0$, the
magnetization approaches a nonzero equilibrium value $m_{eq}$. The approach to this
equilibrium value is analyzed in the  right panel of figure \ref{fig:dyn_planar}. The
difference $m(t)-m_{eq}$ follows a nonuniversal power law $m(t)-m_{eq} \sim t^{-\psi}$.

%%%%%%%%%%%%%%%%%%%%%%%%%%%%%%%%%%%%%%%%%%%%%%%%%%%%%%%%%%%%%%%%%%%%%%%%%%%%%%%%%%%%%%%%%%%%
\section{Quantum phase transitions}
\label{sec:quantum}
%%%%%%%%%%%%%%%%%%%%%%%%%%%%%%%%%%%%%%%%%%%%%%%%%%%%%%%%%%%%%%%%%%%%%%%%%%%%%%%%%%%%%%%%%%%%

In section \ref{sec:classical}, we have seen that rare region effects in classical
systems with generic short-range-correlated disorder are exponentially weak and that
stronger effects require long-range disorder correlations. In this section, we discuss
quantum phase transitions which occur at zero temperature and thus require the inclusion
of the imaginary time direction, as discussed in section \ref{sec:pt}. Quenched
impurities are time independent, the disorder is therefore always correlated in at least
one of the relevant dimensions, and we generically expect rare region effects at a
quantum phase transition to be stronger than at a classical transition.

The main focus of this section will again be on order-disorder transitions between
conventional phases that can be described by Landau-Ginzburg-Wilson theories. However,
the concept of quantum rare region effects or quantum Griffiths singularities is much
broader, and in section \ref{subsec:other_qpt}, we will briefly discuss other examples.

%%%%%%%%%%%%%%%%%%%%%%%%%%%%%%%%%%%%%%%%%%%%%%%%%%%%%%%%%%%%%%%%%%%%%%%%%%%%%%%%%%%%%%%%%%%%
\subsection{Random transverse field Ising model}
\label{subsec:rtim}

One of the prototypical models (if not \emph{the} prototypical model) displaying a quantum
phase transition is the transverse field Ising model. Its Hamiltonian
\be
H= -\sum_{\langle i,j\rangle} J_{ij}\,\hat{S}_i^z \hat{S}_j^z - \sum_i h_i^x \,
\hat{S}_i^x \label{eq:rtim} \ee consists of two noncommuting parts, the first being the
exchange interaction between the $z$-components of the spin-$\frac 1 2$ operators
$\hat{S}_i$ on nearest neighbor sites of a hypercubic lattice in $d$ space dimensions.
The second part describes the interaction of $x$-components of the spins with a magnetic
field in transverse ($x$-) direction.

Let us first briefly consider the clean model with all $J_{ij}\equiv J$ and all $h_i^x
\equiv h^x$. It can be viewed, e.g., as a model for the magnetic behavior of LiHoF$_4$ in
a transverse field \cite{BitkoRosenbaumAeppli96}.\footnote{The actual interaction in
LiHoF$_4$ is of long-range dipolar nature, so the nearest-neighbor interaction in the
model Hamiltonian (\ref{eq:rtim}) constitutes an approximation.} Without the field, the
exchange interaction favors parallel spins pointing in positive or negative
$z$-direction. Thus, for $h^x=0$, the ground state is ferromagnetic. The transverse
magnetic field $h^x$ induces spin flips between the ``up'' and ``down'' states
($2\hat{S}^x= \hat{S}^+ +\hat{S}^-$) and reduces the magnetization in $z$-direction. For
sufficiently strong transverse field $h^x>h_c^x$, this destroys the magnetic long-range
order. The quantum critical point at $h_c^x$ was indeed observed experimentally
\cite{BitkoRosenbaumAeppli96}. The partition function of the transverse field Ising model
(\ref{eq:rtim}) in $d$ space dimensions is equivalent to that of a classical Ising model
in $d+1$ space dimensions \cite{Suzuki76,Sachdev_book99} with imaginary time playing the
role of the extra dimension. This is an example of the quantum-classical mapping
discussed in section \ref{subsec:CPTQPT}. The quantum phase transition of the
$d$-dimensional clean transverse-field Ising model is therefore in the $d+1$ dimensional
classical Ising universality class.

Quenched disorder can be introduced into the transverse field Ising model by making the
interaction $J_{ij}$ and/or the transverse field $h_i^x$ a random function of position.
We will only consider the case where all $J_{ij}$ remain positive. (Introducing negative,
i.e., antiferromagnetic bonds introduces frustration that can lead to qualitatively new
physics such as spin glass behavior \cite{WERAR91,WBRA93}.) Applying the above-mentioned
quantum-classical mapping now leads to a disordered classical Ising model. However,
because the disorder in the quantum model is time-independent, the disorder in the
equivalent classical model is perfectly correlated in one of the dimensions, viz., the
one representing the imaginary time direction. Specifically, the one-dimensional random
transverse-field Ising model (the random transverse Ising chain) maps onto the classical
McCoy-Wu model \cite{McCoyWu68,McCoyWu68a} discussed in section \ref{subsec:McCoyWu} when
only the interactions $J_{ij}$ are random, or on its generalization by Shankar and Murthy
\cite{ShankarMurthy87} when both $J_{ij}$ and $h_i^x$ are random.

Even though the free energy of the random transverse field Ising chain can be obtained
exactly using transfer matrix methods analogous to those applied to the McCoy-Wu model
\cite{McCoyWu68a,ShankarMurthy87}, many other properties including the spontaneous
magnetization and the average spin correlations could not be obtained this way. A better
understanding of this model was achieved in the early and mid 1990s by two methods: (i) a
phenomenological optimal fluctuation theory of rare region effects \cite{ThillHuse95} and
(ii) an asymptotically exact real space renormalization group \cite{Fisher92,Fisher95}.
We will discuss these approaches in the next two subsections.

\subsection{Quantum Griffiths singularities: Optimal fluctuation arguments}
\label{subsec:optimal_rtim}

In this subsection we develop optimal fluctuation arguments for the rare region effects
in the random transverse field Ising model. These arguments, which were introduced by
Thill and Huse \cite{ThillHuse95} and applied to both quantum Ising spin glasses
\cite{GuoBhattHuse96,RiegerYoung96} and the random transverse field Ising model
\cite{YoungRieger96,Young97}, can be viewed as quantum version of the arguments given in
section \ref{subsec:McCoyWu} for the McCoy-Wu model (for an early brief review see also
reference \cite{Rieger98}).

For definiteness, we consider a situation with random interactions $J_{ij}$ but a
homogeneous transverse field $h^x$. The probability distribution of the interactions is
assumed to be binary, $P(J) = (1-p)\, \delta(J-J_0) + p\, \delta(J-cJ_0)$ with $0<c<1$.
Generalizations to other distributions and randomness in the transverse fields are
straight forward. Consider a rare strongly-coupled spatial region of linear size $L_{\rm
RR}$ having only strong bonds. The probability for finding such a region is exponentially
small in its volume
\be
 w(L_{\rm RR}) \sim \exp (-\tilde{p} L_{\rm RR}^d)~.
\label{eq:w(L)_rtim} \ee where $\tilde{p}=-\ln(1-p)$ as before. We are interested in the
Griffiths region of the disordered phase where the applied transverse field $h^x>h_{c}^x$
but $h^x<h_{c,\rm clean}^x$ such that the rare region is locally ordered. The energy
levels of a locally ordered cluster can be described as follows \cite{SenthilSachdev96}.
The two lowest states correspond to the states of a single effective Ising spin with
magnet moment $\sim L_{\rm RR}^d$ in an effective transverse field $h_{\rm eff}(L_{\rm
RR})$. It can be estimated to be $h_{\rm eff}(L_{\rm RR}) \sim h^x \exp[-aL_{\rm RR}^d]$
since the ``up'' and ``down'' states of the effective spin are coupled in perturbation
theory of order $L^d$ in $h^x$. The constant $a$ increases continuously with the
difference between $h_{c,\rm clean}^x$ and $h_{c}^x$.  The energy gap of a rare region of
linear size $L_{\rm RR}$ thus decreases exponentially with its volume
\be
\epsilon(L_{\rm RR}) \sim \exp (- a L_{\rm RR}^d)~. \label{eq:eps(L)_rtim} \ee An
analogous result can be obtained by employing the quantum-classical mapping and
representing the rare region as a one-dimensional classical Ising model
\cite{ThillHuse95,RiegerYoung96}. The exchange interaction in the time-like direction is
then proportional to the spatial volume of the rare region, $J_{\rm eff} \sim L_{\rm
RR}^d$. Together with the fact that the gap in the one-dimensional Ising model depends
exponentially on the interaction, this leads to (\ref{eq:eps(L)_rtim}), see also section
\ref{subsec:McCoyWu}.

Combining equations (\ref{eq:w(L)_rtim}) and (\ref{eq:eps(L)_rtim}) leads to a power-law
density of states for the low-energy excitations of the rare regions,
\be
\rho(\epsilon) \sim \epsilon^{\tilde{p}/a-1}=\epsilon^{d/z'-1}
\label{eq:rho_rtim}
\ee
where the second equality defines the customarily used dynamical exponent $%
z^{\prime }$ \cite{Young97}. It continuously varies with disorder concentration or
distance from the clean critical point. The power-law density of states
(\ref{eq:rho_rtim}) is obtained here because the  exponentially small probability for
finding a rare region is compensated by the exponential dependence of its energy gap on
its volume. In contrast, in generic classical systems, such as the disordered Ising
models of sections \ref{subsec:dil_ising} to \ref{subsec:randomTc_LGW}, the gap depends
on the volume via a power law, and the resulting rare region density of states is
exponentially small.

Many results follow from the power-law density of states (\ref{eq:rho_rtim}).
For instance, a region with a local energy gap $\epsilon $ has a local spin
susceptibility that decays exponentially in imaginary time, $\chi _{\rm{loc}}
(\tau \rightarrow \infty )\propto \exp (-\epsilon \tau )$. Averaging by
means of $\rho $ yields
\begin{equation}
\chi _{\rm{loc}}^{\rm{av}}(\tau \rightarrow \infty )\propto \tau
^{-d/z^{\prime }}.  \label{eq:chi_tau}
\end{equation}%
The temperature dependence of the static average susceptibility is then
\begin{equation}
\chi _{\rm{loc}}^{\rm{av}}(T)=\int_{0}^{1/T}d\tau \ \chi _{\rm{loc}}^{%
\rm{av}}(\tau )\propto T^{d/z^{\prime }-1}.  \label{eq:chi_T}
\end{equation}%
If $d<z^{\prime }$, the local zero-temperature susceptibility diverges, even
though the system is globally still in the disordered phase.
The contribution of the rare regions to the specific heat $C$ can be
obtained from
\begin{equation}
\Delta E=\int d\epsilon ~\rho (\epsilon )~\epsilon ~e^{-\epsilon
/T}/(1+e^{-\epsilon /T})\propto T^{d/z^{\prime }+1}~  \label{eq:cv}
\end{equation}%
which gives $\Delta C\propto T^{d/z^{\prime }}$. To determine the zero temperature
magnetization in a small ordering field $H$ in $z$-direction we note that
all rare regions with $\epsilon < H$ are (almost) fully polarized while
the rare regions with $\epsilon > H$ have a very small magnetization.
Thus,
\be
m(H) \sim \int_0^{H} d\epsilon~\rho(\epsilon) \sim H^{d/z^{\prime}}~.
\label{eq:mh}
\ee
 Other observables can be
determined in a similar fashion.
The power-law density of states (\ref%
{eq:rho_rtim}) in the Griffiths region of a random transverse field Ising
magnet and the resulting power-law singularities (\ref{eq:chi_tau}), (\ref%
{eq:chi_T}), (\ref{eq:cv}), (\ref{eq:mh}) are called the quantum Griffiths singularities
or the Griffiths-McCoy singularities.

\begin{figure}[t]
\centerline{\includegraphics[width=6.4cm]{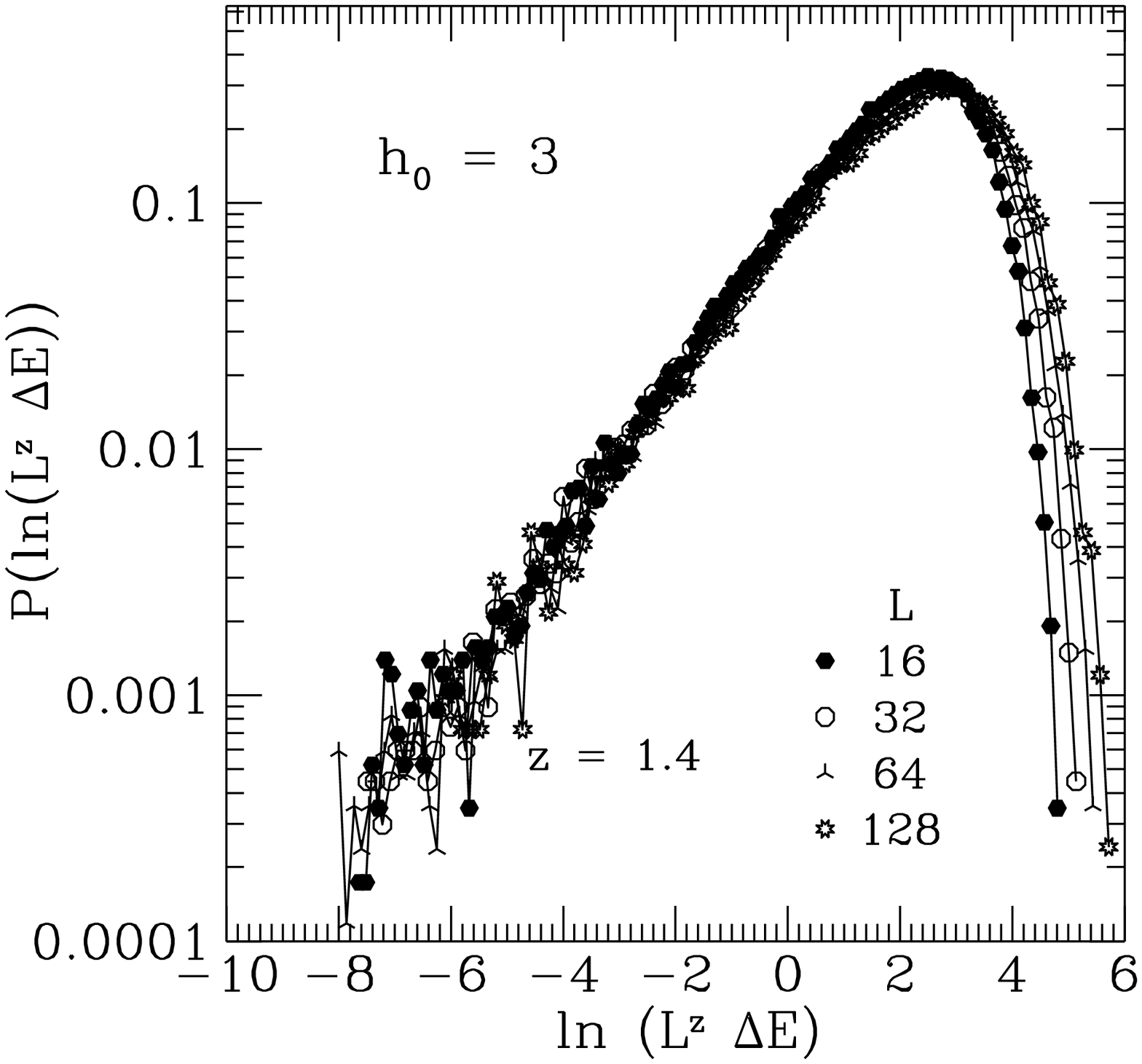}\includegraphics[width=6.9cm]{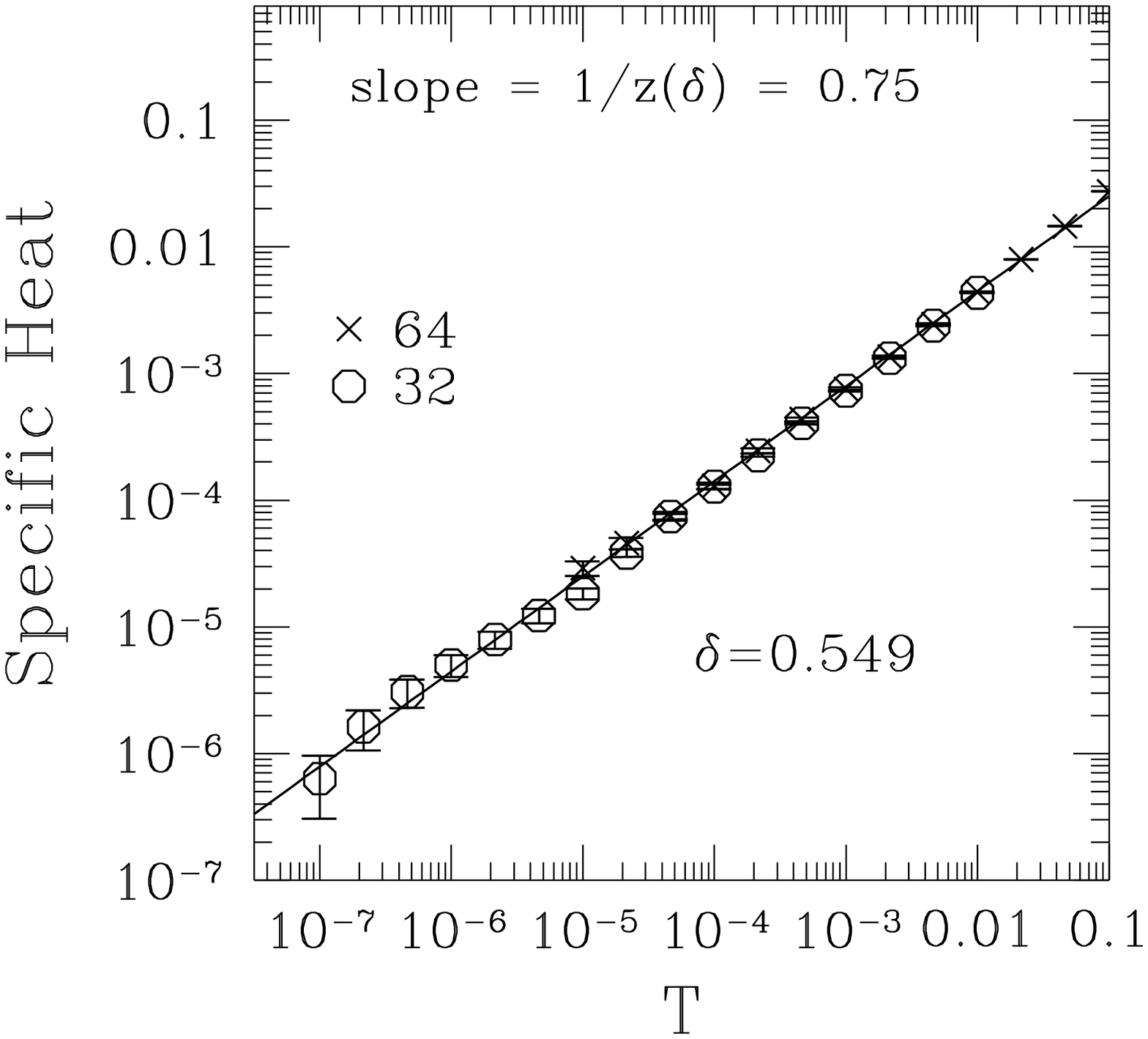}}
\caption{Left: Scaled probability distribution of the energy gap for a one-dimensional
random transverse Ising model in the Griffiths region ($h_0>h_c^x=1$) for different system sizes
$L$. A fit of the low-energy tail to (\ref{eq:rho_rtim}) gives $z^\prime \approx 1.4$ (from \cite{YoungRieger96}).
Right: Specific heat as a function of temperature for the same system. A fit to $\Delta C\propto T^{d/z^{\prime }}$
gives $1/z^\prime \approx 0.75$ (from \cite{Young97})}.
\label{fig:rtim_Griffiths}
\end{figure}

The power-law quantum Griffiths singularities predicted by the optimal fluctuation
arguments above have been verified in numerical simulations. Young and Rieger
\cite{YoungRieger96} mapped the one-dimensional random transverse field Ising chain onto
a system of non-interacting fermions which was then solved numerically. An example of the
resulting density of states in the Griffiths region is shown in the left panel of figure
\ref{fig:rtim_Griffiths}.
 The low-energy tail of this
distribution shows power-law behavior. A fit to the predicted form (\ref{eq:rho_rtim})
gives $z^\prime \approx 1.4$ for the parameter set studied. The right panel of figure
\ref{fig:rtim_Griffiths} shows the temperature dependence of the specific heat for the
same system as calculated by Young \cite{Young97}. As predicted after (\ref{eq:cv}), this
dependence is of power-law type, and a fit gives the exponent $z^\prime \approx 4/3$ in
good agreement with the value from the density of states. These numerical simulations
have later been extended to surface quantities and to the ferromagnetic phase
\cite{IgloiRieger98} as well as to nonlinear susceptibility, higher excitations, and the
energy autocorrelation function \cite{IgloiJuhaszRieger99}.

\subsection{Strong-disorder renormalization group and infinite-randomness critical point
in one dimension}
\label{subsec:sdrg}

While the optimal fluctuation arguments presented in the last subsection provide a simple
qualitative understanding of the behavior of random transverse field Ising model in the
Griffiths region, they do not give quantitative results and they also cannot be applied
to the quantum critical point itself. The behavior of the random transverse field Ising
model was therefore not fully understood before Fisher \cite{Fisher92,Fisher95}
applied a novel real-space or rather energy-space renormalization group method to the problem.
This method which is now often called the ``strong-disorder renormalization group''
has recently been reviewed in detail by Igloi and Monthus \cite{IgloiMonthus05}.
Therefore we will only briefly explain its basic idea and then discuss the results.

The strong-disorder renormalization group was introduced by Ma, Dasgupta and Hu
\cite{MaDasguptaHu79} for the random antiferromagnetic spin chain. Since then it has been
used extensively for various disorder problems including random quantum spin chains
\cite{DasguptaMa80,Fisher94,HYBG96,HymanYang97,WFSL97,MonthusGolinelliJolicoeur97,MonthusGolinelliJolicoeur98,Damle02,%
SaguiaBoechatContinentino02,RefaelKehreinFisher02,SaguiaBoechatContinentino03,YusufYang03b,HoyosMiranda04b}
and ladders \cite{YusufYang02,MLLRI02,YusufYang03a,HoyosMiranda04a} as well as random
walks \cite{FisherLeDoussalMonthus98,LeDoussalMonthusFisher99,Monthus03} and other
nonequilibrium systems
\cite{HooyberghsIgloiVanderzande03,HooyberghsIgloiVanderzande04,LeDoussalMonthus99,FisherLeDoussalMonthus01}.
The main idea of the strong-disorder renormalization group is to take the strongest
coupling in the system and find the ground state of the corresponding part of the
Hamiltonian exactly. The coupling to the rest of the system is treated perturbatively.
One then neglects the excited states involving the strong coupling and derives a new
effective Hamiltonian with the number of degrees of freedom reduced. This basic
renormalization group step is repeated ad infimum.

We now follow reference \cite{Fisher95} and apply this idea to the random transverse field Ising model
(\ref{eq:rtim}) in one dimension. We choose first the largest of the set of couplings,
$\Omega=\max(h_i^x,J_{ij})$. If the largest coupling is a field, say $h_2^x$, the unperturbed Hamiltonian
is $-h_2^x \hat{S}_2^x$ with ground state $|\to_2\rangle$ and excited state
$|\leftarrow_2 \rangle$ separated by a gap $h_2^x$. The coupling of $\hat{S}_2$ to the
rest of the system $-J_{12} \hat{S_1^z} \hat{S_2^z} - J_{23} \hat{S_2^z} \hat{S_3^z}$ is treated
in second order perturbation theory. This yields an effective interaction of the form
$-\bar J_{13} \hat{S_1^z} \hat{S_3^z}$ with the effective exchange constant
\be
\bar J_{13} \approx J_{12} J_{23} / h_2^x~. \label{eq:sdrg_h} \ee We now throw out
$\hat{S}_2$ completely by neglecting its excited state. This leads to a new spin chain
with one less spin and a new coupling $\bar J_{13} < \Omega$. If the largest coupling is
a bond, say, $J_{23}$, the corresponding unperturbed Hamiltonian is $-J_{23} \hat{S_2^z}
\hat{S_3^z}$ with two degenerate ground states $|\uparrow \uparrow \rangle$ and
$|\downarrow \downarrow \rangle$. The transverse fields $h_2^x$ and $h_3^x$ can now be
treated perturbatively yielding an effective field
\be
\bar h_{2}^x \approx h_2^x~ h_3^x / J_{23}
\label{eq:sdrg_J}
\ee
inducing coherent flips of the cluster $\hat{S}_2+\hat{S}_3$. We now throw away the
exited (antiparallel) states of this cluster and treat the cluster as an effective spin
whose moment is the sum of the moments of $\hat{S}_2$ and $\hat{S}_3$. Again, we have a
new spin chain with one less spin degree of freedom and all couplings smaller than
$\Omega$.

The repeated application of this basic renormalization group step constitutes a
aggregation and annihilation process of spin clusters. When the strongest coupling is a
field, the corresponding cluster is annihilated, and when it is a bond, the clusters that
it connects are aggregated into one cluster. At each renormalization group step, the
clusters represent sets of original spins that at correlated at the current cutoff energy
$\Omega$. In the paramagnetic phase, annihilation dominates for $\Omega \to 0$, and no
large clusters are generated. In the ferromagnetic phase, the aggregation dominates for
$\Omega \to 0$ generating arbitrarily large clusters until, at $\Omega=0$ an infinite
cluster will have formed. The quantum critical point is the point where such an infinite
cluster first appears \cite{MonthusGolinelliJolicoeur97,MMHF00}.
The multiplicative structure of the recursion relations (\ref{eq:sdrg_h}) and (\ref{eq:sdrg_J})
is very important because it establishes an exponential relationship between the length scale
and the energy scale: In each RG step, the lengths of clusters or bonds are \emph{added} while
their couplings are \emph{multiplied} to determine the effective renormalized values.
Thus, the length scale $L$ of the clusters and effective bonds scales with the
logarithm of the energy scale
\be
L^\psi \sim \ln(\Omega_0/\Omega)
\label{eq:rtim_length_scaling}
\ee
where $\Omega_0$ is a microscopic energy scale and the critical exponent $\psi$ turns
out to be $\psi=1/2$. Analogously, the typical magnetic moment of a cluster increases as
\be
\mu \sim (\ln(\Omega_0/\Omega))^\phi
\label{eq:rtim_moment_scaling}
\ee
with the exponent $\phi=(1+\sqrt{5})/2$ equal to the golden mean. Thus, the random transverse
field Ising model displays activated dynamical scaling rather than conventional power-law scaling.

Using this approach, Fisher \cite{Fisher92,Fisher95}, derived renormalization group
equations for the probability distributions of the logarithms of the couplings, $\ln J$
and $\ln h^x$. The resulting renormalization group flow displays very interesting and
unusual properties. At the critical point which occurs at $[\ln J]_{\rm av}=[\ln
h^x]_{\rm av}$ where $[ \ldots ]_{\rm av}$ is the disorder average
\cite{ShankarMurthy87}, the probability distributions of $\ln J$ and $\ln h^x$ broaden
without limit with decreasing energy scale $\Omega$. This broadening is crucial for the
renormalization group to work: In the limit of infinitely broad distributions, the
recursion relations (\ref{eq:sdrg_h}) and (\ref{eq:sdrg_J}) become asymptotically exact
for $\Omega \to 0$, since in each renormalization group step the ratio between the
dominating coupling and those treated perturbatively diverges. Because of the diverging
widths of the probability distributions of $\ln J$ and $\ln h^x$, the critical fixed
point in the random transverse field Ising model is also called an infinite randomness
fixed point.

Off criticality, the distance from the critical point can be conveniently defined as
\be
r = \frac {[\ln h^x]_{\rm av} - [\ln J]_{\rm av}} {{\rm var}(\ln h^x) + {\rm var}(\ln J)}
\ee
where ${\rm var}(x)$ is the variance of the probability distribution of $x$. The so
defined $r$ is invariant under renormalization.
In the disordered (paramagnetic) phase, there will be a length scale $\xi$ at which the
renormalization flow qualitatively changes. On scales larger than $\xi$ almost all effective fields are larger
than the effective couplings. Upon further renormalization, only fields will decimated,
leading to longer and weaker bonds but not to longer clusters. Thus, $\xi$ is the
correlation lengths. It is found to scale like
\be
\xi \sim r^{-\nu} \label{eq:rtim_xi} \ee where the exponent $\nu=2$ saturates the Harris
criterion $d\nu \ge 2$ \cite{Harris74}. Analogously, in the ordered (ferromagnetic) phase
for $r$ small and negative, there will be a length scale $\xi \sim |r|^{-\nu}$ beyond
which (almost) all interactions are larger than the fields. Upon further renormalization,
all clusters will be joined together to form the infinite cluster at $\Omega=0$. Thus,
excitations from the ferromagnetic state exist only up to the length scale $\xi$ which is
thus the (connected) correlation length in the ordered phase. The three exponents $\phi,
\psi$ and $\nu$  completely determine the properties of the infinite-randomness critical
point.

The unusual renormalization group flows lead to correspondingly unusual
scaling behavior of observables at the quantum phase transition.
At zero temperature, the magnetization in a small ordering magnetic field $H$ in $z$-direction
can be determined by stopping the renormalization at an energy scale $\Omega_H=H$ and
analyzing the cluster distribution at that scale.\footnote{More precisely,
the field energy scale is $H \mu$ where $\mu$ is the typical cluster moment at scale
$\Omega_H$. This introduces an extra logarithmic correction.}   This leads to the scaling form
\be
m\left (r,\ln(H_0/H)\right ) =  b^{\phi \psi -d} ~ m\left (r b^{1/\nu}, \ln(H_0/H)
b^{-\psi}\right ) \label{eq:rtim_m_scaling} \ee where $b$ is an arbitrary scale factor,
and $H_0$ is a nonuniversal constant. This equation again reflects the activated nature
of the dynamical scaling; the energy scale set by $H$ enters the scaling relations in
logarithmic form rather than in power-law form. At criticality, $r=0$, we can set
$b=[\ln(H_0/H)]^{1/\psi}$ and obtain
\be
m(H) \sim [\ln(H_0/H)]^{\phi -d/\psi}
\ee
The exponent evaluates to
$\phi-d/\psi=\phi-2=(\sqrt{5}-3)/2$. Setting $H=0$ and $b=r^{-\nu}$ we can obtain the
spontaneous magnetization in the ordered phase ($r<0$)
\be
m(r) \sim |r|^{\beta} \ee with $\beta=(d-\phi\psi)\nu=2-\phi=(3-\sqrt{5})/2$. Temperature
dependent quantities can be determined in a similar fashion. At criticality and zero
ordering field $H$, the magnetic susceptibility behaves as
\be
\chi \sim  \frac 1 T [\ln(T_0/T)]^{2\phi-d/\psi}
\ee
where $T_0$ is a nonuniversal constant, and the specific heat is given by
\be
c_v \sim [\ln(T_0/T)]^{-(1+d/\psi)}~.
\ee

The infinite-randomness character of the critical point also has interesting consequences
for the behavior of the spin-spin correlation function $C(x) = \langle S_i^z S_{i+x}^z
\rangle$. Because the coupling constant distributions broaden without limit under renormalization,
$C(x)$ is also very broadly distributed. As a result \emph{typical}
and \emph{average} correlations behave very differently. At the critical point, typical spin pairs are
never in the same renormalization cluster and thus have only weak correlations that fall off
as a stretched exponential
\be
\ln C_{\rm typ} (x) \sim -x^\psi~. \label{eq:rtim_Ctyp} \ee In contrast, the average
correlation function is dominated by rare spin pairs on the same cluster that have
atypically large correlations of order unity. It is thus much larger than the typical one
and falls off as a power law
\be
C_{\rm av} (x) \sim x^{-2(d-\phi\psi)}~. \label{eq:rtim_Cav} \ee This discrepancy between
typical and average properties is one of the main characteristics of an
infinite-randomness critical point. Fisher and Young \cite{FisherYoung98} numerically
studied the end-to-end correlations in an open one-dimensional random transverse-field
Ising model and found good agreement with the prediction of the strong-disorder
renormalization group.

The strong-disorder renormalization group can also be used to analyze thermodynamic observables and
correlation functions off criticality. In the Griffiths region of the disordered phase, Fisher
\cite{Fisher95} found strong power-law quantum Griffiths singularities of the form predicted by
the optimal fluctuation arguments presented in the
last subsection. Moreover, he found an explicit expression for the nonuniversal dynamical exponent
$z^{\prime}$. Close to the true quantum critical point, i.e. for $r\ll 1$ it is given by
$z^{\prime} = 1/|2r|$. The discrepancy between typical and average correlations persists
in the Griffiths region. Typical correlations fall off exponentially with a correlation
length $\xi_{\rm typ}$ that diverges as $\xi_{\rm typ} \sim |r|^{-\nu_{\rm typ}}$ with
$\nu_{\rm typ}=1$. In contrast, the average correlations decay with the much larger true
correlation length $\xi \sim |r|^{-\nu}$ with $\nu=2$.
An analogous Griffiths region with power-law singularities was also
identified in the ordered phase.

Igloi et al.\ \cite{IgloiJuhaszLajko01,Igloi02} extended Fishers analytical solution into
the off-critical region and obtained asymptotically exact results for dynamical
quantities including the dynamical exponent $z^{\prime}$. Static quantities or spatial
correlations, however, cannot be obtained exactly off criticality because the coupling
constant distributions do \emph{not} broaden without limit under renormalization. Rieger
and Igloi \cite{RiegerIgloi99} studied the influence of power-law spatial correlations of
the random couplings on the transition. They found the infinite-randomness scenario still
valid, but sufficiently long-ranged correlations yield new exponents that enhance both
the critical and the quantum Griffiths singularities.

\subsection{Higher dimensions and other generalizations}
\label{subsec:rtim_hd}

\begin{figure}[t]
\centerline{\includegraphics[width=8.5cm]{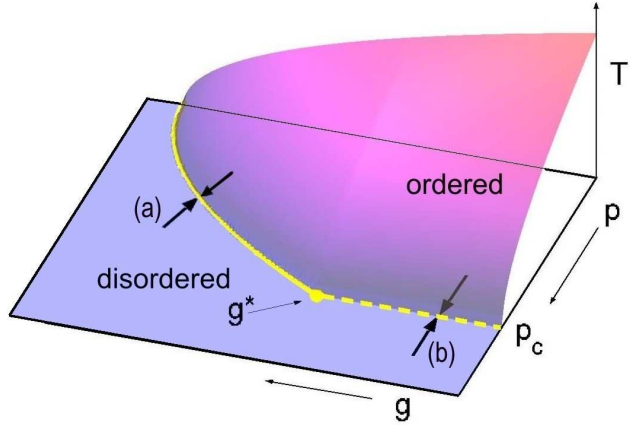}}
\caption{Schematic phase diagram of a general diluted quantum
magnet with impurity concentration $p$, temperature $T$, and quantum fluctuation
strength $g$ (for the transverse-field Ising model $g \sim h^z$.
There is a quantum multicritical point (big dot) at ($p_{c},g^{\ast }$).}
\label{fig:pd_dil_transverse}
\end{figure}

The discovery of an infinite-randomness critical point in the one-dimensional random
transverse-field Ising model lead to a surge of activity in this area starting in
the mid 1990's. Early on, an important question was whether or not the
infinite-randomness critical point is special to one dimension, or whether it also
occurs in higher-dimensional systems. Senthil and Sachdev \cite{SenthilSachdev96} considered a
bond-diluted transverse field Ising model in $d\ge 2$. A sketch of the phase
diagram of a general diluted quantum magnet is shown in figure \ref{fig:pd_dil_transverse}.
The diluted transverse field Ising model has two quantum phase transitions, separated by
a multicritical point at the lattice percolation threshold $p_c$ and a transverse field
$h^x=h^\ast$. The transition at dilutions $p<p_c$ (transition ``a'' in figure
\ref{fig:pd_dil_transverse}) is expected to be in the universality class of the generic
random transverse-field Ising model. Senthil and Sachdev focused on the transition ``b''
across the percolation threshold of the lattice. Using the well known properties of
critical percolation clusters \cite{StaufferAharony_book91}, they showed that this
quantum phase transition shares many properties with the infinite-randomness critical
point in the one-dimensional random transverse Ising model. Specifically, the dynamical
scaling is activated and the magnetization in a small ordering field obeys the scaling
form (\ref{eq:rtim_m_scaling}) with $\psi=D_f$, $\phi=1$ and $\nu=\nu_c$ where $D_f$ is
the fractal dimension of the critical percolation cluster and $\nu_c$ is the correlation
length exponent of the lattice percolation problem. The transition is accompanied by
strong power-law quantum Griffiths singularities both above and below $p_c$. All these
predictions were verified by  quantum Monte-Carlo results of Ikegami et al.\
\cite{IkegamiMiyashitaRieger98}.

To study the generic random transverse field Ising transition in dimensions $d \ge 2$,
Motrunich et al.\ \cite{MMHF00} applied a strong-disorder renormalization group to the
two-dimensional random transverse-field Ising model. In dimensions larger than one, the
renormalization group calculation cannot be carried out analytically because the
renormalization group does not preserve the lattice structure but generates longer and
longer ranged couplings. Therefore, Motrunich et al.\ \cite{MMHF00}, implemented the
renormalization procedure numerically. For two dimensions, they found that the
renormalization group flow for stronger disorder is indeed towards even stronger
disorder, i.e., towards an infinite-randomness fixed point, as in one dimension. As a
result, the same scaling scenario as outlined in section \ref{subsec:sdrg} also applies
to the two-dimensional case, but with different exponent values. Motrunich et al.\
estimated $\psi \approx 0.42$, $\phi \approx 2.5$, and $\nu \approx 1.1$. Other numerical
implementations \cite{LKIR00,KLRKI01} of the strong disorder renormalization group found
similar values. Preliminary results for three dimensions point to the same scenario, but
reliable estimates for the exponents have not yet been obtained. The results of the
numerically implemented strong-disorder renormalization group are in very good agreement
with Monte-Carlo simulations of a two-dimensional random transverse field Ising model
\cite{PYRK98,RiegerKawashima99}.

The infinite-randomness critical fixed points found in the random transverse field Ising
model control more than just Ising ferromagnetic quantum phase transitions. As was first
pointed out by Senthil and Majumdar \cite{SenthilMajumdar96}, Potts models or any random
quantum systems with continuous quantum phase transitions at which a discrete symmetry of
a non-conserved order parameter is broken, will have the same critical behavior as the
Ising case. This applies even to systems that are frustrated on microscopic length
scales, such as quantum Ising spin glasses. Because the coupling constant distributions
become infinitely broad at the infinite-randomness fixed point, in any frustrated loop
the weakest interaction will be infinitely weaker than the others and can thus be
neglected. Thus, for Ising symmetry, frustration becomes irrelevant at the
infinite-randomness critical point.

%%%%%%%%%%%%%%%%%%%%%%%%%%%%%%%%%%%%%%%%%%%%%%%%%%%%%%%%%%%%%%%%%%%%%%%%%%%%%%%%%%%%%%%%%%%%%%%%%
\subsection{Continuous symmetry quantum magnets}
\label{subsec:qrotors}

In the last subsections we have seen that the qualitative zero-temperature behavior of random quantum
Ising systems does not depend much on dimensionality: There is an infinite-randomness
critical point separating a conventional paramagnetic phase from a phase with long-range order
(of course, the critical exponent values at this critical point do depend on dimensionality).

For continuous-symmetry quantum spin systems such as quantum Heisenberg or XY magnets,
the situation is drastically different. In one space dimension, these systems do not show
magnetic long-range order. Instead they can be in a variety of exotic phases that will be
briefly discussed in section \ref{subsec:other_qpt}. In staying with the main topic of
this review, the present subsection is focussed on transitions between a disordered and a
long-range ordered phase that can occur in continuous symmetry quantum magnets for $d\ge 2$.

Let us start by considering a two-dimensional  $S=1/2$ Heisenberg quantum antiferromagnet
with random site dilution. This problem is of direct relevance for antiferromagnetic layered
cuprates with nonmagnetic impurities \cite{CCRBTF91,TPCMW92,CRTCLR95,VMGGL02}.
In the clean limit, the ground state of this system displays
magnetic long-range order. However, the zero-point quantum fluctuations reduce the
staggered magnetization by about 40\% compared to the classical value (see, e.g.,
\cite{Manousakis91}). Upon randomly removing sites (or bonds) with probability $p$,
the tendency towards magnetic order decreases, but the location and character of
the phase transition were initially controversial. Early theoretical investigations
\cite{BehreMiyashita92,YasudaOguchi97,YasudaOguchi99,ChenCastroNeto00}
suggested that magnetic long-range order is destroyed before the impurity concentration
$p$ reaches the classical (geometric) percolation threshold $p_c\approx 0.407$
\cite{StaufferAharony_book91} of the lattice. Kato et al. \cite{KTHKMT00} found evidence of the
critical dilution coinciding with percolation threshold $p_c$ but argued that the infinite percolation
cluster is quantum critical, with power-law spin correlations. However, using
extensive quantum Monte-Carlo simulations, Sandvik \cite{Sandvik01,Sandvik02b} showed
that the infinite percolating cluster at $p_c$ displays long-range order. The transition
at $p_c$ is thus of percolation type. If the quantum fluctuations are enhanced to induce
a quantum phase transition at impurity concentrations $p<p_c$, the zero-temperature phase
diagram takes the general form shown in figure \ref{fig:pd_dil_transverse}. However, at
finite temperatures, no long-range order exists because of the Mermin-Wagner theorem
\cite{MerminWagner66}.

\begin{figure}
\centerline{
\includegraphics[width=8cm]{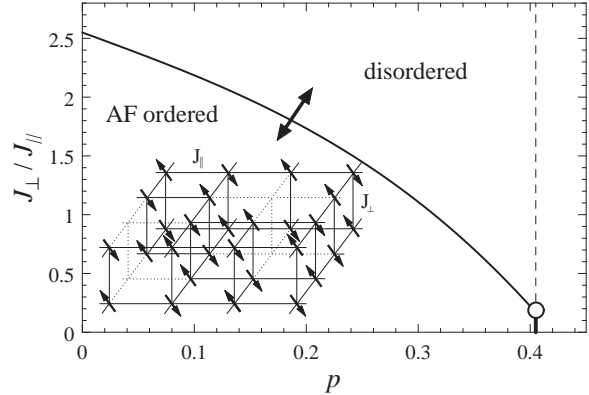}
}
\vspace*{-10pt}
\caption{ Phase diagram \cite{Sandvik02,VajkGreven02} of the diluted bilayer Heisenberg antiferromagnet, as
function of $J_\perp/J_\parallel$ and dilution $p$. The dashed line is the percolation
threshold, the open dot is the multicritical point. Inset: The model: Quantum spins (arrows) reside on the
two parallel square lattices. The spins in each plane interact with the coupling strength
$J_\parallel$. Interplane coupling is $J_\perp$. Dilution is done by removing dimers (from
\cite{SknepnekVojtaVojta04}).}
\label{fig:bilayer}
\end{figure}

An obvious and experimentally relevant way to increase the quantum fluctuations is via
an interlayer coupling in a bilayer Heisenberg quantum antiferromagnet as depicted in
the inset of figure \ref{fig:bilayer}.
The spins in each 2d layer interact via nearest neighbor exchange $J_\parallel$, and the
interplane coupling is $J_\perp$ with the ratio $g \sim J_\perp/J_\parallel$ controlling
the strength of the quantum fluctuations. The clean version of this model has been
studied extensively
\cite{Hida90,MillisMonien93,SandvikScalapino94,ShevchenkoSandvikSushkov00}. For
$J_\perp\gg J_\parallel$, neighboring spins from the two layers form singlets, and the
ground state is paramagnetic. In contrast, for $J_\parallel\gg J_\perp$ the system
develops N\'eel order. Both phases are separated by a quantum phase transition at
$J_\perp / J_\parallel \approx 2.525$.

Quenched disorder can be introduced by randomly removing spins. We emphasize that in
order to achieve random-mass type disorder, {\em pairs} (dimers) of adjacent spins, one
from each layer have to be removed. In this way, no random Berry phases arise, because
the Berry phase contributions from the two spins of each unit cell exactly cancel
\cite{ChakravartyHalperinNelson89,Sachdev_book99}. In contrast, for site dilution, the
physics would completely change: The random Berry phases (which have no classical
analogue) are equivalent to impurity-induced moments \cite{SachdevVojta01}, and those
become weakly coupled via bulk excitations. For all $p<p_p$ the ground state of the
system shows long-range order, independent of $J_\perp / J_\parallel$! Thus, site
dilution actually changes the bulk phases, i.e., it introduces disorder effects more
complicated than random-mass disorder.  These are beyond the scope of this review.

The Hamiltonian of the  dimer-diluted bilayer Heisenberg model is:
\begin{equation}
\label{eq:1} H=J_\parallel \sum_{{\langle i,j\rangle} \atop a=1,2}\epsilon_i\epsilon_j
{\mathbf{\hat{S}}}_{i,a}\cdot{\mathbf{\hat{S}}}_{j,a}+J_\perp\sum_i\epsilon_i{\mathbf{\hat{S}}}_{i,1}\cdot{\mathbf{\hat{S}}}_{i,2},
\end{equation}
and $\epsilon_i$=0 ($\epsilon_i$=1) with probability $p$ ($1-p$). The zero-temperature
phase diagram of this model has been studied in detail by Sandvik \cite{Sandvik02} and
Vajk and Greven \cite{VajkGreven02}; it is shown in Fig.~\ref{fig:bilayer}. For small
$J_\perp$, magnetic order survives up to the percolation threshold $p_p\approx 0.4072$,
and a multicritical point exists at $p=p_p$ and $J_\perp / J_\parallel \approx 0.16$.
Thus, the dimer-diluted bilayer Heisenberg antiferromagnet has two quantum phase
transitions, the generic transition for $p<p_c$ (corresponding to transition ``a'' in the
generic phase diagram, figure \ref{fig:pd_dil_transverse}) and a quantum percolation
transition at $p=p_c, J_\perp < 0.16 J_\parallel$ (corresponding to transition ``b''in
figure \ref{fig:pd_dil_transverse}).

The low-energy properties
of bilayer quantum antiferromagnets are represented by a (2+1)-dimensional O(3) quantum rotor
model  with the rotor coordinate $\mathbf{\hat{n}}_i$ corresponding to
$\mathbf{\hat{S}}_{i,1} - \mathbf{\hat{S}}_{i,2}$ and the angular momentum $\mathbf{\hat{L}}_i$
representing $\mathbf{\hat{S}}_{i,1} + \mathbf{\hat{S}}_{i,2}$ \cite{ChakravartyHalperinNelson89,Sachdev_book99}.
This quantum rotor model in turn maps onto a 3d classical Heisenberg model
in which the disorder is perfectly correlated in the imaginary time direction. At zero temperature,
the rare
regions are thus equivalent to classical Heisenberg magnets that are infinite in the
time-like direction but finite in the two space directions.

A first qualitative understanding of the rare region effects can again be obtained from
optimal fluctuation arguments similar to that used in section \ref{subsec:optimal_rtim}.
The probability for finding a rare region of linear size $L_{\rm RR}$ is exponentially
small in its volume, $w(L_{\rm RR}) \sim \exp (-\tilde{p} L_{\rm RR}^d)$ with $d=2$.
However, the energy gap of a locally ordered rare region depends on its \emph{spatial}
volume via a power law (rather than exponentially as in the Ising case) as can be
obtained from a renormalization group analysis of the (0+1) dimensional Heisenberg model
at its strong coupling fixed point or by an explicit large-$N$ calculation
\cite{VojtaSchmalian05}. These arguments thus predict the rare region effects in the
Heisenberg case to be exponentially weak and the critical point to be conventional with
power-law dynamical scaling.

Sknepnek, Vojta and Vojta \cite{SknepnekVojtaVojta04} performed large scale Monte-Carlo
simulations of a classical three-dimensional Heisenberg model with linear defects to test
these predictions. They focused on the generic transition for $p<p_c$ and analyzed the
scaling behavior of the so-called Binder ratio of the order parameter moments, given by
\begin{equation}
\label{eq:Binder} g_{av}=\left[ 1-\frac{\langle |\mathbf{m}|^4\rangle}{3\langle
|\mathbf{m}|^2\rangle^2}\right]_{av},
\end{equation}
where $\left[\ldots\right]_{av}$ denotes the disorder average and $\langle\ldots\rangle$ is
the Monte-Carlo average for each sample.
This quantity has scale dimension 0. Thus, its finite-size scaling form is given by
\begin{eqnarray}
\label{eq:Binder_power} g_{av}&=&\tilde{g}_C (rL^{1/\nu},L_\tau/L^z)\qquad {\rm or}
\\ \label{eq:Binder_activated} g_{av}&=&\tilde{g}_A (rL^{1/\nu},\log(L_\tau)/L^\psi)
\end{eqnarray}
for conventional scaling or for activated scaling, respectively. Here $L$ and $L_\tau$
are the linear system sizes in the space and time-like directions. Two important
characteristics immediately follow from these scaling forms: (i) For fixed $L$, $g_{av}$
has a peak as a function of $L_{\tau}$. The peak position $L_{\tau}^{\rm max}$ marks the {\em optimal} sample shape,
where the ratio $L_{\tau}/L$ roughly behaves like the corresponding ratio of the
correlation lengths in time and space directions, $\xi_{\tau}/\xi$. At the critical
temperature $T_c$, the peak value $g_{av}^{\rm max}$ is independent of $L$. Thus, for
power law scaling, plotting $g_{av}$ vs. $L_\tau/L_\tau^{max}$ at $T_c$ should collapse
the data, without the need for a value of $z$. In contrast, for activated scaling the
$g_{av}$ data should collapse when plotted as a function of
$\log(L_\tau)/\log(L_\tau^{\rm max})$. (ii) For samples of the optimal shape
($L_\tau=L_\tau^{max}$), plots of $g_{av}$ vs. temperature for different $L$ cross at
$T_c$.

\begin{figure}
\centerline{
\includegraphics[width=8cm]{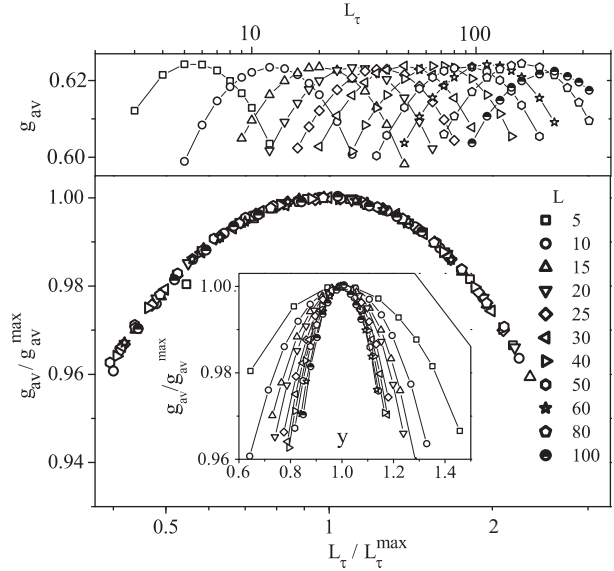}}
\caption{ Upper panel: Binder ratio $g_{av}$ as a function of $L_\tau$ for various $L$
($p=\frac 1 5$). Lower panel: Power-law scaling plot $g_{av}/g_{av}^{max}$ vs.
$L_\tau/L_\tau^{max}$  Inset: Activated scaling plot $g_{av}/g_{av}^{max}$ vs.
$y=\log(L_\tau)/\log(L_\tau^{max})$  (from
\cite{SknepnekVojtaVojta04}).}
\label{fig:Binder}
\end{figure}

A typical result of these calculations is presented in figure \ref{fig:Binder}.
This figure shows the Binder ratio at the critical point for a system with impurity
concentration $p=1/5$ and various system sizes $L$ and $L_\tau$. As can be seen in the
main panel, the data scale very well when analyzed according to conventional power-law
scaling while the inset shows that they do not scale at all when plotted according to
activated scaling. Analogous data where obtained for impurity concentrations 1/8, 2/7,
and 1/3. Sknepnek et al. \cite{SknepnekVojtaVojta04} thus concluded that the quantum
critical point in the dimer-diluted bilayer Heisenberg antiferromagnet exhibits conventional
power-law scaling with exponentially weak rare region effects, in agreement with the
optimal fluctuation arguments above.
They also determined the critical exponents  at this quantum critical point and
found universal (i.e., disorder-independent) values of $z=1.310(6)$, $\nu=1.16(3)$,
$\beta/\nu=0.56(5)$, and $\gamma/\nu=2.15(10)$.

Sandvik \cite{Sandvik02} and Vajk and Greven \cite{VajkGreven02} studied the scaling
behavior at the multicritical point at $p=p_p$ and $J_\perp / J_\parallel \approx 0.16$.
They found that this point also exhibits conventional power-law scaling. Their value of
the dynamical exponent $z \approx 1.3$ agrees within the error bars with that of the
generic transition discussed above. Whether or not the dynamic exponents at the generic
transition and at the multicritical point are indeed identical and whether or not this
agreement is accidental is not resolved so far.

More recently, Vojta and Schmalian \cite{VojtaSchmalian05b} considered the quantum phase
transition across the percolation threshold (transition ``b'' in the generic phase
diagram, figure \ref{fig:pd_dil_transverse}). Combining the well known properties of the
lattice percolation transition \cite{StaufferAharony_book91} with a scaling analysis of
finite-size quantum Heisenberg clusters at the \emph{stable} low-energy fixed point, they
derived a complete scaling theory of this transition. In contrast to the corresponding
quantum Ising case \cite{SenthilSachdev96}, the critical behavior is of power-law type.
The dynamical exponent is identical to fractal dimension of the percolation cluster,
$z=D_f$, and the correlation length exponent is identical to that of the lattice
percolation problem, $\nu=\nu_c$. In fact, these results apply not only to the
dimer-diluted bilayer model but general diluted Heisenberg quantum antiferromagnets. The
conclusion $z=D_f$ has been confirmed by Monte Carlo simulations of single-layer
\cite{YuRoscildeHaas05} and bilayer \cite{Sandvik02} models.

We can thus conclude that all quantum phase transitions of the dimer-diluted bilayer
quantum Heisenberg antiferromagnet (the generic transition for $p<p_c$, the percolation
transition at $p_c$ and the multicritical point) exhibit conventional power-law scaling
and exponentially weak rare region effects in agreement with the simple extremal
statistics arguments. This conclusion is also supported by preliminary results of a
strong-disorder renormalization group calculation for the Heisenberg case  \cite{MMHF00}
that show that the infinite-randomness fixed point is unstable and the renormalization
group flow is towards smaller disorder strength. Lin et al.\ \cite{LMRI03} used the
strong-disorder renormalization group to study the stable low-energy (as opposed to
critical) fixed points of random Heisenberg models in two and three dimensions. They
found conventional power-law scaling and no infinite randomness behavior. This result has
recently been confirmed by quantum Monte-Carlo simulations and exact numerical
diagonalizations \cite{LWLR06}.

%%%%%%%%%%%%%%%%%%%%%%%%%%%%%%%%%%%%%%%%%%%%%%%%%%%%%%%%%%%%%%%%%%%%%%%%%%%%%%%%%%%%%%%%%%%%
\subsection{Itinerant quantum magnets}
\label{subsec:itinerant}

In this subsection, we turn from systems of localized spins to  quantum phase transitions
in itinerant (metallic) spin systems. This field was pioneered by Hertz \cite{Hertz76}.
Starting from a microscopic theory of interacting electrons, he derived
Landau-Ginzburg-Wilson order parameter field theories for the ferromagnetic and
antiferromagnetic quantum phase transitions of both clean and dirty metals. He then
studied these Landau-Ginzburg-Wilson theories using renormalization group methods. Later,
Millis \cite{Millis93} worked out the experimentally important finite-temperature
behavior in the vicinity of the quantum critical point. In the case of the ferromagnetic
transition, the specific conclusions of the Hertz-Millis theory turned out to be
incorrect because generic scale invariance in the of form additional fermionic soft modes
invalidates the Landau-Ginzburg-Wilson approach
\cite{VBNK96,VBNK97,BelitzKirkpatrickVojta97,BelitzKirkpatrickVojta05}. This was briefly
discussed at the end of section \ref{subsec:CPTQPT}. The effective long-range
interactions created by the additional soft modes will likely modify the rare region
effects, too. However, these extra complications at the dirty ferromagnetic transition
are presently not fully understood, and therefore we will not discuss them in detail.

Physically, a crucial difference between itinerant magnets and systems of localized spins
is that magnetic excitations are damped in the former while they are undamped in the
latter. This is caused by the coupling between the magnetic modes and the gapless
particle-hole excitations in the metal. Technically, it is reflected in the frequency
dependence of the Gaussian term of the Landau-Ginzburg-Wilson theory. For definiteness,
let us consider the itinerant antiferromagnetic transition. The Landau-Ginzburg-Wilson
free energy functional of the clean transition reads
\cite{Hertz76}\setcounter{footnote}{1}\footnote{This simple bosonic Landau-Ginzburg
Wilson theory only applies for dimensions $d>2$. In two dimensions, fermionic soft modes
lead to singular vertices even for the antiferromagnet \cite{AbanovChubukovSchmalian03}.}
\be
\fl S= \int d^dx d^dyd\tau d\tau^{\prime} \,
\phi(\mathbf{x},\tau)\Gamma(\mathbf{x},\tau,\mathbf{y},\tau^\prime)
\phi(\mathbf{y},\tau^\prime) +u \int d^dx d\tau \, \phi^4(\mathbf{x},\tau)~.
\label{eq:Hertz_AFM} \ee Here, $\Gamma(\mathbf{x},\tau,\mathbf{y},\tau^\prime)$ is the
bare two-point vertex whose Fourier transformation is
\be
\Gamma(\mathbf{q},\omega_n) = r+ \mathbf{q}^2 +|\omega_n| \label{eq:Gamma_AFM} \ee with
$r$ being the (bare) distance from criticality and $\omega_n$ bosonic Matsubara
frequencies. We have suppressed constant prefactors. The nonanalytic frequency dependence
proportional to $|\omega_n|$ in the dynamical part of $\Gamma$ reflects the overdamping
of the dynamics due to the coupling of the order parameter to fermionic particle-hole
excitations. In contrast, undamped dynamics would lead to an $\omega_n^2$ term (see also
equation (\ref{eq:QLGW})).

Weak, random-mass disorder can be introduced by making $r$ a random function of position,
$r \to r + \delta r({\bf x})$ \cite{Hertz76,KirkpatrickBelitz96}. The rare regions in
this system are large spatial regions where the local $r$ is smaller than its average
value. Let us consider a single such region that is locally in the ordered phase, while
the bulk is still in the disordered phase. The order parameter on such a region is
correlated in space, i.e., it behaves coherently over the whole island. The crucial
difference between itinerant magnets and the localized spin systems considered in
previous sections is in the dynamics of the rare region. Without damping, the order
parameter slowly fluctuates giving rise to the quantum Griffiths effects discussed in
section \ref{subsec:optimal_rtim}. Damping or dissipation will potentially hinder the
dynamics of the rare region. To study this phenomenon, Millis, Morr and Schmalian
\cite{MillisMorrSchmalian01,MillisMorrSchmalian02} explicitly calculated the tunneling
rate of a locally ordered rare region in an itinerant Ising magnet. They found that for
sufficiently large rare regions, the tunneling rate vanishes. This means, these rare
regions completely stop to tunnel, and  their order parameter becomes static. The same
result can also be obtained from quantum-to-classical mapping \cite{Vojta03a}. In the
equivalent classical system, a rare region corresponds to a quasi-one-dimensional Ising
model which is of finite size in the space directions but infinite in the time-like
direction. The linear frequency dependence in the two-point vertex $\Gamma$ is equivalent
to a long-range interaction in imaginary time of the form $(\tau-\tau')^{-2}$. Each rare
region is thus equivalent to an one-dimensional Ising model with a $1/r^2$ interaction.
This model is known to have a phase transition \cite{Thouless69,Cardy81}. Thus, true
static order can develop on those rare regions which are locally in the ordered
phase.\footnote{At the first glance, this result seems to contradict basic statistical
mechanics that tells us that spontaneous symmetry breaking cannot occur in a finite-size
system. The resolution lies in the fact that the damping is caused by the underlying
electronic system which must be infinite to be gapless.} Very recently, Schehr and Rieger
\cite{SchehrRieger05} analyzed a related problem, viz. a random transverse field Ising
model coupled to an Ohmic dissipative bath, by a version of the strong-disorder
renormalization group. They arrive at an analogous conclusion: Sufficiently large,
locally ordered clusters freeze, and develop a static order parameter.

As a result, the phenomenology of rare region effects in an itinerant Ising magnet is
very different from that of a localized magnet. Because the large rare regions do not
fluctuate the low-energy behavior does not display power-law quantum Griffiths effects,
contrary to earlier predictions that did not correctly take into account the dissipation
\cite{CastroNetoCastillaJones98}. Instead, the global phase transition is smeared
\cite{Vojta03a} by the same mechanism as the transition in a classical Ising model with
planar defects discussed in section \ref{subsec:planar}: Once static order has developed
on a few isolated rare regions, an infinitesimally small interaction or an
infinitesimally small symmetry-breaking field are sufficient to align them. Consequently,
a macroscopic order parameter develops inhomogeneously, with different spatial part of
the system ordering at different values of the control parameter.

The leading zero-temperature behavior in the tail of the smeared quantum phase transition
can be determined by optimal fluctuation arguments similar to section
\ref{subsec:optimalfluc}. Here, we just summarize the results for dilution-type (Poisson)
disorder, i.e., $\delta r(\mathbf{x})=0$ everywhere except on randomly distributed
finite-size islands (impurities) of spatial density $\tilde{p}$ where $\delta
r(\mathbf{x})=W>0$. More details can be found in reference \cite{Vojta03a}. The
probability $w$ for finding a region of linear size $L_{\rm RR}$ devoid of any impurities
is given by $w \sim \exp( -\tilde{p} L_{\rm RR}^d)$ (up to pre-exponential factors). Such
a rare region develops static order at a distance $r_c(L_{\rm RR}) < 0$ from the {\em
clean} critical point. Finite size scaling yields $|r_c(L_{\rm RR})| \sim L_{\rm
RR}^{-\phi}$ where $\phi$ is the finite-size scaling shift exponent of the clean system.
Thus, the probability for finding a rare region which becomes critical at $r_c$ is given
by $w(r_c) \sim \exp  (-B ~|r_c|^{-d/\phi})$ for  $r\to 0-$. The total order parameter
$m$ is obtained by integrating over all rare regions which are ordered at $r$, i.e., all
rare regions having $t_c>t$. This leads to an exponential tail of $m$ as a function of
the distance $r$ from the {\em clean} critical point:
\begin{equation}
m(r) \sim \exp( -B |r|^{-d/\phi} ) \qquad (\textrm{for } r\to 0-)~. \label{eq:qpt-m-dilute}
\end{equation}
This is identical to the classical result (\ref{eq:m-dilute}) if one replaces the number
of uncorrelated dimensions $d_{\rm ran}$ by the space dimensionality $d$.

\begin{figure}
\centerline{
\includegraphics[width=7cm]{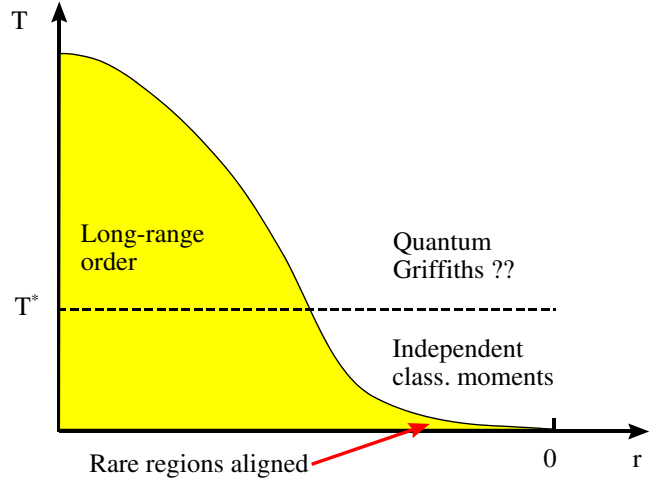}}
\caption{Schematic phase diagram in the vicinity of a smeared quantum phase transition.
For details see text.}
\label{fig:smeared_pd}
\end{figure}

We now turn to the experimentally important finite-temperature behavior in the vicinity
of the smeared quantum phase transition. A schematic phase diagram is shown in figure
\ref{fig:smeared_pd}.
 At any finite temperature, the static order on the rare
regions is destroyed, and a finite interaction of the order of the temperature is necessary to
align them. This means a sharp phase transition is recovered. To estimate the transition temperature
we note that the interaction between two rare regions depends exponentially on their (typical)
spatial distance $|\mathbf{x}|$ which itself depends exponentially on $r$. This leads to a
double-exponential dependence of the critical temperature on $r$
\begin{eqnarray}
 \log( -a \log T_c) &\sim& |r|^{-d/\phi} ~.
\end{eqnarray}
In other words, the long-range ordered phase develops an double-exponential tail towards
to clean quantum critical point at $r=0$. (Note that the inclusion of a long-range
spatial interaction such as the RKKY interaction would transform this into a simple
stretched exponential. Moreover, weak competing interactions not contained in the model
(\ref{eq:Hertz_AFM}) may induce spin glass order in the tail region.) Above this very low
temperature different rare regions are not aligned anymore but act as independent
classical moments leading to super-paramagnetic behavior with a Curie-type susceptibility
\cite{MillisMorrSchmalian02}.

What happens at even higher temperatures depends on nonuniversal microscopic parameters
and is subject of an ongoing debate. If the damping of the magnetic modes by the
electrons is weak, i.e., if the prefactor of the $|\omega_n|$-term in the Gaussian vertex
(\ref{eq:Gamma_AFM}) is parametrically small compared to other scales, a subleading
$\omega_n^2$ term (which describes undamped dynamics) becomes dominating at higher
energies. This gives rise to a crossover temperature $T^\ast$. Power-law quantum
Griffiths behavior can occur in a transient temperature window \emph{above} $T^\ast$ but
below a microscopic cutoff scale (like the Fermi temperature). However, if the damping is
not weak, $T^\ast$ should roughly coincide with the microscopic cutoff, and quantum
Griffiths effects should not be observable. For more details see the discussion on
experiments in section \ref{subsec:experiments}.

So far, our exploration of itinerant quantum magnets has focused on the case of Ising
symmetry where we found a smeared phase transition and static (frozen) rare regions. For
continuous spin symmetry, the results are markedly different, as can be readily
understood from quantum-to-classical mapping arguments \cite{VojtaSchmalian05}. A rare
region in an itinerant magnet with continuous spin symmetry is equivalent to a
quasi-one-dimensional classical $O(N)$ model ($N\ge2$). The interaction in the time-like
direction has the form $(\tau-\tau')^{-2}$ as in the Ising case.  One-dimensional
continuous-symmetry $O(N)$ models with $1/\mathbf{x}^{2}$ interaction do not have a phase
transition, but they are exactly \emph{at} their lower critical
dimension\cite{Joyce69,Dyson69,Bruno01}. Therefore, in the quantum problem, an isolated
rare region of linear size $L_{\rm RR}$ cannot independently undergo a phase transition.
Its energy gap depends exponentially on its volume, $\epsilon(L_{\rm RR})\sim \exp
(-aL_{\rm RR}^{d})$. Equivalently, the susceptibility of such a region diverges
exponentially with its volume. Combining this exponential law for the gap with an the
exponentially small probability for finding a rare region leads to exactly the same
power-law quantum Griffiths singularities as in the random transverse field Ising model
\cite{VojtaSchmalian05}. Loh et al.\ \cite{LohTripathiTurlakov05} studied a related
problem, viz., a single spin-$S$ impurity with XY rotational symmetry in a metal. Their
results are completely compatible with the quantum Griffiths scenario outlined above.

This scenario is correct as long as different rare regions are effectively independent.
Dobrosavljevic and Miranda \cite{DobrosavljevicMiranda05} pointed out that in realistic
metallic systems the long-range RKKY interaction leads to an additional damping of the
rare regions that destabilizes the quantum Griffiths behavior and leads to static frozen
rare regions even in the Heisenberg case. However, in contrast to the Ising case, the
freezing requires an interaction between the rare regions and is therefore expected to
occur at extremely low temperatures only.

%%%%%%%%%%%%%%%%%%%%%%%%%%%%%%%%%%%%%%%%%%%%%%%%%%%%%%%%%%%%%%%%%%%%%%%%%%%%%%%%%%%%%%%%%%%%
\subsection{Quantum Griffiths effects in other systems}
\label{subsec:other_qpt}

In accordance with the main theme of this review, our discussion of quantum Griffiths
effects so far has been restricted to order-disorder quantum phase transitions between
conventional phases, i.e., we have assumed that the phases themselves are not
qualitatively changed by the quenched disorder. In the present subsection we will briefly
mention, without any pretense of completeness, other examples of quantum rare region
effects.

One-dimensional Heisenberg quantum antiferromagnets do not display long-range order even
at zero temperature. The behavior of the clean chain depends on the spin value
\cite{Haldane83a,Haldane83b}: For integer spin, there is a Haldane gap for magnetic
excitations and short-range magnetic order. In the case of half-integer spin, there is
quasi long-range spin order with algebraically decaying correlations and no gap. This
suggests that the half-integer chains are more susceptible towards weak disorder. Indeed,
the ground state of the random antiferromagnetic spin-$1/2$ chain is a random-singlet
state, that is controlled by an infinite-disorder fixed point \cite{Fisher94}. Numerical
simulations mostly agree with this prediction
\cite{HeneliusGirvin98,IgloiJuhaszRieger00}. In contrast, the dimerized spin-$1/2$ chain
has a gap in the absence of quenched disorder. As a result, the dimerized chain is stable
against disorder \cite{HYBG96} and the spin correlations remain short-ranged (random
dimer phase). For strong disorder or weak dimerization, the random dimer phase is
gapless, with a susceptibility that diverges as some nonuniversal inverse power of
temperature. This phase can be understood as a Griffiths phase belonging to the critical
point at zero dimerization.

The clean antiferromagnetic spin-1 chain has a Haldane gap. Consequently, it is stable
against weak disorder. Strong-disorder renormalization group studies
\cite{HymanYang97,MonthusGolinelliJolicoeur97,MonthusGolinelliJolicoeur98,SaguiaBoechatContinentino02}
predict a random-singlet phase for sufficiently strong disorder. It is separated from the
gaped phase by a gapless region of Griffiths singularities, the so-called gapless Haldane
phase. While the numerical simulations
\cite{Nishiyama98,Hida99,TodoKatoTakayama00,BergkvistHeneliusRosengren02,LCRI05} support
the overall structure of the phase diagram, they give somewhat conflicting results on
what phases can be reached for which bare disorder distribution and on the location of
the critical points. For the spin-$3/2$ chain, even the question of the phase diagram has
not yet been settled. A strong disorder renormalization group study by Refael et al.\
\cite{RefaelKehreinFisher02} gave two random singlet phases with different effective
spins ($S_{\rm eff}=3/2$ for strong disorder and $S_{\rm eff}=1/2$ for weak disorder). A
different implementation of the renormalization group by Saguia et al.\
\cite{SaguiaBoechatContinentino03} gives a random singlet phase only for strong disorder.
Weak disorder is an irrelevant perturbation. The latter result also agrees with recent
numerical simulations \cite{CLRI04}.
In addition to spin chains, disordered spin ladder systems also display random singlet,
Griffiths and other exotic phases \cite{YusufYang02,MLLRI02,YusufYang03a,YusufYang03b,
HoyosMiranda04a}.

Interesting quantum Griffiths effects have also been identified in inhomogeneously
diluted quantum antiferromagnets. As was discussed in section \ref{subsec:qrotors}, in a
(homogeneously) diluted two-dimensional Heisenberg quantum antiferromagnet, the ordered
phase persists to the classical percolation threshold of the lattice, and the quantum
phase transition that eventually destroys the long-range order is of percolation type. Yu
et al.\ \cite{YuRoscildeHaas05,YuRoscildeHaas06} introduced \emph{inhomogeneous} bond
dilution by using different occupation probabilities for different types of bonds
(``dimer'' and ``ladder'' bonds). In this way, they increase the quantum fluctuations and
the local tendency towards valence-bond solid order. If the two occupation probabilities
are sufficiently different, magnetic long-range order is destroyed before the classical
percolation threshold is reached. In the resulting quantum disordered phase, the
low-temperature susceptibility diverges algebraically with a non-universal exponent. This
phase can be understood as a quantum Griffiths phase.

Dobrosavljevic, Miranda and coworkers
\cite{MirandaDobrosavljevic01,AguiarMirandaDobrosavljevic03,TanaskovicMirandaDobrosavljevic04}
introduced the concept of an \emph{electronic} Griffiths phase, a phase that occurs in
strongly correlated electron systems in the vicinity of a disorder-driven metal insulator
transition. These systems possess localized magnetic moments, that are either present
from the outset (localized $f$-electrons in heavy fermion materials) or are
self-generated due to the interplay between disorder and correlations. The electronic
Griffiths phase is characterized by a power-law distribution of the Kondo temperatures
$T_K$ of these moments. Integrating over this distribution leads to power-law
singularities in many observables including specific heat and magnetic susceptibility.
More details can be found in the recent review \cite{MirandaDobrosavljevic05}.

Other rare region effects in quantum systems include the Lifshitz tails in the density of
states in doped semiconductors \cite{Lifshitz64,Lifshitz64b}. They are probably the
oldest known rare region effects but not directly related to a phase transition.
Analogous effects were also found in dirty superconductors
\cite{LarkinOvchinnikov71,BalatskyTrugman97,LamacraftSimons00,MeyerSimons01}. Motrunich
et al.\ \cite{MotrunichDamleHuse01} investigated Griffiths effects and quantum critical
points in dirty one-dimensional superconductors; and Vafek et al.\
\cite{VafekBeasleyKivelson05} discussed rare region effects at the flux-driven
superconductor-metal transition in ultrasmall cylinders (see also reference
\cite{LZRRCW01}). Finally,  Mildenberger et al.\ \cite{MENMD06} identified a Griffiths
phase in the thermal quantum Hall effect.

%%%%%%%%%%%%%%%%%%%%%%%%%%%%%%%%%%%%%%%%%%%%%%%%%%%%%%%%%%%%%%%%%%%%%%%%%%%%%%%%%%%%%%%%%%%%
\section{Nonequilibrium phase transitions}
\label{sec:noneq}
%%%%%%%%%%%%%%%%%%%%%%%%%%%%%%%%%%%%%%%%%%%%%%%%%%%%%%%%%%%%%%%%%%%%%%%%%%%%%%%%%%%%%%%%%%%%

The nonequilibrium behavior of many-particle systems has attracted considerable attention
in recent years. Of particular interest are continuous phase transitions between
different nonequilibrium states. These transitions are characterized by large scale
fluctuations and collective behavior over large distances and times very similar to the
behavior at equilibrium critical points. Examples of nonequilibrium transitions can be
found in population dynamics and epidemics, chemical reactions, growing surfaces, and in
granular flow and traffic jams (for recent reviews see, e.g., references
\cite{SchmittmannZia95,MarroDickman99,Dickman97,ChopardDroz_book98,Hinrichsen00,Odor04,TauberHowardVollmayrLee05}).
In this section we demonstrate that quenched spatial disorder at such nonequilibrium
phase transitions leads to rare region effects that are analogous to those at equilibrium
classical and quantum phase transitions.

\subsection{Contact process}
\label{subsec:CP}

A prominent class of nonequilibrium phase transitions separates active fluctuating states
from inactive, absorbing states where fluctuations cease entirely. Recently, much effort
has been devoted to classifying possible universality classes of these absorbing state
phase transitions \cite{Hinrichsen00,Odor04}. The generic universality class is directed
percolation (DP) \cite{GrassbergerdelaTorre79}. According to a conjecture by Janssen and
Grassberger \cite{Janssen81,Grassberger82}, all absorbing state transitions with a scalar
order parameter, short-range interactions, and no extra symmetries or conservation laws
belong to this class. Examples include transitions in catalytic reactions
\cite{ZiffGulariBarshad86}, interface growth \cite{TangLeschhorn92}, or turbulence
\cite{Pomeau86}. In the presence of conservation laws or additional symmetries, other
universality classes can occur, e.g., the parity conserving class
\cite{TakayasuTretyakov92,ZhongAvraham95,CardyTauber96} or the $Z_2$-symmetric directed
percolation (DP2) class \cite{GrassbergerKrauseTwer84,KimPark94,Menyhard94,Hinrichsen97}.

The contact process \cite{HarrisTE74} is a prototypical system in the directed
percolation universality class. It can be interpreted, e.g., as a model for the spreading
of a disease. The contact process is defined on a $d$-dimensional hypercubic lattice.
Each lattice site $\mathbf{x}$ can be active (occupied by a particle) or inactive
(empty). In the course of the time evolution, active sites can infect their neighbors, or
they can spontaneously become inactive. Specifically, the dynamics is given by a
continuous-time Markov process during which particles are created at empty sites at a
rate $\lambda n/ (2d)$ where $n$ is the number of active nearest neighbor sites.
Particles are annihilated at rate $\mu$. The ratio of the rates $\lambda$ and $\mu$
controls the behavior of the system. Without loss of generality we will set the
annihilation rate $\mu$ to unity, unless noted otherwise.

For small birth (infection) rate $\lambda$, annihilation dominates, and the absorbing state without
any particles is the only steady state (inactive phase). For large $\lambda$,
there is a steady state with finite particle density (active phase).  The two phases are
separated by a nonequilibrium phase transition in the directed percolation universality class
at some critical value $\lambda_c^0$.
The central quantity in the contact process is the average density
of active sites at time $t$
\begin{equation}
\rho(t) = \frac 1 {L^d} \sum_{\mathbf{x}} \langle  n_\mathbf{x}(t) \rangle
\end{equation}
where $n_\mathbf{x}(t)$ is the particle number at site $\mathbf{x}$ and time $t$, $L$ is
the linear system size, and $\langle \ldots \rangle$ denotes the average over all
realizations of the Markov process. The longtime limit of this density (i.e., the steady
state density)
\begin{equation}
\rho_{\rm stat} = \lim_{t\to\infty} \rho(t)
\end{equation}
is the order parameter of the nonequilibrium phase transition.

Close to the nonequilibrium phase transition, the density shows scaling behavior very
similar to that of equilibrium transitions discussed in section \ref{subsec:scaling}.
Specifically, the order parameter
$\rho_{\rm stat}$ varies according to the power law
\begin{equation}
\rho_{\rm stat} \sim (\lambda-\lambda_c)^\beta \sim r^\beta
\end{equation}
where $r=(\lambda-\lambda_c)/\lambda_c$ is the dimensionless distance from the critical
point, and $\beta$ is the critical exponent of the particle density. In addition to the
average density, we also need to characterize the length and time scales of the density
fluctuations. Close to the transition, correlation length $\xi_\perp$ and correlation
time $\xi_t$ diverge as
\begin{equation}
\xi_\perp \sim |r|^{-\nu_{\perp}}~, \quad \xi_t \sim \xi_\perp^z,
\label{eq:powerlawscaling}
\end{equation}
i.e., the dynamical scaling is of power-law form with dynamical exponent
$z$.\footnote{Following the convention in the literature on the contact process, we
denote the spatial correlation length and its critical exponent with a subscript
$\perp$.} Consequently the scaling form of the density as a function of $r$, the time $t$
and the linear system size $L$ reads
\begin{equation}
\rho(r,t,L) = b^{-\beta/\nu_\perp} \rho(r b^{1/\nu_\perp},t b^{-z}, L b^{-1})~.
\end{equation}
Here, $b$ is an arbitrary dimensionless scaling factor.

Two important quantities arise from initial conditions consisting of a single active site
in an otherwise empty lattice. The survival probability $P_s$ describes the probability
that an active cluster survives when starting from such a single-site seed. For directed
percolation, the survival probability scales exactly like the density
\begin{equation}
P_s(r,t,L) = b^{-\beta/\nu_\perp} P_s(r b^{1/\nu_\perp},t b^{-z}, L b^{-1})~.
\end{equation}
Thus, for directed percolation, the three critical exponents $\beta$, $\nu_\perp$ and $z$
completely characterize the critical point.\footnote{At more general absorbing state
transitions, e.g., with several
    absorbing states, the survival probability scales with an exponent $\beta'$ which may
    be different from $\beta$ (see, e.g., \cite{Hinrichsen00})}
The pair connectedness function $C(\mathbf{x'},t',\mathbf{x},t)=\langle
n_{\mathbf{x'}}(t') \, n_{\mathbf{x}}(t) \rangle$ describes the probability that site
$\mathbf{x}'$ is active at time $t'$ when starting from an initial condition with a
single active site at $\mathbf{x}$ and time $t$. For a clean system, the pair
connectedness is translationally invariant in space and time. Thus, it only depends on
two arguments $C(\mathbf{x},t',\mathbf{x},t)=C(\mathbf{x'}-\mathbf{x},t-t')$. Because $C$
involves a product of two densities, its scale dimension is $2\beta/\nu_\perp$, and the
full scaling form reads (as long as hyperscaling is valid)
\begin{equation}
C(r,\mathbf{x},t,L) = b^{-2\beta/\nu_\perp} C(r b^{1/\nu_\perp}, \mathbf{x}b^{-1}, t
b^{-z}, L b^{-1})~.
\end{equation}
The total number of particles $N$ when starting from a single seed site can be obtained
by integrating the pair connectedness function  $C$ over all space. This leads to the
scaling form
\begin{equation}
N(r,t,L) = b^{-2\beta/\nu_\perp +d} N(r b^{1/\nu_\perp},t b^{-z}, L b^{-1})~.
\end{equation}

At the critical point, $r=0$, and in the thermodynamic limit, $L\to\infty$, the
above scaling relations lead to the following predictions for the time dependencies of
observables: The density and the survival probability asymptotically decay like
$\rho(t) \sim t^{-\delta}$ and $P_s(t) \sim t^{-\delta}$ with $\delta=\beta/(\nu_\perp z)$.
In contrast, the number of particles in a cluster starting from a single seed site
increases like $N(t) \sim t^\Theta$
where $\Theta=d/z - 2\beta/(\nu_\perp z)$ is the so-called critical initial slip
exponent. Highly precise estimates of the critical exponents for clean one-dimensional directed
percolation have been obtained by series expansions \cite{Jensen99}: $\beta=0.276486$,
$\nu_\perp=1.096854$, $z=1.580745$, $\delta=0.159464$, and $\Theta=0.313686$.

%%%%%%%%%%%%%%%%%%%%%%%%%%%%%%%%%%%%%%%%%%%%%%%%%%%%%%%%%%%%%%%%%%%%%%%%%%%%%%%%%%%%%%%%%%%%%%
\subsection{Contact process with uncorrelated disorder}
\label{subsec:CP_dis}

Quenched spatial disorder can be introduced into the contact process by making the birth
rate $\lambda$ or/and the death rate $\mu$ random functions of the lattice site $\mathbf{x}$.
In this subsection, we discuss point defects, i.e., spatially uncorrelated disorder.

The investigation of disorder effects on spreading transitions in the directed
percolation universality class actually has a long history, but a coherent picture has
been slow to emerge. According to the Harris criterion \cite{Harris74}, directed
percolation is unstable against weak disorder because the clean spatial correlation
length critical exponent $\nu_\perp$ violates the inequality $d\nu_\perp > 2$ in all
dimensions $d>4$. (The exponent values are $\nu_\perp \approx 1.097$ (1D), 0.73 (2D), and
0.58 (3D) \cite{Hinrichsen00}). A field-theoretic renormalization group study
\cite{Janssen97} confirmed the instability of the directed percolation critical fixed
point. Moreover, no new critical fixed point was found. Instead the renormalization group
displays runaway flow towards large disorder, indicating unconventional behavior. Early
Monte-Carlo simulations \cite{Noest86} showed significant changes in the critical
exponents while later studies \cite{MoreiraDickman96,DickmanMoreira98} of the
two-dimensional contact process with dilution found logarithmically slow dynamics in
violation of power-law scaling. In addition, slow dynamics was found in whole parameter
regions in the vicinity of the phase transition
\cite{Noest86,Noest88,BramsonDurrettSchonmann91,WACH98,CafieroGabrielliMunoz98}.

\subsubsection{Griffiths singularities: Optimal fluctuation arguments.}
\label{subsubsec:DCP_OF}

A first qualitative understanding of disorder and rare region effects in the disordered
contact process can  be gained by optimal fluctuation arguments similar to that in
section \ref{subsec:OFT}. For definiteness, we assume the infection rate depends on the
source site only and its probability distribution is binary,
\begin{equation}
P[\lambda({\mathbf x})] = (1-p)\, \delta[\lambda({\mathbf x})-\lambda] + p\,
\delta[\lambda({\mathbf x}) - c\lambda]~. \label{eq:impdist}
\end{equation}
The impurity density $p$ and their strength $c$ are constants between 0 and 1. The
impurities locally \emph{reduce} the birth rate; the nonequilibrium transition will
therefore occur at a value $\lambda_c$ that is larger than the clean critical birth rate
$\lambda_c^0$.

The inactive phase, $\lambda<\lambda_c$, can be divided into two regions. For birth rates
below the clean critical point, $\lambda<\lambda_c^0$, the behavior is conventional
because non of the rare regions are locally in the active phase. The system approaches
the absorbing state exponentially fast in time. The decay time increases with $\lambda$
and diverges as $|\lambda-\lambda_c^0|^{-z\nu_\perp}$ where $z$ and $\nu_\perp$ are the
exponents of the clean critical point \cite{Bray88,DickisonVojta05}.
In contrast, in the Griffiths region, i.e., for birth rates between the clean and the dirty critical points,
$\lambda_c^0 < \lambda < \lambda_c$, rare spatial regions devoid of impurities are locally in the
active phase. Because they are of finite size, they cannot support a non-zero steady
state density but their decay is very slow because it requires a rare, exceptionally
large density fluctuation.

Following Noest \cite{Noest86,Noest88}, the contribution of these rare regions to the
time evolution of the total density can be estimated as follows. The probability $w$ for
finding a rare region of linear size $L_{\rm RR}$ devoid of impurities is given by
$w(L_{\rm RR}) \sim \exp( -\tilde p L_{\rm RR}^d)$ where $\tilde p= - \ln(1-p)$ is a
nonuniversal constant. The long-time decay of the density is dominated by these rare
regions. To exponential accuracy, the rare region contribution to the density can be
written as
\begin{equation}
\rho(t) \sim \int dL_{\rm RR} ~L_{\rm RR}^d ~w(L_{\rm RR}) \exp\left[-t/\tau(L_{\rm
RR})\right] \label{eq:rrevo}
\end{equation}
where $\tau(L_{\rm RR})$ is the decay time of a rare region of size $L_{\rm RR}$. Let us
first discuss the behavior at the clean critical point, $\lambda_c^0$, i.e., at the
boundary between the conventional inactive phase and the Griffiths region. At this point,
the decay time of a single, impurity-free rare region of size $L_{\rm RR}$ scales as
$\tau(L_{\rm RR}) \sim L_{\rm RR}^z$ as follows from finite size scaling
\cite{Barber_review83}. Here, $z$ is the clean dynamical exponent. A saddle-point
analysis leads to a stretched exponential time dependence,
\begin{equation}
\ln \rho(t) \sim - \tilde{p}^{z/(d+z)}~ t^{d/(d+z)}~, \label{eq:stretched}
\end{equation}
in contrast to the simple exponential decay as for $\lambda<\lambda_c^0$.
Inside the Griffiths region, i.e., for $\lambda_c^0<\lambda<\lambda_c$, the decay time of
a single rare region depends exponentially on its volume $\tau(L_{\rm RR}) \sim \exp(a
L_{\rm RR}^d)$ because a coordinated fluctuation of the entire rare region is required to
take it to the absorbing state \cite{Noest86,Schonmann85}.  The constant $a$ vanishes at
the clean critical point $\lambda_c^0$ and increases with increasing $\lambda$. Repeating
the saddle point analysis of the integral (\ref{eq:rrevo}) for this case, we obtain a
power-law decay of the density
\begin{equation}
\rho(t) \sim  t^{-\tilde p/a}  = t^{-d/z'} \label{eq:griffithspower}
\end{equation}
where $z'=da/\tilde p$ is a customarily used nonuniversal dynamical exponent in the
Griffiths region.  Thus, the optimal fluctuation arguments predict strong power-law
Griffiths effects for the disordered contact process. The reason is the exponential
increase of the decay time $\tau(L_{\rm RR})$ with the size of the rare region that
compensates for their exponential rarity.

\subsubsection{Strong-disorder renormalization group.}
\label{subsubsec:CP_SDRG}

An important step towards understanding spatial disorder effects on the directed
percolation transition has recently been made by Hooyberghs et al.
\cite{HooyberghsIgloiVanderzande03,HooyberghsIgloiVanderzande04}. These authors used the
Hamiltonian formalism \cite{Alcaraz94} to map the one-dimensional disordered contact
process onto a random quantum spin chain. They then applied a version of the Ma-Dasgupta-Hu
strong-disorder renormalization group \cite{MaDasguptaHu79} discussed in section \ref{subsec:sdrg}
and showed that the transition is controlled by an infinite-randomness critical point,
at least for sufficiently strong disorder. The basic ingredients of this analysis can be
explained by directly discussing the birth and death rates without reference to the
equivalent quantum system.

Let us consider a one-dimensional contact process where both the annihilation rate $\mu_i$ at site $i$
and the infection rate $\lambda_{ij}$ between neighboring sites $i$ and $j$ are random
functions of the lattice sites with broad probability distributions.
 Following the strong-disorder renormalization group approach, we identify the
largest rate in the system $\Omega = \max(\mu_i,\lambda_{ij})$. If the largest
rate is an annihilation rate, say $\mu_2$, site 2 is almost always empty and can be decimated
out. However, ``virtual occupations'' of site 2 generate a renormalized infection rate between
sites 1 and 3 that can be estimated as follows: To propagate the infection from site 1 to site 3, first it has to
propagate from site 1 to site 2 which happens with rate $\lambda_{12}$. However, site 3
can only be reached, if the infection propagates from site 2 to site 3 \emph{before} site
2 decays. Since the probability distributions are broad, $\mu_2 \gg \lambda_{23}$, and
the probability for this to happen is $\lambda_{23}/\mu_2$. Thus, the
renormalized infection rate between sites 1 and 3 is
\be
\bar\lambda_{13} = \lambda_{12} \lambda_{23} /\mu_2~.
\label{eq:DCP_lambda}
\ee
If the largest rate is an infection rate, say $\lambda_{23}$, which connects sites 2 and
3, the two sites will (almost) always be in the same state, either both active or both
inactive. They can therefore be replaced by a new single site with twice the weight
(moment). To estimate the renormalized annihilation rate, we note that total
annihilation of the cluster can occur via two processes: (i) site 2 decays (with rate $\mu_2$),
then site 3 decays \emph{before} site 2 gets reinfected. Since $\lambda_{23} \gg \mu_3$, the
probability for this is $\mu_3/\lambda_{23}$. Thus
the total rate for process (i) is $\mu_2 \mu_3 /\lambda_{23}$. Process (ii) consists of site 3 decaying first,
followed by site 2 and makes an identical contribution to the rate. Thus, the
renormalized death rate of the cluster is
\be
\bar\mu = 2\mu_2 \mu_3/\lambda_{23}~.
\label{eq:DCP_mu}
\ee
Except for an extra factor of 2, the recursion relations (\ref{eq:DCP_lambda}) and (\ref{eq:DCP_mu})
are completely equivalent to the recursion relations (\ref{eq:sdrg_h}) and (\ref{eq:sdrg_J}) in the
strong disorder renormalization group analysis of the random transverse field Ising
chain. Consequently, Hooyberghs et al. \cite{HooyberghsIgloiVanderzande03,HooyberghsIgloiVanderzande04}
also found equivalent results. The critical point in the disordered one-dimensional
contact process is of infinite-randomness type, i.e., under renormalization, the relative
with of the probability distributions of $\mu_i$ and $\lambda_{ij}$ increase without
limit.

The scaling behavior at such an infinite randomness critical point is different from
that of the clean contact process. Most importantly, the dynamics is extremely slow.
The power-law
scaling (\ref{eq:powerlawscaling}) gets replaced by activated dynamical scaling
\begin{equation}
\ln(\xi_t) \sim \xi_\perp^\psi, \label{eq:activatedscaling}
\end{equation}
characterized by a new exponent $\psi$. This exponential relation between time and length
scales implies that the dynamical exponent $z$ is formally infinite. In contrast, the
static scaling behavior remains of power law type.

Moreover, at an infinite-randomness fixed point the probability distributions of
observables become extremely broad, so that averages are dominated by rare events such as
rare spatial regions with large infection rate. In such a situation, averages and typical
values of a quantity do not necessarily agree. Nonetheless, the scaling form of the
\emph{average} density  is obtained by simply
replacing the power-law scaling combination $t b^{-z}$ by the activated combination $\ln(t)
b^{-\psi}$ in the argument of the scaling function:
\begin{equation}
\rho(r,\ln(t),L) = b^{-\beta/\nu_\perp} \rho(r b^{1/\nu_\perp},\ln(t) b^{-\psi}, L
b^{-1})~. \label{eq:rho_activated}
\end{equation}
Analogously, the scaling forms of the average survival probability and the average number
of sites in a cluster starting from a single site are
\begin{eqnarray}
P_s(r,\ln(t),L) &=& b^{-\beta/\nu_\perp} P_s(r b^{1/\nu_\perp},\ln(t) b^{-\psi},L
b^{-1}) \label{eq:Ps_activated}\\ N(r,\ln(t),L) &=& b^{-2\beta/\nu_\perp +d} N(r
b^{1/\nu_\perp},\ln(t) b^{-\psi},L b^{-1}) ~~. \label{eq:N_activated}
\end{eqnarray}

These activated scaling forms lead to logarithmic time dependencies at the critical point
(in the thermodynamic limit). The average density and the survival probability
asymptotically decay like $\rho(t) \sim [\ln(t)]^{-\bar\delta}$ and
$P_s(t) \sim [\ln(t)]^{-\bar\delta}$
with $\bar\delta=\beta/(\nu_\perp \psi)$ while the average number of particles in a
cluster starting from a single seed site increases like
$N(t) \sim [\ln(t)]^{\bar\Theta}$
with $\bar\Theta=d/\psi-2\beta/(\nu_\perp \psi)$.
Within the strong disorder renormalization group approach the
critical exponents of the disordered one-dimensional contact process can be calculated
exactly. The numerical values are $\beta=0.38197$, $\nu_\perp=2$, $\psi=0.5$,
$\bar\delta=0.38197$, and $\bar\Theta=1.2360$. In spatial dimensions larger than one,
the renormalization group can only be implemented numerically. As was discussed in
section \ref{subsec:rtim_hd}, this has been done for the strong-disorder renormalization
group of the two-dimensional random transverse field Ising model \cite{MMHF00}.
Using the analogy between
the renormalization group of the Ising model and the contact process, Hooyberghs et al.
\cite{HooyberghsIgloiVanderzande04} therefore concluded that the disordered contact
process in two dimensions will also show an infinite randomness critical point.

In addition to the infinite-randomness behavior at the critical point, the
strong-disorder renormalization group also yields strong power-law rare region effects
in the Griffiths phase in agreement with the optimal fluctuation arguments presented
in section \ref{subsubsec:DCP_OF}.

\subsubsection{Computer simulations.}

Early Monte-Carlo simulations of the disordered contact process
\cite{Noest86,Noest88,MoreiraDickman96,DickmanMoreira98} gave somewhat inconclusive
results ranging from a modification of the critical exponents to logarithmical violations
of scaling.  This was partially due to the fact that a consistent theory was not
available at that time. Hooyberghs et al. \cite{HooyberghsIgloiVanderzande04} performed
both Monte-Carlo and density matrix renormalization group (DMRG) studies of rather small
systems ($L \le 24$). They concluded that the phase transition is indeed controlled by
the infinite randomness fixed point discussed in section \ref{subsubsec:CP_SDRG}, but
only if the initial (bare) disorder is large enough. For weaker initial disorder they
predict nonuniversal continuously varying exponents with either power-law or activated
dynamical scaling.

\begin{figure}[t]
\centerline{\includegraphics[width=7cm]{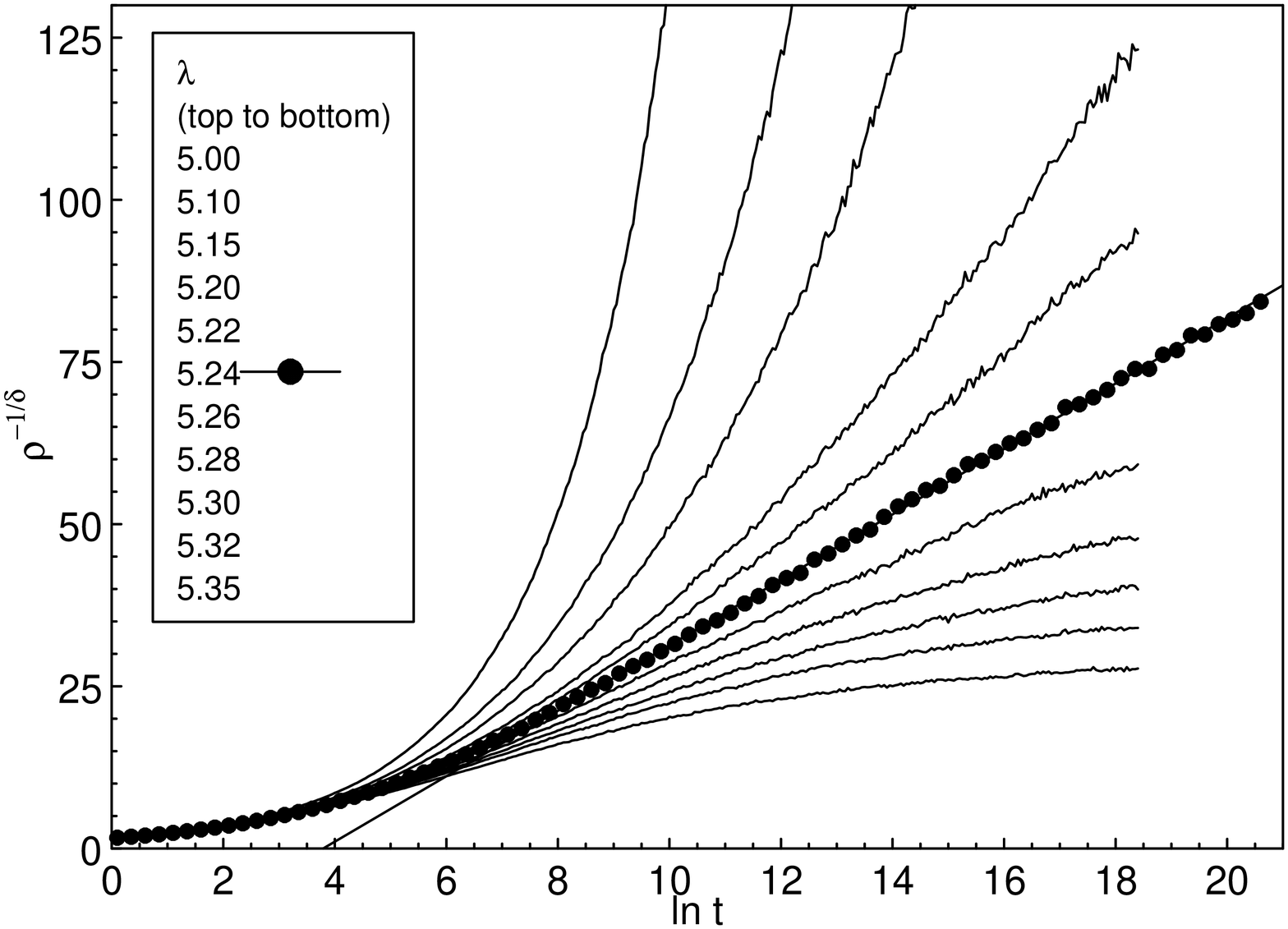}\includegraphics[width=7cm]{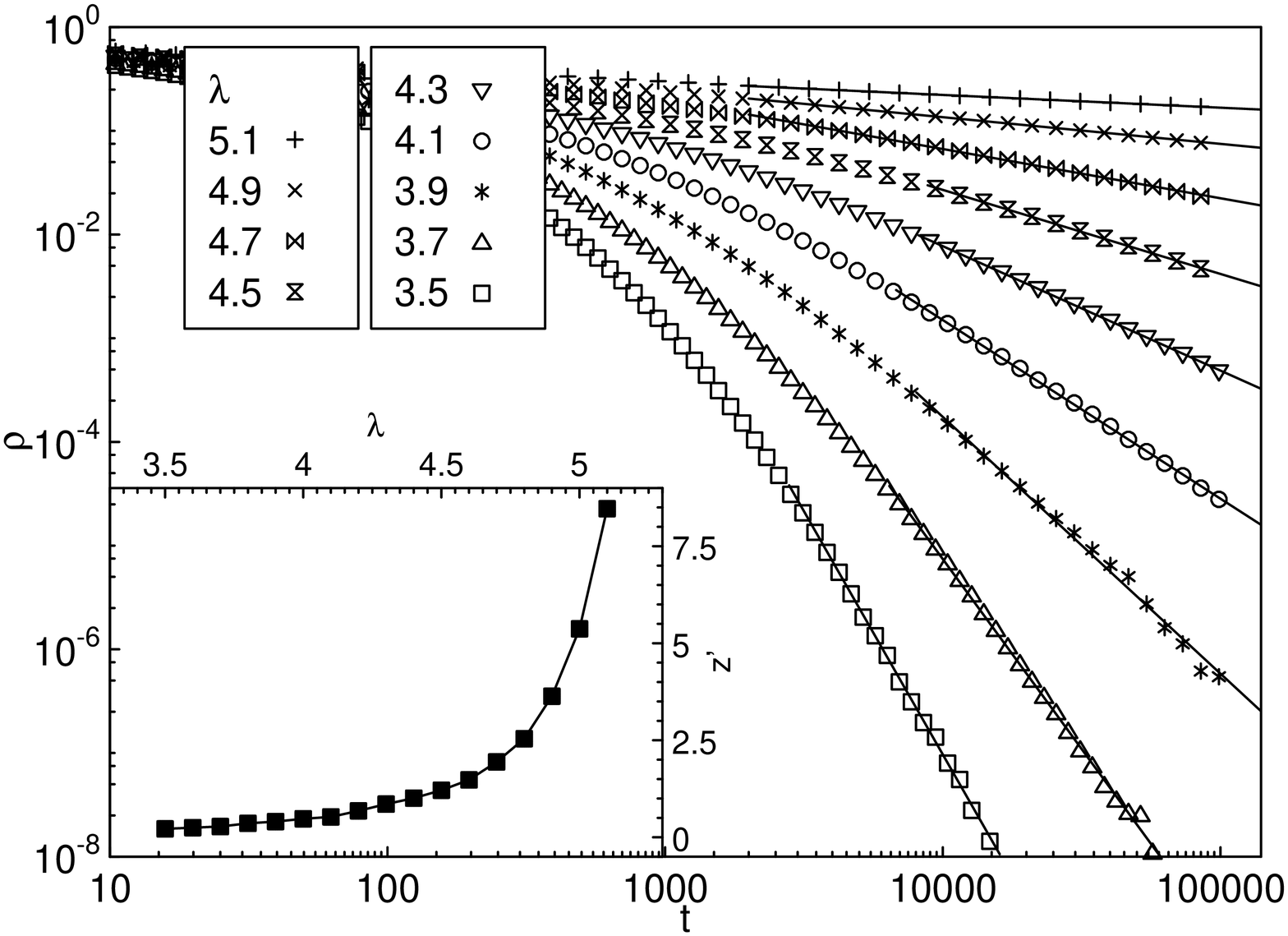}}
\caption{Left:$\rho^{-1/\bar\delta}$ vs. $\ln(t)$ for a system of $10^4$ sites with $p=0.3$
and $c=0.2$. The data are averages over 480 runs, each with a different disorder realization.
The filled circles mark the critical birth rate $\lambda_c=5.24$, and the
straight line is a fit of the long-time behavior to eq.\ (\ref{eq:logdecay}).
Right: Log-log plot of the density time evolution in the Griffiths region for systems
with $p=0.3, c=0.2$ and several birth rates $\lambda$. The system sizes are $10^7$ sites
for $\lambda=3.5, 3.7$ and $10^6$ sites for the other $\lambda$ values. The straight
lines are fits to the power law $\rho(t) \sim t^{-1/z'}$ predicted in eq.\
(\ref{eq:griffithspower}). Inset: Dynamical exponent $z'$ vs. birth rate $\lambda$ (from
\cite{VojtaDickison05}).} \label{fig:criticalp03c02}
\end{figure}

Recently, Vojta and Dickison \cite{VojtaDickison05} have performed a detailed large-scale
Monte-Carlo study of the one-dimensional contact process with binary disorder
distribution (\ref{eq:impdist}). They studied system sizes up to $L=10^7$ and very
long-times up to $t=10^9$ which is three orders of magnitude in $t$ longer than previous
simulations. A first set of calculations started from a full lattice and followed
the time evolution of the average density.  To test the strong-disorder renormalization group
theory, the time dependence of the density can be plotted according to the predicted
activated scaling law
\begin{equation}
\rho(t) \sim [\ln(t)]^{-\bar\delta}, \label{eq:logdecay}
\end{equation}
with $\bar\delta=0.38197$. The left panel of figure \ref{fig:criticalp03c02} shows the resulting graph for
a system of $10^4$ sites with $p=0.3$ and $c=0.2$.
The evolution of the density at the critical birthrate $\lambda=5.24$ follows eq.\
(\ref{eq:logdecay}) over almost six orders of magnitude in time. Therefore, the critical
point of the disordered contact process is indeed of infinite-randomness type. Analogous
calculation were performed for two other sets of parameters ($p=0.3$ with $c$ varying
from 0.2 to 0.8 as well as $c=0.2$ with $p$ varying from 0.2 to 0.5). For all parameters
(including the cases of weak disorder), the critical point is characterized by the
logarithmic density decay (\ref{eq:logdecay}) with a universal exponent
$\bar\delta=0.38197$.
In addition to the critical point, Vojta and Dickison \cite{VojtaDickison05}
also studied the Griffiths region between the
clean critical birthrate, $\lambda_c^0=3.298$ and the dirty critical birthrate
$\lambda_c=5.24$.
The right panel of figure \ref{fig:criticalp03c02} shows a double-logarithmic plot of the density time evolution
for birth rates $\lambda = 3.5 \ldots 5.1$ and $p=0.3, c=0.2$.
For all birth rates $\lambda$ shown, $\rho(t)$ follows  (\ref{eq:griffithspower}) over
several orders of magnitude in $\rho$  (except for the largest $\lambda$ where we could
observe the power law only over a smaller range in $\rho$ because the decay is too slow).
The nonuniversal dynamical exponent $z'$ can be obtained by fitting the long-time
asymptotics of the curves  to eq.\ (\ref{eq:griffithspower}).
The inset of the figure shows $z'$ as a function of the birth rate
$\lambda$. As predicted, $z'$ increases with increasing $\lambda$ throughout the
Griffiths region with an apparent divergence around $\lambda = \lambda_c= 5.24$.

These large-scale simulations of the one-dimensional contact process with point defects
thus provide strong evidence that the critical point is of infinite-randomness type with
universal critical exponents (even for weak bare disorder). The critical point is
accompanied by strong Griffiths singularities characterized by non-universal power-law
decay of the density. These results are in excellent agreement with the general
classification of dirty phase transitions suggested in section \ref{subsec:rrclass}.

\subsubsection{Contact process on a randomly diluted lattice.}

\begin{figure}
\centerline{\includegraphics[width=8cm]{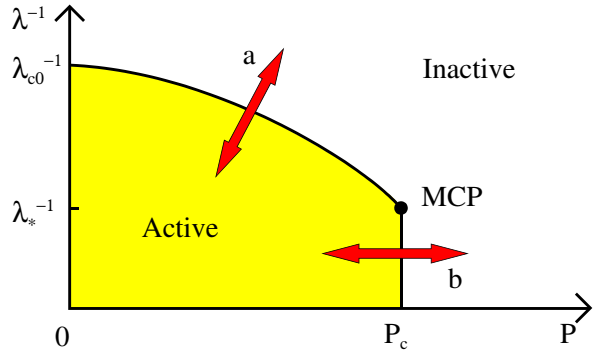}}
\caption{Schematic phase diagram of a site diluted contact process
 as function of impurity concentration $p$ and birth rate $\lambda$. There is a multicritical point at $p=p_c$ and
 $\lambda=\lambda_*$ (from \cite{VojtaLee06}). }
\label{fig:cp_pd}
\end{figure}

Quenched spatial disorder can also be introduced into the contact process by site or bond
dilution of the underlying lattice. In space dimensions $d \ge 2$, this leads to the
phase diagram shown in figure \ref{fig:cp_pd} which is very similar to that of a diluted
quantum magnet discussed in section \ref{subsec:rtim_hd}.
(In contrast, in $d=1$, any dilution immediately destroys the active phase.)
For small impurity concentrations below the percolation threshold of the lattice,
$p<p_c$, the active phase survives, but the critical birth rate increases with $p$ (to
compensate for the missing neighbors). Right at the percolation threshold the active
phase survives on the infinite percolation cluster for $\lambda>\lambda_*$. For $p>p_c$, no
active phase can exist because the lattice consists of disconnected clusters of finite
size that do not support a steady state density of active sites.

The contact process on a site-diluted lattice therefore has two nonequilibrium phase
transitions, separated by a multicritical point at $(p_c, \lambda_*)$. For $p<p_c$, the transition (marked by
``a'' in Fig.\ \ref{fig:cp_pd}) is expected to be in the universality class of the generic
disordered contact process discussed in the last subsections. In contrast, the phase transition
``b'' across the percolation threshold of the lattice is expected to show different critical
behavior.

Vojta and Lee \cite{VojtaLee06} showed that the interplay of geometric criticality and
dynamic fluctuations of the contact process does indeed lead to a novel universality
class for this nonequilibrium phase transition. Even though the transition is driven
entirely by the geometry of the lattice, the dynamical fluctuations of the contact
process enhance the singularities in all quantities involving dynamic correlations. By
combining results from classical percolation theory \cite{StaufferAharony_book91} with
the finite-size properties of a supercritical contact process, Vojta and Lee
\cite{VojtaLee06} determined the critical behavior analytically. They found activated
dynamical scaling $\ln \xi_t \sim \xi_\perp^\psi$ with the critical exponent $\psi$ being
equal to the fractal dimension of the critical percolation cluster, $\psi =D_f$. As a
result, the critical behavior is very similar to that of the infinite-randomness critical
point discussed above. Specifically, the long-time decay of the density $\rho$ of active
sites at $p=p_c$ is logarithmically slow, $\rho(t)  \sim [\ln(t/t_0)]^{-\bar\delta}$. The
exponent $\bar\delta=\beta_c/(\nu_c D_f)$ is determined by $D_f$ together with the order
parameter and correlation length exponents, $\beta_c$ and $\nu_c$, of the lattice
percolation transition. In contrast to the enhanced dynamical singularities, the
exponents of static quantities like the steady state density $\rho_{st}$  and the spatial
correlation length $\xi_\perp$ are identical the corresponding lattice percolation
exponents, $\rho_{st} (p) \sim |r|^{\beta_c}$ and $\xi_\perp \sim |r|^{\nu_c}$ where
$r=p-p_c$ measures the distance from the percolation threshold. Off criticality, i.e.,
away from the percolation threshold, there are strong rare region effects characterized
by a non-exponential density decay $\rho(t)  \sim (t/t_0)^{-d/z'}$ for $p>p_c$ and
$\rho(t)-\rho_{st} \sim e^{ -\left[(d/z'') \ln(t/t_0) \right]^{1-1/d}}$ for  $p<p_c$. The
nonuniversal exponents $z'$ and $z''$ diverge as $\xi_\perp^{D_f}$ for $p \to p_c$.

%%%%%%%%%%%%%%%%%%%%%%%%%%%%%%%%%%%%%%%%%%%%%%%%%%%%%%%%%%%%%%%%%%%%%%%%%%%%%%%%%%%%%%%%%%%%
\subsection{Contact process with extended defects}
\label{subsec:CP_extended}

So far, our discussion of disorder effects on the phase transition in the contact process
has been restricted to uncorrelated or short-range correlated disorder (i.e., point
defects). In this subsection we consider spatially extended (linear or planar) defects.
From the general arguments in section \ref{sec:rr}, we expect the disorder correlations
to enhance the impurity effects.

Vojta \cite{Vojta04} studied a contact process with quenched spatial disorder perfectly
correlated in $d_{\rm cor} >0$ space dimensions but uncorrelated in the remaining $d_{\rm
ran}=d-d_{\rm cor}$ space dimensions. In such a system, the rare regions are infinite in
$d_{\rm cor}$ dimensions but finite in $d_{\rm ran}$ dimensions. This is a crucial
difference from systems with uncorrelated disorder, where the rare regions are finite.
Because the contact process has a phase transition in all dimensions larger than zero, an
infinite rare region can undergo a real phase transition {\em independently} of the rest
of the system. Those rare regions that are locally in the ordered phase will have a true
nonzero stationary density, even if the bulk system is still in the inactive phase.
The resulting global phase transition is thus smeared by the same mechanism that smears
the equilibrium transition in an Ising model with planar defects discussed in section
\ref{subsec:planar} or the itinerant Ising magnet of section \ref{subsec:itinerant}. As
the birth rate $\lambda$ is increased, the order parameter develops very inhomogeneously
in space with different parts of the system becoming active independently at different
values of $\lambda$.

The leading behavior in the tail of the smeared transition can again be determined by
optimal fluctuation arguments \cite{Vojta04}. Here we summarize the results for a binary
probability distribution $P[\lambda({\mathbf x}_{\rm ran})] = (1-p)\,
\delta[\lambda({\mathbf x}_{\rm ran})-\lambda] + p\, \delta[\lambda({\mathbf x}_{\rm
ran}) - c\lambda]$ with constants $p$ and $c$ between zero and one. ${\mathbf x}_{\rm
ran}$ is the projection of the position vector ${\mathbf x}$ on the uncorrelated
directions. The probability $w$ for finding a rare region of linear size $L_{\rm RR}$
devoid of impurities is, up to pre-exponential factors, given by $w \sim \exp(-\tilde p
L_{\rm RR}^{d_{\rm ran}})$ with  $\tilde p = -\ln(1-p)$. According to finite-size scaling
\cite{Barber_review83}, such a region undergoes a true phase transition to the active
phase at $\lambda_c(L_{\rm RR}) =  \lambda_c^0 + A L_{\rm RR}^{-\phi}$ where $\phi$ is
the clean ($d$-dimensional) finite-size scaling shift exponent and $A$ is a constant. If
the total dimensionality $d=d_{\rm cor}+d_{\rm ran}<4$, hyperscaling is valid and $\phi =
1/\nu_\perp$ which we assume from now on.
The total (average) density $\rho$ at a certain $\lambda>\lambda_c^0$ is obtained by summing
over all active rare regions. In this way, the stationary density develops an
exponential tail,
\begin{equation}
\rho(\lambda) \sim \exp \left( -B (\lambda-\lambda_c^0)^{-d_{\rm ran}\nu_{\perp}}
\right)~, \label{eq:cpext_rho}
\end{equation}
reaching to the clean critical point $\lambda_c^0$. Here, $B= \tilde p A^{d_{\rm
ran}\nu_\perp}$ is a constant. Analogous arguments can be made for the survival
probability $P(\lambda$) of a single seed site. If the seed site is on an active rare
region it will survive with a probability that depends on $\lambda$ with a power law. If
is not on an active rare region, the seed will die. To exponential accuracy the survival
probability is thus also given by (\ref{eq:cpext_rho}).

We now turn to the dynamics in the tail of the smeared transition. The long-time behavior of the
density is dominated by the slow decay of the rare regions while the bulk contribution decays exponentially.
Proceeding in analogy to section \ref{subsec:dynsmeared}, Vojta \cite{Vojta04} showed
that the leading long-time decay of the density at the clean critical point, $\lambda=\lambda_c^0$,
 follows a stretched exponential,
\begin{equation}
\ln \rho(t) \sim -\tilde p^{z/(d_{\rm ran}+z)} t^{d_{\rm ran}/(d_{\rm ran}+z)}~
\label{eq:CP_stretched}
\end{equation}
where $z$ is the clean dynamical exponent of the $d$-dimensional bulk system. For
$\lambda>\lambda_c^0$, the behavior changes. At times below a crossover time $t_x\sim
(\lambda-\lambda_c^0)^{-(d_{\rm ran}+z)\nu_\perp}$ the density decay is still given by
the stretched exponential (\ref{eq:CP_stretched}). For times larger than $t_x$ the system
realizes that some of the rare regions are in the active phase and contribute to a finite
steady state density. The approach of the average density to this steady state value is
characterized by a nonuniversal power-law.
\begin{equation}
\rho(t) - \rho(\infty) \sim t^{-\zeta}~. \label{eq:CP_power}
\end{equation}

To test the predictions of the optimal fluctuation theory, Dickison and Vojta
\cite{DickisonVojta05} have performed large-scale Monte-Carlo simulations of a contact
process on a square lattice. The disorder consists of linear defects, i.e., the local
birthrate $\lambda(\mathbf{x})$ depends only on one of the coordinates and is perfectly
correlated in the other. The disorder distribution is binary.

\begin{figure}[t]
\centerline{\includegraphics[width=4.8cm,angle=-90]{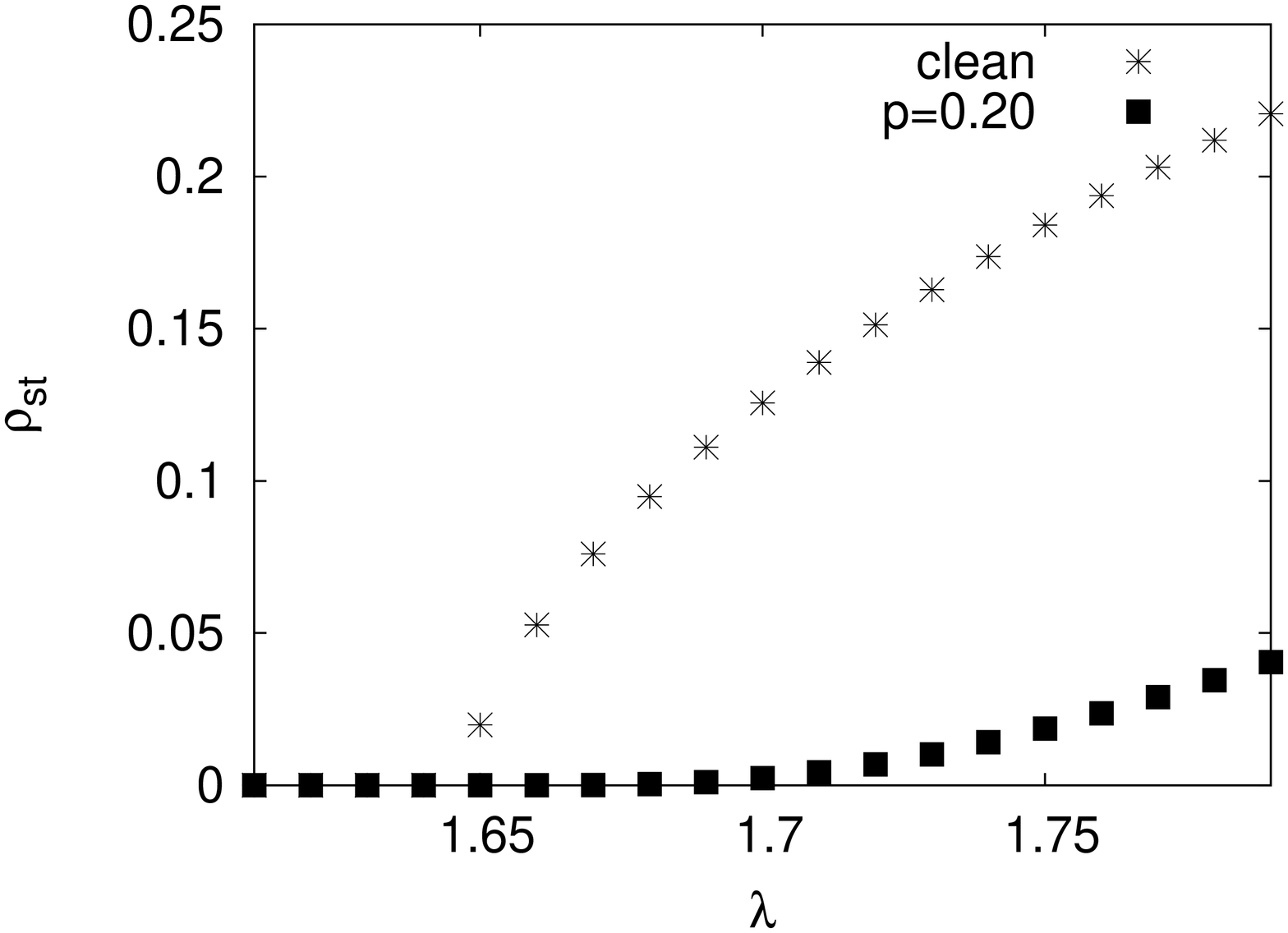}\includegraphics[width=4.8cm,angle=-90]{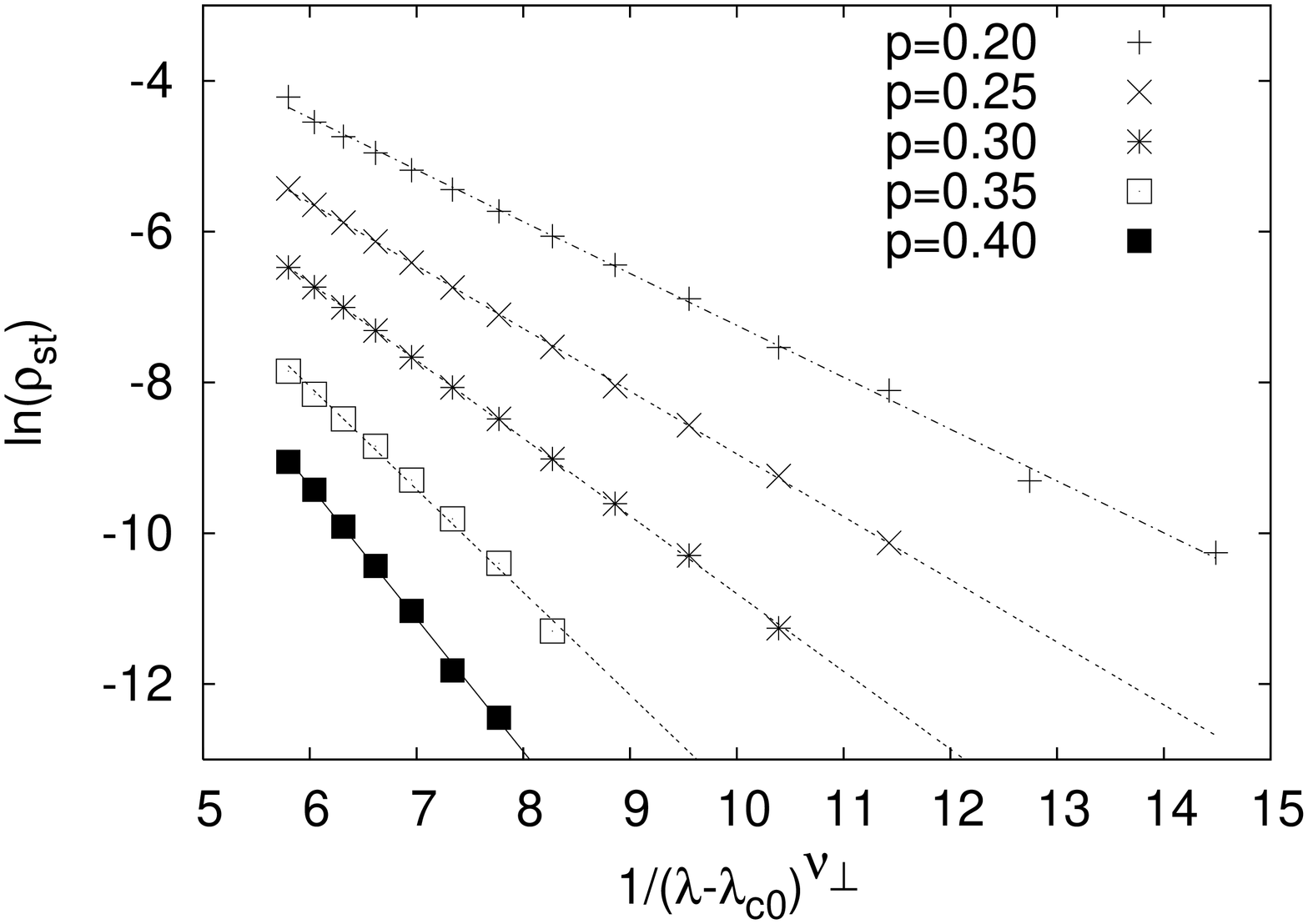}}
\caption{Left: Stationary density $\rho_{\rm st}$ as a function of birth rate $\lambda$
for a clean system and a system with impurity concentration $p=0.2$ and strength $c=0.2$.
The data are averages over 200 disorder realizations and the system size is $L=1000$.
Right: Logarithm of the stationary density $\rho_{\rm st}$  as a function of
$(\lambda-\lambda_c^0)^{-\nu_\perp}=(\lambda-\lambda_c^0)^{-0.734}$ for several impurity
concentrations $p$ and $L=3000$. The straight lines are fits to eq.\ (\ref{eq:cpext_rho})
(from \cite{DickisonVojta05}).} \label{fig:cpext_density}
\end{figure}

The left panel of figure \ref{fig:cpext_density} shows a comparison of the stationary density $\rho_{\rm st}$ as a
function of $\lambda$ between the clean system and a dirty system with $p=0.2, c=0.2$.
The clean system ($p=0$) has a sharp phase transition with a power-law singularity of the
density, $\rho_{\rm st} \sim (\lambda-\lambda_c^0)^\beta$ at the clean critical point
$\lambda_c^0\approx 1.65$ with $\beta\approx 0.58$ in agreement with the literature
\cite{MoreiraDickman96}. In contrast, in the dirty system, the density increases much more slowly
with $\lambda$ after crossing the clean critical point. This suggests either a critical
point with a very large exponent $\beta$ or exponential behavior.
To test the prediction (\ref{eq:cpext_rho}) of the optimal fluctuation theory, the right
panel of figure \ref{fig:cpext_density} shows $\ln \rho_{\rm st}$ as a function of
$(\lambda-\lambda_c^0)^{-\nu_\perp}$ for several impurity concentrations $p$. (The clean
two-dimensional spatial correlation length exponent is $\nu_\perp=0.734$
\cite{VoigtZiff97}.) The data show that the density tail is indeed a stretched
exponential, following the law (\ref{eq:cpext_rho}) over at least two orders of magnitude
in $\rho_{\rm st}$. Fits of the data to eq.\ (\ref{eq:cpext_rho}) can be used to
determine the decay constants $B$. As predicted by extremal statistics theory, the decay
constants depend linearly on $\tilde p = -\ln(1-p)$.

\begin{figure}[t]
\centerline{\includegraphics[width=6cm,angle=-90]{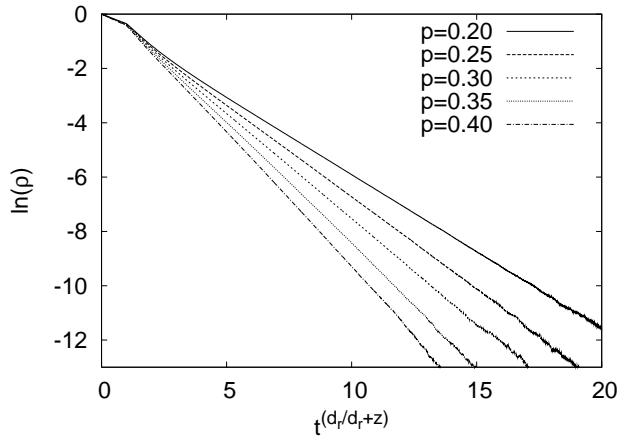}}
\caption{Logarithm of the density at the clean critical point $\lambda_c^0$ as a function
of $t^{1/(1+z)}=t^{0.362}$ for $c=0.2$ and several impurity concentrations ($p=0.2,\ldots,0.4$ from
top to bottom) and system size $L=3000$. The long-time behavior follows a stretched exponential $\ln
\rho = -E t^{0.362}$ (from \cite{DickisonVojta05}).} \label{fig:cp_stretched}
\end{figure}

In addition to the stationary density, Dickison and Vojta \cite{DickisonVojta05}
also studied its time evolution in the tail of the smeared transition.
Figure \ref{fig:cp_stretched} shows the time dependence of the density right at the
clean critical point.
The data follow the predicted stretched exponential behavior (\ref{eq:CP_stretched}) with
$z=1.76$ \cite{VoigtZiff97} over more than three orders of magnitude in $\rho$.
Additional calculations have been performed for $\lambda>\lambda_c^0$. In this parameter
region, the stationary density is nonzero.  The approach of the density to this nonzero
stationary value follows a nonuniversal power-law as predicted in (\ref{eq:CP_power}).

\subsection{Other nonequilibrium systems}

While the influence of quenched spatial disorder on the directed percolation universality
class has been studied in great detail, other dirty absorbing state transitions have
received much less attention. Hooyberghs et al.\ \cite{HooyberghsIgloiVanderzande04}
studied disorder effects on the transition of a generalized contact process with two
absorbing states \cite{Hinrichsen97}. This model is in the DP2
\cite{GrassbergerKrauseTwer84} or parity conserving \cite{TakayasuTretyakov92}
universality class (the two coincide in one space dimension \cite{Hinrichsen00}).
According to the Harris criterion, the clean transition is unstable with respect to weak
disorder because the clean correlation length exponent is $\nu_\perp\approx 1.83$ (see,
e.g., \cite{Hinrichsen00}) violating the inequality $d\nu_\perp >2$. Hooyberghs et al.\
applied the strong-disorder renormalization group to a weakly disordered generalized
contact process with quenched spatial disorder. They found that the renormalization group
scheme does not work; the flow is \emph{not} towards larger disorder. They concluded that
the transition is unlikely to be controlled by an infinite disorder fixed point. Very
recently, Odor and Menyhard \cite{OdorMenyhard06} reported numerical simulations of a
cellular automaton in the parity conserving universality class. They found that weak
disorder does not change the critical exponents from their clean values. How this can be
reconciled with the violation of the Harris criterion remains a question for future
research.

Let us also briefly mention another important nonequilibrium model displaying rare region
effects, even though it is not directly related to an order-disorder phase transition,
viz. Sinai's model of diffusion with random local bias \cite{Sinai82}. In the absence of
a bias, the mean-square displacement in this model grows logarithmically slowly in time,
while for small bias it grows with a nonuniversal power-law. Igloi, Turban and Rieger
\cite{IgloiTurban96,IgloiRieger98} pointed out a connection between the Ising model and
directed walks. Later, Igloi and Rieger \cite{IgloiRieger98b} established an exact
relation between the statistical properties of anomalous diffusion and the properties of
random quantum spin chains. At about the same time Fisher, Le Doussal and Monthus used a
version of the strong-disorder renormalization group to determine the exact asymptotic
long-time behavior of the Sinay model
\cite{FisherLeDoussalMonthus98,LeDoussalMonthusFisher99}. These papers started a stream
of work on this topic. A detailed discussion is beyond the scape of this review, for more
details see, e.g., reference \cite{IgloiMonthus05}.

%%%%%%%%%%%%%%%%%%%%%%%%%%%%%%%%%%%%%%%%%%%%%%%%%%%%%%%%%%%%%%%%%%%%%%%%%%%%%%%%%%%%%%%%%%%%
\section{Discussion and conclusions}
\label{sec:conclusions}
%%%%%%%%%%%%%%%%%%%%%%%%%%%%%%%%%%%%%%%%%%%%%%%%%%%%%%%%%%%%%%%%%%%%%%%%%%%%%%%%%%%%%%%%%%%%

In this section we summarize the main theoretical ideas presented in the review. To this
end, we organize the examples presented in sections \ref{sec:classical},
\ref{sec:quantum}, and \ref{sec:noneq} according to the proposed classification of rare
region effects, and we relate our results to those of other methods. We also discuss the
generality of the rare region classification, its limitations, the experimental
situation, and some related questions.

\subsection{Summary: Classification of rare region effects}

 This topical review has been
devoted to rare region effects at order-disorder transitions between conventional phases.
In section \ref{subsec:rrclass} we have introduced a classification of these effects
according to the effective dimensionality $d_{\rm RR}$ of the rare regions
\cite{VojtaSchmalian05}. Its basic idea is as follows: The probability $w$ for finding a
rare region of volume $V_{\rm RR}$ (in the random directions) decreases exponentially
with its size, $w \sim \exp(-pV_{\rm RR})$. This must be compared to how rapidly the
contribution of a \emph{single} region to thermodynamic quantities increases with its
size. As was explained in section \ref{subsec:rrclass}, three cases can be distinguished.

\emph{Class A:} If the effective dimensionality of the rare regions is below the lower
   critical dimension of the problem, $d_{\rm RR}<d_c^-$, the contribution of a rare region to
   thermodynamic quantities only increases like a power of its size. Therefore, Griffiths
   effects are expected to be exponentially weak and the dirty critical point will display
   power-law scaling.

\emph{Class B:} For $d_{\rm RR}=d_c^-$, rare region contributions increase exponentially
   with the size of the rare region resulting in strong power-law Griffiths singularities.
    The dirty critical point is expected to be of infinite randomness type with activated
    scaling.

\emph{Class C:} If $d_{\rm RR}>d_c^-$, an isolated rare region can undergo the phase
    transition independent of bulk and develop a static order parameter.
    This completely destroys the global phase transition by smearing.

In the following we relate the explicit results discussed in sections
\ref{sec:classical}, \ref{sec:quantum}, and \ref{sec:noneq} to this classification. We
start with the classical systems of section \ref{sec:classical}.
The classical randomly diluted Ising model introduced in section \ref{subsec:dil_ising}
as well as the classical random-$T_c$ Landau-Ginzburg-Wilson theory of section
\ref{subsec:randomTc_LGW} have spatially uncorrelated disorder. The rare regions are of
finite size in all directions and their effective dimensionality is $d_{\rm RR}=0$. Since
the lower critical dimension of the Ising model is $d_c^-=1$, these systems fall into
class A. In agreement with the classification, their Griffiths singularities are
exponentially weak, as was discussed in sections \ref{subsec:OFT} and
\ref{subsec:randomTc_LGW}. In contrast, the properties of the dirty critical point are
still controversial. On the one hand, high precision numerical simulations (see, e.g.,
\cite{BFMM98,CMMPV03}) yield conventional power-law critical behavior and no indications
of exotic properties, in agreement with early perturbative renormalization results
\cite{HarrisLubensky74,GrinsteinLuther76}. Moreover, the exponent values are in
\emph{quantitative} agreement with re-summed high-order perturbative renormalization
group calculations
\cite{Mayer89,FolkHolovatchYavorskii00,PelissettoVicari00,Varnashev00,PakhninSokolov00}.
On the other hand, Dotsenko et al. \cite{DHSS95,DotsenkoFeldman95} analyzed the stability
of the perturbative fixed points with respect to replica symmetry breaking and found them
unstable, suggesting unconventional behavior. However, two-loop field theoretical studies
\cite{PrudnikovPrudnikovFedorenko01,Fedorenko03} did not find such instabilities with
respect to replica-symmetry breaking. Clearly, more work is necessary to determine
whether the critical point in the Ising model with uncorrelated disorder indeed follows
all of the predictions of class A.

If the disorder is perfectly correlated in $d_{\rm cor}$ dimensions, the rare regions are
infinite in the corresponding directions but finite in the remaining $d_{\rm
ran}=d-d_{\rm cor}$ directions. Thus, there effective dimensionality is $d_{\rm
RR}=d_{\rm cor}$. The McCoy-Wu model discussed in section \ref{subsec:McCoyWu}, a
two-dimensional Ising model with linear defects, falls into class B because $d_{\rm
RR}=d_c^-=1$. In agreement with the classification, this model displays strong power-law
Griffiths singularities and an infinite-randomness critical point. Using Monte-Carlo
simulations \cite{PYRK98} and a numerical implementation of the strong-disorder
renormalization group \cite{MMHF00}, the same behavior was also found in two dimensions.
As was discussed in section \ref{subsec:McCoyWu}, perturbative renormalization group
calculations \cite{Dorogovtsev80,BoyanovskyCardy82,BlavatskaFerberHolovatch03} cannot
capture the nonperturbative rare region physics, they predict conventional behavior in
disagreement with the  exact results. In this context, an investigation of
replica-symmetry breaking in systems with extended defects would be of great interest.

In an Ising model with planar defects, the rare regions have dimension $d_{\rm RR}=2$.
This is above the lower critical dimension, and the model falls into class C. The
explicit results discussed in section \ref{subsec:planar} show that the global phase
transition is indeed smeared \cite{Vojta03b} as was also confirmed by large-scale Monte
Carlo simulations \cite{SknepnekVojta04}.

In the case of zero-temperature quantum phase transitions, the imaginary time direction
has to be included when calculating the effective rare region dimension $d_{\rm RR}$.
Therefore, the rare regions in the random transverse field Ising model introduced in
section \ref{subsec:rtim} are one-dimensional, $d_{\rm RR}=1$, and the model falls into
class B. The strong disorder renormalization group results
\cite{Fisher92,Fisher95,MMHF00} and Monte-Carlo simulations
\cite{YoungRieger96,Young97,PYRK98} discussed in sections \ref{subsec:rtim} to
\ref{subsec:rtim_hd} are in complete agreement with this classification: The critical
point is of infinite-randomness type with activated scaling. It is accompanied by strong
power-law Griffiths singularities.

At the quantum phase transition of the diluted bilayer Heisenberg quantum antiferromagnet
discussed in section \ref{subsec:qrotors}, the effective dimensionality of the rare
regions in $d_{\rm RR}=1$, too. However, the lower critical dimension of the Heisenberg
universality class is $d_c^-=2$. Therefore, this system falls into class A. This agrees
with the results of large-scale Monte-Carlo simulations \cite{SknepnekVojtaVojta04} that
show a conventional critical point with power-law scaling and universal exponents.

In systems with dissipative order parameter dynamics like the itinerant magnets of
section \ref{subsec:itinerant}, the behavior gets modified because the effective
interaction in imaginary time direction takes a long-ranged $1/\tau^2$ form. The rare
regions therefore correspond to one-dimensional classical systems systems with $1/\mathbf
x^2$ interaction. For Ising symmetry, the rare regions can undergo the phase transition
independently from the bulk, and the system is in class C.  This results in a smeared
global phase transition \cite{Vojta03a}. One-dimensional Heisenberg models with
$1/\mathbf x^2$ interaction are exactly at their lower critical dimension. Consequently,
disordered itinerant Heisenberg magnets fall into class B and show power-law quantum
Griffiths effects \cite{VojtaSchmalian05} (at least as long as long-range interactions
between the rare regions can be neglected \cite{DobrosavljevicMiranda05}). As is to be
expected, a straight-forward perturbative analysis \cite{KirkpatrickBelitz96} of the
transition in a disordered itinerant magnet does not predict any unconventional
properties. However, Narayanan et al.\ \cite{NVBK99a,NVBK99b} included the rare regions
in a perturbation renormalization group calculation in an approximate way. They found
that the rare regions destabilize the conventional fixed points but they could not
resolve the ultimate fate of the transition.

In section \ref{sec:noneq}, we have demonstrated that the rare region classification also
applies to nonequilibrium phase transitions in the directed percolation universality
class. In the contact process with uncorrelated disorder (point defects), the effective
dimension of the rare regions is $d_{\rm RR}=1$ including the time dimension. This is
identical to the lower critical dimension of directed percolation, and thus the system is
in class B. In agreement with the classification,  the strong-disorder renormalization
group as well as Monte-Carlo simulations yield an infinite-randomness critical point and
strong power-law Griffiths effects. If the defects are extended, the rare region
dimension is $d_{\rm RR}>1$ placing the system in class C. The classification thus
predicts a smeared transition in agreement with the explicit results in section
\ref{subsec:CP_extended}.

In summary, all results presented in sections \ref{sec:classical}, \ref{sec:quantum}, and
\ref{sec:noneq}  fit into the general classification of rare region effects at weakly
disordered order-disorder transitions, as introduced in section  \ref{subsec:rrclass}.

%%%%%%%%%%%%%%%%%%%%%%%%%%%%%%%%%%%%%%%%%%%%%%%%%%%%%%%%%%%%%%%%%%%%%%%%%%%%%%%%%%%%%%%%%%%%%%
\subsection{Experiments}
\label{subsec:experiments}

This subsection is devoted to experimental observations of the rare region effects
discussed in this review.

\subsubsection{Classical phase transitions.}
As explained in section \ref{sec:classical}, the thermodynamic rare region effects at
generic classical (i.e., thermal) phase transitions are exponentially weak. As was first
argued by Imry \cite{Imry77} they are probably unobservable in experiments, because the
probability for finding sufficiently large rare regions is extremely small even in a
macroscopic sample of about $10^{23}$ particles. Indeed, to the best of our knowledge,
the thermodynamic Griffiths singularities derived in sections \ref{subsec:dil_ising} to
\ref{subsec:randomTc_LGW} have never been experimentally verified.

In contrast to the thermodynamics, the long-time dynamical behavior of generic classical
systems is predicted to be dominated by the rare regions, see section
\ref{subsec:dynamics}. Lloyd and Mitchell \cite{LloydMitchell89} investigated the
dynamics above the Neel temperature $T_N=210$K of the dilute Heisenberg antiferromagnet
KMn$_{0.3}$Ni$_{0.7}$F$_3$ using inelastic neutron scattering. Below the transition
temperature $T_N^0=246$K of pure KNiF$_3$ they found very slow relaxation compatible with
a long-time tail of stretched exponential form. However, the data were not sufficient for
a quantitative confirmation of (\ref{eq:C(t)_Griffiths_Heisenberg}).

In recent years, there have been a number of systems in which experiments have revealed
unusual cluster or domain structures. Many of these phenomena have been related related
to Griffiths phases, at least at a qualitative level. Binek, Kleemann, and coworkers
\cite{BinekKleemann94,BABK95,BBRK96} found domain-like antiferromagnetic correlations in
the Ising-type antiferromagnets FeCl$_2$ and FeBr$_2$ when exposed to an axial field $H$
at temperatures below the Neel temperature on zero field. They attributed theses
structures to fluctuating distributions of demagnetizing fields and called the
corresponding phase a \emph{field-induced Griffiths phase}. However, it should be
emphasized that the character of this domain phase is fundamentally different from the
Griffiths phases discussed in this review. The latter are caused by \emph{static}
disorder fluctuations while the demagnetizing fields fluctuates in time. Later, Binek and
Kleemann also studied the diluted ferromagnet K$_2$Cu$_{1-x}$Zn$_x$F$_4$ and the diluted
antiferromagnet Fe$_{1-x}$Zn$_x$F$_4$ \cite{BinekKleemann95} by measuring the magnetic
susceptibility. They found deviations from the Curie-Weiss behavior of the corresponding
clean systems, but no relation to the exponential Griffiths singularities discussed in
sections \ref{subsec:dil_ising} to \ref{subsec:randomTc_LGW} was established. Similar
results were also reported for Rb$_2$Co$_{1-x}$Mg$_x$F$_4$
\cite{BinekKleemannBelanger98}. Colla et al.\ \cite{CWCC04} studied the dielectric
response of a disordered antiferroelectric,
Pb$_{0.97}$La$_{0.03}$(Zr$_{0.60}$Sn$_{0.30}$Ti$_{0.10}$)O$_3$. They found nonanalytic
contributions compatible with a Griffiths scenario in the nonlinear dynamic
susceptibility but not in the dependence on dc fields.

An important class of materials whose properties have been related to rare region effects
are the manganites. They are members of a broader class of compounds where the electronic
correlations play an important role. The manganites have rich phase diagrams exhibiting a
variety of phases with unusual spin, charge, lattice and orbital order; for details see,
e.g., the reviews \cite{SalamonJaime01,Dagotto_book03} and the references therein. They
display many exotic properties, most notably unexpectedly large magnetotransport
properties. Moreover, even in the best available crystals, manganites are intrinsically
inhomogeneous, a phenomenon sometimes called nanoscale phase separation. Salamon and
coworkers \cite{SalamonLinChun02,SalamonChun03} proposed a connection between the unusual
properties of doped LaMnO$_3$ and the Griffiths phase that arises when disorder
suppresses the magnetic transition (see also reference \cite{BMMMD01} for a theoretical
account). They found that the magnetic susceptibility in La$_{0.7}$Ca$_{0.3}$MnO$_3$
strongly increases \emph{before} the system reaches the Curie temperature. They
interpreted this upturn as a Griffiths singularity and analyzed it using the
susceptibility matrix approach of Bray \cite{Bray87}. However, an explicit verification
of the exponential classical Griffiths singularity was not found. Hartinger et al.\
\cite{HMLK05} and Deisenhofer et al.\ \cite{DBNH05} found inhomogeneous states in
La$_{1-x}$Sr$_{x}$MnO$_3$ that they attributed to Griffiths-like phases. The latter paper
also pointed out that the disorder in the manganites is likely to be correlated. Because
these disorder correlations enhance the rare region effects they may be responsible for
making the Griffiths singularities in the manganites accessible to experiment.

A Griffiths phase has also been predicted to occur in diluted magnetic semiconductors
above the ferromagnetic transition temperature \cite{GalitskiKaminskiDasSarma04}. Recent
measurements on Mn$_x$Ge$_{1-x}$ \cite{LSTW05} are compatible with a scenario in which
ferromagnetic long-range order occurs below some $T_c$  and short-range order occurs
between $T_c$ and $T_c^\ast$. (For $x=0.05$, $T_c \approx 12$K and $T_c^\ast \approx
112$K). However, the explicit form of the Griffiths singularities was not studied.

\subsubsection{Quantum phase transitions.}
While thermodynamic Griffiths singularities at generic classical transitions are very
weak, looking for rare region effects at zero-temperature quantum phase transitions is
a-priori more promising, because quantum Griffiths effects are generically stronger than
classical ones.

When the prototypical transverse-field Ising magnet LiHoF$_4$ is diluted by randomly
replacing magnetic Ho$^{3+}$ with nonmagnetic Y$^{3+}$, the long-range dipolar character
of the interaction leads to frustration. In a transverse magnetic field, the result is
therefore a quantum spin glass rather than a random quantum ferromagnet.
LiHo$_{1-x}$Y$_x$F$_4$ has been studied extensively since the seminal work of Wu et al.\
\cite{WERAR91,WBRA93}, but its complex properties are beyond the scope of this review.

 In recent years, there has been a lot of interest in the possibility of
quantum Griffiths effects in the so-called heavy-fermion materials. These are
intermetallic compounds of rare earth or actinide elements such as Cerium, Ytterbium or
Uranium. The localized magnetic moment of the $f$-electrons of these atoms hybridizes
with the delocalized conduction electrons generating quasi-particles with effective
masses of 100 to 1000 electron masses. Many of these systems can undergo magnetic quantum
phase transitions as a function of pressure or doping or magnetic field. Moreover, many
show non-Fermi liquid behavior, i.e., anomalous temperature dependencies of specific
heat, magnetic susceptibility and resistivity, as was first shown for
Y$_{0.8}$U$_{0.2}$Pd$_3$ in 1991 \cite{SMLG91}. Since then, similar behavior has been
found in a huge number of systems (for a detailed experimental review, see reference
\cite{Stewart01}).

Castro Neto, Castilla, and Jones \cite{CastroNetoCastillaJones98} suggested that the
anomalous temperature dependencies are a manifestation of power-law quantum Griffiths
effects in the vicinity of a magnetic quantum phase transition. However, this early
version of the theory neglected the damping of the magnetic modes due to the itinerant
electrons. As was discussed in section \ref{subsec:itinerant} of this review, this
damping qualitatively changes the low-temperature behavior. For Ising symmetry, which is
relevant for the heavy-Fermion systems as most of them have sizeable spin anisotropies,
the damping causes the rare regions to freeze leading to a smeared transition rather than
quantum Griffiths effects \cite{MillisMorrSchmalian01,MillisMorrSchmalian02,Vojta03a}.
Later, Castro Neto and Jones refined their theory by including the damping
\cite{CastroNetoJones00,CastroNetoJones05}. They found that for sufficiently weak damping
there will be a crossover temperature $T^\ast$. Power-law quantum Griffiths behavior can
occur in a transient temperature window \emph{above} $T^\ast$ but below a microscopic
cutoff scale (like the Fermi temperature).
For the experimentally important example of the heavy fermion materials, different
approaches disagree concerning the value of this crossover temperature. Castro Neto and
Jones \cite{CastroNetoJones00,CastroNetoJones05,CastroNetoJones05b} suggest that the
damping of the locally ordered rare regions is weak. Therefore, quantum Griffiths
behavior should occur in a broad temperature range and can be the source of the non-Fermi
liquid behavior observed in at least some of the heavy-Fermion systems. In contrast,
Millis, Morr and Schmalian \cite{MillisMorrSchmalian02,MillisMorrSchmalian05} argue that
the carrier-spin coupling in heavy fermions is large, restricting quantum Griffiths
effects to a narrow temperature window (if any). A complete resolution of this question
will likely come from experiments that give more direct access the the strength of the
damping term.

Recently, Johnston et al.\ \cite{JBZBSK05} observed slow non-exponential nuclear magnetic
relaxation in NMR experiments on the $d$-electron heavy fermion system LiV$_2$O$_4$ with
a small amount of magnetic defects. One possible explanation for this observation
consists in the defects inducing magnetic order in large but finite volumes around the
defect site. Johnston et al.\ \cite{JBZBSK05} described these magnetic clusters  by
Griffiths physics.

Let us also briefly mention some experiments on random quantum spin chains. Early
examples include the organic crystals NMP-TCNQ \cite{TheodorouCohen76}, Qn(TCNQ)$_2$
\cite{TippieClark81}, and copper-doped TMMC \cite{ESBA79}. The former two systems show
low-temperature susceptibilities proportional $T^{-\alpha}$ with $\alpha<1$ that have
been interpreted as evidence for rare region physics. The latter system, TMMC, realizes
an almost classical spin chain, presumably due to the large value of the spin involved
($S=5/2$ for Mn$^{2+}$). More recently, Payen et al.\ \cite{PJSBHA00} reported
susceptibility and magnetization for the doped $S=1$ Haldane spin system
Y$_2$BaNi$_{1-x}$Zn$_x$O$_5$. Below 4K they observed power-law temperature dependencies
with non-universal exponents suggesting a gapless phase due to the disorder. Power-law
temperature dependencies have also been reported for the $S=5/2$ spin chain system
MnMgB$_2$O$_5$ above a spin glass freezing temperature of about 600mK \cite{FGCR04}.
Masuda et al.\ \cite{MZUCP04} reported inelastic neutron scattering on the $S=1/2$ spin
chain compound BaCu$_2$(Si$_{0.5}$Ge$_{0.5}$)$_2$O$_7$. They found excellent agreement
with the results of a strong-disorder renormalization group \cite{MotrunichDamleHuse01}.

\subsubsection{Nonequilibrium phase transitions.}
 For nonequilibrium phase transitions, the experimental situation is even less
satisfactory. The (clean) directed percolation universality class is ubiquitous in theory
and simulations, but experimental verifications are strangely lacking
\cite{Hinrichsen00b}. To the best of our knowledge, the only (partial) verification so
far has been found in the spatio-temporal intermittency in ferrofluidic spikes
\cite{RuppRichterRehberg03}. While disorder is often suggested as a reason for the
striking absence of clean directed percolation scaling in at least some of the
experiments, a systematic search for the infinite-randomness behavior discussed in
section \ref{subsec:CP_dis} and the accompanying power law Griffiths effects has not been
performed.

\vspace*{5mm}

 In summary, even though rare region effects have been invoked in the
analysis of a variety of experiments, clearcut experimental verifications of Griffiths
singularities discussed in sections \ref{sec:classical}, \ref{sec:quantum}, and
\ref{sec:noneq} have not yet been achieved.

%%%%%%%%%%%%%%%%%%%%%%%%%%%%%%%%%%%%%%%%%%%%%%%%%%%%%%%%%%%%%%%%%%%%%%%%%%%%%%%%%%%%%%%%%%%%%%
\subsection{Conclusions}

Quenched spatial disorder can have dramatic consequences for the properties of continuous
phase transitions and critical points. An important role is played by rare strong
disorder fluctuations and the rare spatial regions where they occur. This topical review
has focused on transitions of the order-disorder type at which the quenched disorder does
not qualitatively change the properties of the phases but locally modifies the tendency
towards one or the other. For this random-$T_c$ or random-mass type disorder (arguably
the simplest type of quenched disorder), we have set up a general framework for
understanding rare region effects in the vicinity of the ``dirty phase transitions''.
According to the classification introduced in section \ref{subsec:rrclass}, the overall
phenomenology of the transition is determined by the effective dimensionality of the rare
regions.

Let us briefly point out the limitations of this approach. Even for order-disorder
transitions between conventional phases, the effects of long-range spatial interactions
are not understood. First, the very concept of a rare region may require modifications
when the order parameter of a locally ordered island generically develops power-law
tails. Second, long-range interactions can significantly increase the coupling between
the rare regions. As a result, the rare region effects in systems with long-range
interactions are likely to be more complex than those discussed here. This is
particularly important in metallic systems, were the presence of electronic soft modes
often self-generates long-range order parameter interactions
\cite{BelitzKirkpatrickVojta05}. For instance, it has been suggested
\cite{DobrosavljevicMiranda05} that the long-range RKKY interaction in metals further
enhances the rare region effects in itinerant Heisenberg magnets.

Often, the presence of quenched disorder leads to the formation of new phases, such as
the spin glass phases in classical magnets or the random-singlet states of certain
antiferromagnetic spin chains, as briefly mentioned in section \ref{subsec:other_qpt}.
These states, and phase transitions between them, are beyond the rare region
classification in its present form.

In recent years, there has been strong interest in ``exotic'' phases and phase
transitions that do not follow Landau's order parameter paradigm (for a recent
pedagogical overview, see, e.g., \cite{AletWalczakFisher05}. The influence of disorder on
such transitions has not been studied in any detail, and it is likely very different from
the behavior discussed here.

Finally, Griffiths singularities are normally associated with continuous phase
transitions, and we have also considered this case. Concerning first-order phase
transitions, it has been known for a long-time that quenched disorder generally weakens
the first-order nature. Building on earlier work by Imry and Wortis \cite{ImryWortis79},
Aizenman and Wehr \cite{AizenmanWehr89} proved, that in dimensions $d\le 2$, a
first-order transition turns into a continuous one upon introduction of arbitrarily weak
(random-mass type) disorder. However, for $d>2$, first order transitions can persist in
the presence of quenched disorder. Not much is known about pretransitional phenomena at
such first-order transitions (see, however, reference \cite{Timonin04}). Exploring this
in more detail would also be important from an experimental point of view, as some of the
transition mentioned in the last section (e.g., in the manganites) appear to be of first
order.

%%%%%%%%%%%%%%%%%%%%%%%%%%%%%%%%%%%%%%%%%%%%%%%%%%%%%%%%%%%%%%%%%%%%%%%%%%%%%%%%%%%%%%%%%%%%%%%
\section*{Acknowledgements}
%%%%%%%%%%%%%%%%%%%%%%%%%%%%%%%%%%%%%%%%%%%%%%%%%%%%%%%%%%%%%%%%%%%%%%%%%%%%%%%%%%%%%%%%%%%%%%%

This work would have been impossible without the contributions of many friends and
colleagues. In particular, I would like to thank my collaborators on some of the topics
discussed in this review: D. Belitz, M. Dickison, T.R. Kirkpatrick, M.Y. Lee, J.
Schmalian, R. Sknepnek, and M. Vojta. I have also benefitted from discussions with A.
Castro-Neto, P. Coleman, A. Chubukov, K. Damle, V. Dobrosavljevic, P. Goldbart, M.
Greven, S. Haas, J.A. Hoyos, A. Millis, D. Morr, M. Norman, P. Phillips, H. Rieger, A.
Sandvik, Q. Si, G. Steward, J. Toner, U. T\"auber, and P. Young. Finally, I'd like to
thank F. Igloi, J.A. Hoyos, H. Rieger, J. Schmalian, and P. Young for the critical
reading of the manuscript.

This work has been supported in part by the NSF under grant nos. DMR-0339147 and
PHY99-07949, by Research Corporation and by the University of Missouri Research Board.
Parts of this work have been performed at the Aspen Center for Physics and the Kavli
Institute for Theoretical Physics, Santa Barbara.

\section*{References}
\bibliographystyle{iop}
\bibliography{rareregions}

\end{document}